\documentclass[fleqn]{2023SCGE}
\usepackage{graphicx}

\usepackage{mwe,tikz,ulem}
\usepackage[percent]{overpic}
\usepackage{graphicx}
\usepackage{xspace}
\usepackage[numbers]{natbib}
\usepackage[utf8x]{inputenc}

\usepackage[gen]{eurosym}
\usepackage{microtype}
\usepackage{color}
\usepackage{txfonts}
\usepackage{makeidx}
\usepackage[columns=1]{idxlayout}

\usepackage{enumitem}
\usepackage{zhlipsum}

\newcommand{\extp}{\textsl{eXTP}\xspace}

\newcommand{\xmm}{\textsl{XMM-Newton}\xspace}
\newcommand{\lfa}{SFA\xspace}
\newcommand{\gpd}{PFA\xspace}
\newcommand{\cgsflux}{\ensuremath{\mathrm{erg}\,\mathrm{s}^{-1}\,\mathrm{cm}^{-2}}}

\newcommand{\arcmin}{{$^\prime$}}
\newcommand{\arcsec}{{$^{\prime\prime}$}}

\begin{document}
\ensubject{subject}

\ArticleType{Article}
\SpecialTopic{SPECIAL TOPIC: }
\Year{ }
\Month{ }
\Vol{ }
\No{ }
\DOI{ }
\ArtNo{ }

\title{Observatory Science with eXTP}{extp WG5}

\author[1,2]{Ping Zhou}{pingzhou@nju.edu.cn}
\author[3]{Jirong Mao}{jirongmao@ynao.ac.cn}
\author[4]{Liang Zhang}{zhangliang@ihep.ac.cn}
\author[5,6]{Alessandro Patruno}{patruno@ice.csic.es}
\author[7]{Enrico Bozzo}{Enrico.Bozzo@unige.ch}
\author[4]{Yanjun Xu}{}
\author[8]{\\ Andrea Santangelo}{}
\author[9]{Silvia Zane}{}
\author[4]{Shuang-Nan Zhang}{}
\author[4]{Hua Feng}{}
% ...

\address[1]{School of Astronomy and Space Science, Nanjing University, Nanjing, 210023, China}
\address[2]{\  \ Key Laboratory of Modern Astronomy and Astrophysics (Nanjing University), Ministry of Education, Nanjing, 210093, China}
\address[3]{\  \ Yunnan Observatories, Chinese Academy of Sciences, Kunming 650216, China}
\address[4]{\  \ State Key Laboratory of Particle Astrophysics, Institute of High Energy Physics, Chinese Academy of Sciences, Beijing 100049, China}
\address[5]{\  \ Institute of Space Sciences (ICE, CSIC), Campus UAB, Carrer de Can Magrans s/n, 08193 Barcelona, Spain}
\address[6]{\  \ Institut d'Estudis Espacials de Catalunya (IEEC), E-08034 Barcelona, Spain}
\address[7]{\  \ Department of Astronomy, University of Geneva, chemin d'Ecogia  16, 1290, Versoix, Switzerland}
\address[8]{\  \ Institut f\"{u}r Astronomie und Astrophysik, Kepler Center for Astro and Particle Physics, Eberhard Karls Universit\"{a}t Tübingen, Sand 1, 72076 T\"{u}bingen, Germany}
\address[9]{\  \ Mullard Space Science Laboratory, University College London, Holmbury St Mary, Dorking, Surrey RH5 6NT, UK}
% ...


\AuthorMark{Zhou P., Mao J.R., Zhang L., Patruno A., Bozzo E.}


\AuthorCitation{Zhou P., Mao J.R., Zhang L., Patruno A., Bozzo E., et al.}


\abstract{Scheduled for launch in 2030, the enhanced X-ray Timing and Polarization (eXTP) telescope is a Chinese space-based mission aimed at studying extreme conditions and phenomena in astrophysics. eXTP will feature three main payloads: Spectroscopy Focusing Array (SFA), Polarimetry Focusing Array (PFA), and a Wide-field Camera (W2C).  This white paper outlines observatory science, incorporating key scientific advances and instrumental changes since the publication of the previous white paper \citep{intzand19}. We will discuss perspectives of eXTP on the research domains of flare stars, supernova remnants, pulsar wind nebulae, cataclysmic variables, X-ray binaries, ultraluminous X-ray sources, AGN, and pulsar-based positioning and timekeeping.  }



\maketitle


\begin{multicols}{2}

\section{Introduction}\label{sec:Introduction}

The enhanced X-ray Timing and Polarimetry (eXTP) mission, led by China and scheduled to launch in 2030 \citep{WP-intru}, is a space observatory designed to study high-energy astrophysics and fundamental physics \citep{WP-intru}. Its primary objectives are to resolve critical questions of matter under extreme density \citep{WP-WG1}, gravity \citep{WP-WG2}, and magnetic fields \cite{WP-WG3}, as well as to advance time-domain and multi-messenger astronomy \citep{WP-WG4}.  Beyond its core focus, eXTP will also bring remarkable opportunities to observe various X-ray sources -- from stellar-sized objects to supermassive black holes -- while delivering scientific insights across multiple scientific categories.

In the new baseline design, the scientific payload of eXTP consists of three main instruments: the Spectroscopic Focusing Array (SFA), the Polarimetry Focusing Array (PFA) and the Wide-band and Wide-field Camera (W2C). The technical specifications and expected performance of these payloads can be found in \citep{WP-intru}

The SFA consists of five SFA-T (where T denotes Timing) X-ray focusing telescopes covering the energy range $0.5$--$10\, \mathrm{keV}$, featuring a total effective area of $\ge 2750\,{\rm cm^2}$ at $1.5\, \mathrm{keV}$ and $\ge 1670\,{\rm cm^2}$ at $6\,\mathrm{keV}$. The designed angular resolution of the SFA is $\le 1^\prime$ (HPD) with a $18^{\prime}$ field of view (FoV). The SFA-T are equipped with silicon-drift detectors (SDDs), which combine good spectral resolution ($\le $ 180~eV at 6 keV) with very short dead time and a high time resolution of $\le 10\,{\mu\mathrm {s}}$. They are therefore well-suited for studies of X-ray emitting compact objects at the shortest time scales. The SFA array also includes one unit of the SFA-I (where I signifies Imaging) telescope equipped with pn-CCD detectors (p-n junction charged coupled device), to enhance the capabilities of imaging and improve sensitivities to faint, point-like sources. The expected FoV of SFA-I is $18^\prime \times 18^\prime$. Therefore, the overall sensitivity of SFA could reach around $3.3\times 10^{-15}\,{\rm ergs\,cm^{-2}\,s^{-1}}$ for an exposure time of $1\, \mathrm{Ms}$. Since it is not excluded that the SFA might in the end include six SFA-T units, simulations presented here have taken this possibility into consideration. 

The PFA features three identical telescopes, with an angular resolution better than $30^{\prime\prime}$ (HPD) in a $9.8^{\prime} \times 9.8^{\prime}$ FoV, and a total effective area of $250\,{\mathrm{ cm^{2}}}$ at $3\, \mathrm{keV}$ (considering the detector efficiency). Polarization measurements are carried out by gas pixel detectors (GPDs) working at 2 -- 8\,keV with an expected energy resolution of $\le 1.8$~keV at 6\,keV and a time resolution better than $10\,{\mathrm {\mu{s}}}$ \cite{2001Natur.411..662C, 2003NIMPA.510..176B,2007NIMPA.579..853B,2013NIMPA.720..173B,eXTP2019}. The instrument reaches an expected minimum detectable polarization (MDP) at $99\%$ confidence level ($\mathrm{MDP}_{99}$) of about $2\%$ in $1\,\mathrm {Ms}$ for a milliCrab-like source.

The W2C is a secondary instrument of the science payload, featuring a coded mask camera with a FoV of approximately {1500} {square degrees} (Full-Width Zero Response, FWZR). The instrument achieves a sensitivity of $4\times 10^{-7}\,{\rm ergs\, cm^{-2}\,s^{-1}}$ (1\,s exposure) across the 30–600\,keV energy range, with an angular resolution of $20^{\prime}$, a position accuracy of $5'$, and an energy resolution better than $30\%$ at 60\,keV.

eXTP will be launched into a highly elliptical orbit with perigee and apogee altitudes of $\sim 5000$ km and $\ge 100$,000 km, respectively, while a low-Earth orbit remains a viable alternative. This high orbit allows continuous target monitoring observations, which are critical for studying variable X-ray sources that require uninterrupted long-duration (up to $\sim 30$~hr) observations. 

With its large collecting area, superb timing resolution, and imaging polarization capabilities, eXTP will be a powerful X-ray observatory for broad scientific purposes.
In addition to its primary goals (see \citep{WP-WG1, WP-WG2,WP-WG3,WP-WG4}), eXTP will tackle fundamental questions that span a wide range of scientific topics that shape observatory science. 

This white paper outlines the objectives of eXTP’s observatory science, incorporating key scientific advances since the previous white paper on observatory science \citep{intzand19}. It will show how eXTP will address key questions in various research areas, from stars (Section~\ref{sec:star}), supernova remnants (SNRs) and pulsar wind nebulae (PWNe; Section~\ref{sec:snr}), cataclysmic variables (CVs;  Section~\ref{sec:cv}), X-ray binaries (Section~\ref{sec:xrb}), ultraluminous X-ray sources (ULXs; Section~\ref{sec:ulx}), active galactic nuclei (AGN; Section~\ref{sec:agn}), to pulsar-based timing and time-keeping (Section~\ref{sec:pos}). In line with the current configuration of the mission, the revised goals emphasize the capabilities of the SFA and PFA, while incorporating observational (e.g., by IXPE and NICER) and theoretical progress over the past few years. Moreover, this paper excludes the discussion of time-domain and multi-messenger science objectives originally presented in \cite{intzand19}, as these topics now constitute a new science working group and will be presented in a separate white paper \citep{WP-WG4}.

\section{Flare stars} \label{sec:star}


\noindent All stars across the Hertzsprung-Russell diagram emit X-rays. 
For late-type stars (i.e., from late F to M types), their X-ray emission originates from the release of free magnetic energy in stellar coronae. 
Consequently, X-ray observations of stars provide critical insights into flare dynamics and stellar dynamos (through the X-ray activity--rotation relation; e.g., refs. \cite{Noyes1984ApJ279.763,Pizzolato2003A&A397.147,Wright2011ApJ743.48}). 
Additionally, these observations are invaluable for exploring star--planet interactions and determining the habitable zones around various types of stars. 

\begin{figure}[H]
\centering
\includegraphics[width=0.48\textwidth]{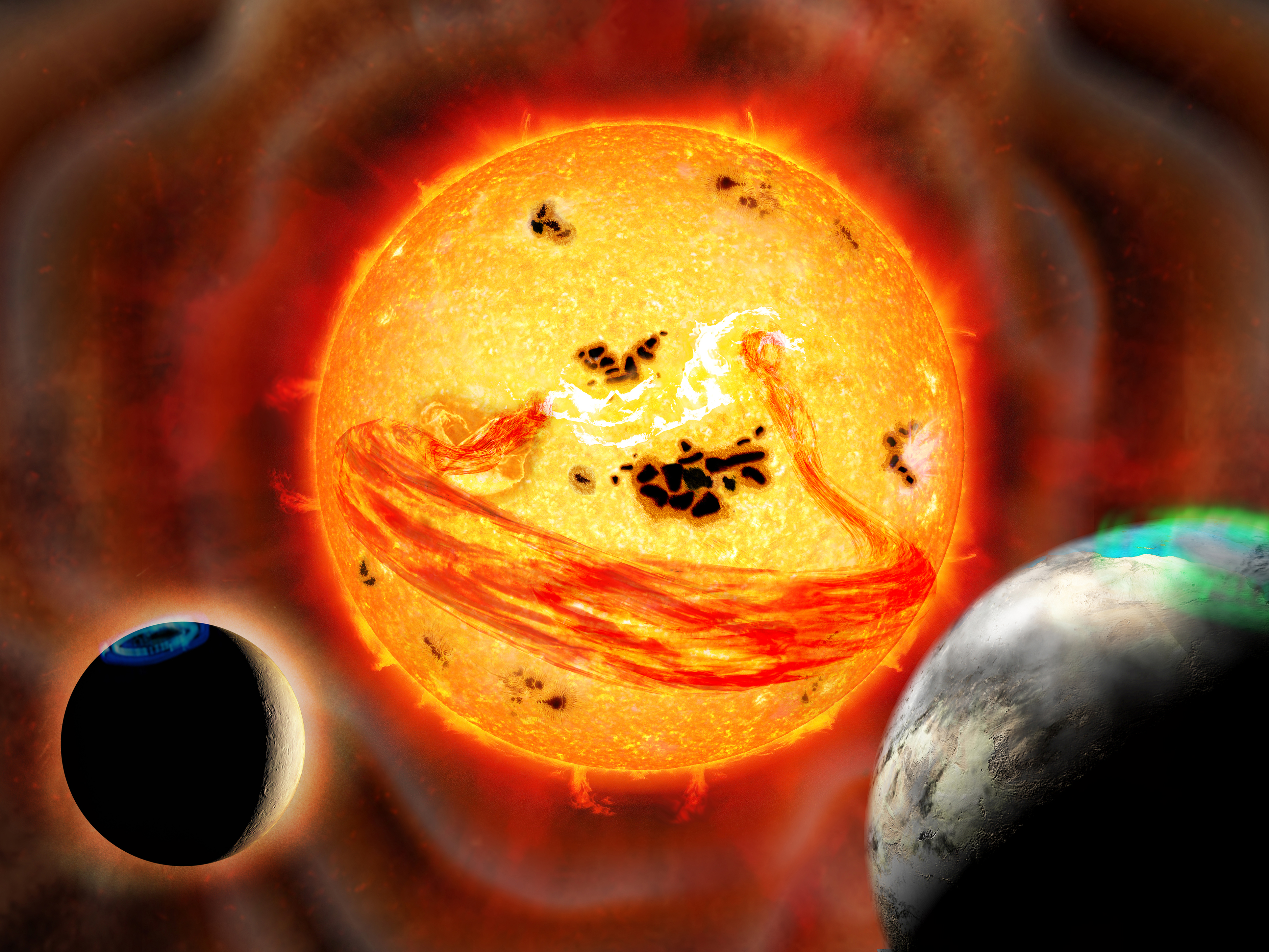}
\caption{Artist’s impression of a supermassive filament released by a superflare on EK Draconis approaching young planets. 
(Credit: National Astronomical Observatory of Japan) }
\label{Stars_Fig0}
\end{figure}

Among late-type stars, rapidly rotating M-dwarfs are thought to be particularly magnetically active and frequently produce superflares. 
The superflares explosively release magnetic energy with a total bolometric output of $\gtrsim 10^{33} \ \mathrm{erg}$ in a single event (e.g., refs. \cite{Maehara2012Nat485.478,Yang2017ApJ849.36,Gunther2020AJ159.60,Zhao2024ApJ961.130}), which increases stellar luminosity by orders of magnitude compared to quiescent levels, stronger than any solar flare ever detected (see Figure~\ref{Stars_Fig0} for an artist's impression). 

Stellar flares are also possibly accompanied by coronal mass ejections (CMEs),  which are large-scale eruptions of plasma and magnetic structures from solar and stellar atmospheres (see, e.g., ref. \cite{Chen2011LRSP8.1}). 
The occurrence of solar flares and CMEs often shows a significant positive correlation, indicating that they may originate from the same physical process (e.g., ref. \cite{Yashiro2009IAUS257.233,Cheng2010ApJ712.752}). 
Solar and stellar CMEs can notably impact the density, temperature, and magnetic field configuration in the corona. 
In addition, frequent or powerful CMEs may expose orbiting planets to high levels of radiation and charged particles, which potentially influence the atmospheres of the planets and their habitability (e.g., refs. \cite{Lammer2007AsBio7.185,Linsky2019LNP955}). 

However, unlike the Sun, although there have been reports of CME candidates on some active late-type stars, there has not yet been a fully confirmed CME event (see, e.g., ref. \cite{Namekata2022arXiv}). 
In addition to CME-induced effects like type II and moving type IV radio bursts,
direct observational signatures of stellar coronae, such as Doppler shifts (or asymmetries) of emission lines and signals of coronal dimmings, are also considered feasible and promising (e.g., refs. \cite{Namekata2022arXiv,Tian2022SC}). 
The soft X-ray band of eXTP can potentially capture signatures from stellar CMEs, and further constrain their properties, thereby facilitating more definitive detections of these events in the future. 

In practice, long-term monitoring of stellar flares can be effectively conducted using smaller ground-based or space-based telescopes, 
and mainly focusing on certain magnetically active stars. 
Given that stellar flares can typically last for $\sim 1-10$ hours in the soft X-ray band (e.g., ref. \cite{Zhao2024ApJ961.130}), there is a sufficient time window for triggering follow-up observations with eXTP once a flare has been detected. 
This strategy eliminates the need for eXTP to engage in continuous monitoring, thereby significantly reducing the observational cost for eXTP, while also ensuring a high likelihood of capturing a sufficient number of stellar flares and their accompanying phenomena. 
For each solar-type star, the estimated occurrence frequencies are $>0.64 \ \mathrm{h}$, $3.7 \ \mathrm{h}$, and $22 \ \mathrm{h}$ for stellar flares with soft X-ray energies of  $10^{28}$--$10^{29} \ \mathrm{erg}$, $10^{29}$--$10^{30} \ \mathrm{erg}$, and $10^{30}-10^{31} \ \mathrm{erg}$, respectively \cite{Zhao2024ApJ961.130}.
Flares with such energy ranges from nearby stars ($\lesssim 1 \ \mathrm{kpc}$) are easily detectable by eXTP according to the sensitivities of its payloads. 
eXTP will not respond to all stellar flare alerts but only observe those with important scientific values.
Accordingly, the threshold for triggering eXTP follow-up observations can be set 
to a high level (e.g., corresponding soft X-ray energy $\gtrsim 10^{30} \mathrm{erg}$) for the investigation of high-energy flares and accompanying mass motions. 
Nearby M-dwarfs like Proxima Centauri also offer intriguing targets for X-ray follow-up observations due to their high magnetic activity and relevance to habitability studies.

Soft X-ray observations serve as a powerful tool for examining the characteristics of the stellar corona and transient eruptions such as flares and CMEs. 
By combining these observations with physical models, we can gain deeper insights into the mechanisms driving these phenomena and the role of magnetic fields in shaping stellar activity. 
Below, we list a few important scientific questions about stars and stellar magnetic activities that will greatly benefit from the eXTP observations.

\begin{figure*}
\centering
\includegraphics[width=1.0\textwidth]{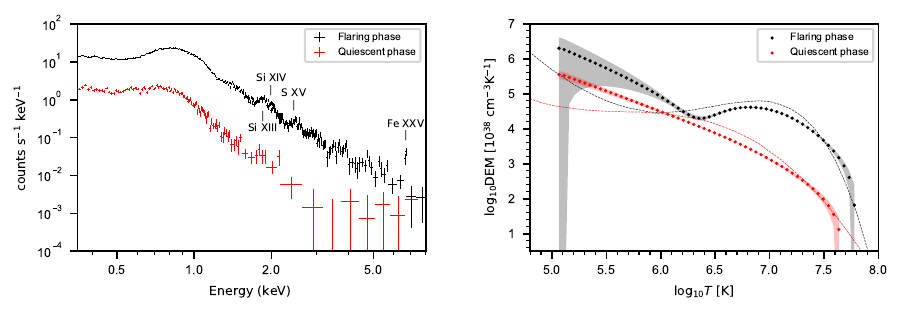}
\label{Stars_Fig1}
\caption{Left panel: Simulated 10-ks SFA spectra of Proxima Centauri at quiescent (red) and flaring (black) phases. 
Right panel: The fitting results (diamond points) and the $1\sigma$ confidence belt of the spectra. The flaring phase and quiescent phase are marked in black and red, respectively. Dashed lines show the models given by  \cite{Fuhrmeister2022AA663.A119}. The spectral fit employs a different DEM model (Regularization method) from the input model to investigate how the fit results vary with different input models.}
\end{figure*}

$\bullet$ {\it What are the properties of quiescent and flaring stellar coronae? }


\indent Due to its large effective area, SFA can collect more X-ray photons from nearby stars than other X-ray telescopes with the same exposure time. 
Based on the differential emission measure (DEM) of Proxima Centauri at both the quiescent and flaring phases obtained by \cite{Fuhrmeister2022AA663.A119}, we simulate 10-ks spectra for each (Figure~\ref{Stars_Fig1}a). 
The flaring spectrum shows a flux $\sim 10$ times higher than the quiescent one. 
The spectral fit with another DEM model can still distinguish these two phases by finding an excess of high-temperature components in the flaring phase. 
The enhancement of high-temperature components (usually with a temperature of $1-10 \ \mathrm{MK}$) in the DEM during a flare (see, e.g., ref. \cite{Coffaro2020A&A636.A49}) primarily reflects localized heating in the corona, which causes the low-temperature plasma to rapidly heat up, shifting the emission measure from lower to higher temperatures. 
In addition, high-temperature emission may also result from chromospheric material that is heated and evaporates into the corona (e.g., ref. \cite{Jin2020IAUS354.426}). 

X-ray observations, combined with the correlation between X-ray emissions and magnetic fields (e.g., ref. \cite{Kochukhov2020}), provide valuable insight into stellar magnetism. 
Additionally, the uneven distribution of X-ray sources across stellar surfaces can induce rotational modulations in the observed X-ray light curves, offering a powerful diagnostic tool for studying the coverage of active regions and flares (e.g., ref. \cite{Coffaro2020A&A636.A49}). 
Comparative analysis of the DEMs between quiescent and flaring phases enables the diagnosis of temperature and density in the flaring atmosphere, thereby providing essential constraints for hydrodynamic models of stellar flares. 

In addition, strong emission lines including Si{\,\small XIII}, Si{\,\small XIV}, S{\,\small XV}, and even Fe{\,\small XXV} can be observed in the current 10-ks spectrum, suggesting a probability of abundance measurement. 
Ar and Ca abundances can probably be further constrained with longer exposures.

\indent $\bullet$ {\it How do X-ray emissions from stellar flares correlate with emissions at other bands? }


\indent In history, simultaneous multi-wavelength observations play a key role in helping form the standard model of solar flares, in which magnetic reconnection releases magnetic energy and powers all kinds of flare radiation (e.g., ref. \cite{Su2013NatPh...9..489S,Kowalski2024LRSP21.1}). 
Based on the scaling laws combining the statistical results of stellar flares at optical (e.g., refs. \cite{Maehara2012Nat485.478,Shibayama2013ApJS209.5,Maehara2015EPS67.59,Yang2017ApJ849.36,Tu2020ApJ890.46,Tu2021ApJS253.35,2022AJ....164..213G}) and soft X-ray (e.g., refs. \cite{Getman2021ApJ916.32,Zhao2024ApJ961.130}) bands with solar flares, stellar flares are considered to have the same physical process as solar flares (e.g., ref. \cite{Benz2010ARAA48.241}). 
However, these deductions require validation by direct observations. 
Multi-wavelength monitoring of stellar flares offers a promising approach to this challenge. 
With its large effective area, the new-generation X-ray observatory eXTP has an outstanding photon-collecting ability that enables spectroscopic observations with high temporal resolution. 
With a combination of continuous and simultaneous photometric and spectroscopic observations, eXTP will provide a strong constraint towards understanding the physical mechanisms driving stellar flares. 

\indent There is also a possibility that flares on different types of stars have different mechanisms. 
Therefore, magnetically active stars of different types, e.g., solar-type stars, K- and M-dwarfs, pre-main-sequence stars, giants and subgiants, and interacting close binaries (such as RS CVn-type binaries) need to be included in the target library of the joint observational campaigns. 

\indent Furthermore, eXTP can also help constrain planetary habitability. 
The high-energy radiation from stellar flares can both promote life by triggering organic chemical reactions and hinder habitability by stripping atmospheres and damaging prebiotic molecules (e.g., Refs. \cite{2018SciA....4.3302R,2021NatAs...5..298C}). 
Therefore, stellar flare models established through multi-wavelength observations will provide important insights for more accurately evaluating planetary habitability.

\indent $\bullet$ {\it What are the observable X-ray signals of stellar CMEs? }


\indent 
Plasma ejection reduces the density of the source region, leading to a decreased emission of EUV and soft X-rays, which is known as coronal dimming (see, e.g., ref. \cite{Dissauer2019ApJ874.123}). 
With their excellent temporal resolution resulting from the large effective areas, the SFA instruments aboard eXTP make coronal dimming a promising method for stellar CME detection. 
For stars with a coronal temperature similar to that of the Sun, the characteristic lines of the corona are mostly at the EUV band. 
However, young or rapidly rotating stars tend to have higher coronal temperature and thus have stronger soft X-ray emissions (e.g., ref. \cite{Tian2022SC}), making it possible to detect coronal dimmings in the X-ray band of eXTP. 
Therefore, these stars are ideal targets for searching for CMEs via coronal dimming. 
The mass and velocity of CMEs are related to the amplitude and speed of the dimming of coronal emission, respectively (see, e.g., ref. \cite{Mason2016ApJ830.20}). 
The higher temporal resolution of SFA and SFA-I makes it possible to distinguish finer and weaker changes of flux in time, which enables eXTP to detect CME events less massive than former X-ray telescopes, and subsequently increases the detection rate within the same exposure time, since smaller CMEs are statistically more frequent (see, e.g., refs. \cite{Yashiro2009IAUS257.233,Zhao2024ApJ961.130}). 
Such a higher detection rate will significantly enlarge the sample set of potential stellar CME candidates. 

Moreover, the coronal dimming method for detecting CMEs requires an accurate diagnosis of the quiescent stellar corona, which serves as a basis for the calculation of the properties of potential CMEs. 
Thanks to the excellent photon-collecting ability of the SFA, the quiescent emission level from stellar coronae can be measured with shorter exposures. Consequently, the library of magnetically active stars with accurate quiescent coronal emission levels will also be significantly extended, thus establishing a robust basis for detecting stellar CMEs via the coronal dimming method. 

$\bullet$ {\it Are there any new X-ray activity probes of the activity--rotation relation? }

The X-ray activity--rotation relation provides important insights into stellar magnetic dynamo processes (e.g., refs. \cite{2003A&A...397..147P, 2011ApJ...743...48W}).
Typically, stellar X-ray activity is described as the ratio of X-ray luminosity to bolometric luminosity (i.e., $R_X = L_X/L_{\rm bol}$), while the Rossby number, defined as the ratio of the rotation period to the convective turnover time, is used to trace stellar rotation.
The relation is generally described by two regimes: a saturated region for rapidly rotating stars and a power-law decay region for slowly rotating stars (e.g., ref. \cite{2011ApJ...743...48W}). 
However, this standard picture has been challenged by many recent studies (see, e.g., refs. \cite{2014ApJ...794..144R,2018A&A...618A..48M,2019A&A...628A..41P,2024ApJS..273....8H}).

Considering that the activity proxy $R_X$ is only based on stellar X-ray luminosity, it is highly intriguing to explore the activity-rotation relation with new probes, such as coronal temperature, density, or emission measure ($EM=\int n_e n_i dl$), where $n_e$ and $n_i$ represent the electron and ion densities, respectively.
These parameters provide a more direct insight into the properties of the stellar corona.
\extp can establish an important X-ray spectral sample, with the SFA and SFA-I instruments, for nearby bright stars spanning different spectral types, rotation periods, ages, and metallicities. 
By using the new X-ray spectral-based activity proxies, we can build entirely novel activity--rotation relations, which may significantly advance our understanding of the relationship between the corona's structure and stellar physical properties.

\section{Supernova remnants and pulsar wind nebulae} \label{sec:snr}

Supernovae result from energetic explosions of massive stars or white dwarfs.
The ejected materials of supernovae interact with the interstellar medium and create supernova remnants (SNRs), each of which releases a huge amount of supernova mechanical energy (typically $10^{50}$--$10^{51}$~erg). The core-collapse explosion of a massive star leaves behind a neutron star or a black hole. 
In some cases, the result can be a rapidly rotating, highly magnetized neutron star that powers
a wind of charged particles, which interact with the surrounding magnetic fields and gas to form pulsar wind nebulae (PWNe). 
If the pulsar wind nebula (PWN) is young, it may still be embedded within an SNR. In such cases, the combined PWN/SNR system is classified as a subclass of SNRs, either as a plerion (or crab-like SNR) or, if a distinct shell is present, as a composite SNR \citep{vink20}.

SNRs and PWNe are among the brightest extended sources in the X-ray sky. They are important heating sources in galaxies and ideal laboratories for studying high-energy phenomena. Through the diffusive shock acceleration mechanism, fast shocks of SNRs accelerate CR particles---relativistic protons, atomic nuclei, and electrons and, for PWNe, also positrons.  CR electrons and positrons radiate synchrotron emission with a maximum emission at the characteristic energy 
$E_{\rm ch}=18.9 (B_\perp/100~{\rm \mu G})(E/100~{\rm TeV})^2$~keV \citep{ginzburg65},
where $E$ and $B_\perp$ are the electron energy and the magnetic field component perpendicular to the motion of electrons. 
The characteristic cooling timescale of relativistic electrons due to synchrotron radiation is highly dependent on the electron/positron energy $\tau\approx 12.5\left(B_\perp/100~{\rm \mu G}\right)^{-2}\left(E/100~{\rm  TeV}\right)~{\rm yr}$. As X-ray synchrotron typically comes from 10--100~TeV electrons, X-ray synchrotron radiation is only expected at or very near CR acceleration sites.

X-ray synchrotron emission is polarized and its polarization properties are powerful tools for studying the orientation and turbulence of magnetic fields.  Moreover, the measurements of the synchrotron spectrum help to study the CR acceleration and magnetic field strength in comparison with $\gamma$-ray measurements. X-ray emission of most SNRs is dominated by thermal emission, which is composed of atomic lines and continuum emission (bremsstrahlung, radiative recombination continuum, emission lines). The observations of X-ray spectra help to unveil the physical and chemical properties of the gas.

Thanks to the excellent imaging polarization capabilities and large effective area of eXTP, the SNRs and PWNe will embrace new opportunities. Below, we list a few important scientific questions of SNRs and PWNe that will greatly benefit from the eXTP observations.

\begin{table*}
  \caption{Summary of published X-ray polarization measurements of SNRs by IXPE. \label{tab:snrs}}

  \begin{tabular}{lrrllll}\hline\hline\noalign{\smallskip}
    SNR & Age (yr) & $B~({\rm \mu G})$$^a$ & PD  & Region size (\arcmin) &$B$ orientation & References \\
    \noalign{\smallskip}\hline\noalign{\smallskip}
    Cas A              &  $\sim$ 350           & $\sim 250$ & $\sim 2.5$\%/$\sim$15\% & 5/0.7 &radial &\cite{vink22}/\cite{mercuri25}\\
    Tycho's SNR        & 453                   & $\sim 200$ & $\sim 9$\% & 10 &radial & \cite{ferrazzoli23}\\
    SN 1006            & 1019                  & $\sim 80$  & $\sim 20$--40\% & 1.5--4 &radial &   \cite{zhou23,zhou25}          \\
    RX J1713.7-3946           & $\sim$1500            & $\sim 20$  & 26--30\%    & $\sim 2$ & tangential & \cite{ferrazzoli23}\\\
    RX J0852.0-4622/`` Vela jr''   & $\sim 2000$          &  $\sim 10$  & 10--20\%    & 1--5 &tangential & \cite{prokhorov24} \\
    \noalign{\smallskip}\hline\noalign{\smallskip}    
  \end{tabular}
  \\
{\noindent \footnotesize $^a$\ Based on reference \citep{helder12}, but also taking into account leptonic $\gamma$-ray modeling.}\\
\end{table*}

$\bullet$ {\it What are the mechanisms of magnetic turbulence and amplification in SNRs? }

Young SNRs can amplify the magnetic fields near the shocks by up to two orders of magnitude, as proved by measurements of the narrow widths of the nonthermal rims and the variability of X-ray structures \citep{vink03a,uchiyama07}.
Strong and turbulent magnetic fields near young SNRs play an important role in accelerating CRs to very-high energy regimes, as they can trap the charged particles near the shock for continuing acceleration. Magnetic amplification is therefore closely related to the maximum energy of the CR particles that SNRs can accelerate \citep{blasi13}, an important topic of CR research.

Despite a great deal of theoretical and observational progress, there are still many uncertainties regarding magnetic field amplification in and near SNRs. The compression of the interstellar medium alone only amplifies the magnetic fields by a factor of a few, so other processes are needed to explain the high amplification level. Several mechanisms have been proposed, such as amplification through CR-induced instabilities upstream of the shock \citep{bell04} and due to turbulence dynamo processes downstream of the shock for SNRs evolving inside inhomogeneous gas environment \citep{inoue13,xu17}. 
The upstream magnetic field turbulence is diminished by shock compression, which amplifies the tangential component. But downstream turbulence induced by density inhomogeneities ---- which may in itself be induced by magnetic field turbulence \cite{bykov24}---may stretch again the magnetic fields in the radial directions \cite{inoue13}.

The above mentioned mechanisms induce magnetic field turbulence, whereas the turbulence properties themselves are determined by the CR and interstellar/circumstellar medium properties and the distance from the shock front.
Magnetic field turbulence and amplification, on one hand, and CR acceleration, on the other hand, are, therefore, closely intertwined.

Polarization measurements provide a valuable way to study magnetic field turbulence and amplification since the polarization degree (PD) and polarization angle (PA) reflect the orderliness and orientation of the magnetic fields.
Since the polarization direction can vary across the region, it is crucial to resolve the polarization structures on the smallest possible scales to avoid the depolarization.

Imaging X-ray polarization observation has been available only in recent years since the launch of the Imaging X-ray Polarization Explorer \citep{weisskopf22}. Compared to radio polarization, X-ray polarization probes amplified magnetic fields immediately behind the shock, due to the aforementioned short cooling times of the very-high-energy CR electrons.

IXPE has conducted a small survey of young SNRs with bright nonthermal X-ray emission, including Cas A \citep{vink22}, Tycho \citep{ferrazzoli23}, SN 1006 \citep{zhou23,zhou25}, RX~J1713 \citep{ferrazzoli24}, Vela Jr.\ \citep{prokhorov24}.
These observations have revealed a great variety in polarization properties among young SNRs: (1) those expanding into a tenuous medium show a higher PD, and (2) the three youngest SNRs exhibit radially oriented magnetic fields, whereas the two older ones exhibit tangential fields (see Table~\ref{tab:snrs} for a summary). Furthermore, the X-ray polarization properties of SN~1006 are found to differ significantly from radio measurements, highlighting the distinct ability of X-ray polarimetry to probe amplied fields \citep{zhou25}. Together, these studies demonstrate the unique value of imaging X-ray polarimetry in investigating the magnetic fields of SNRs \citep{slane24}.

Due to the small effective area of IXPE, none of the SNRs have their X-ray polarization resolved above the $3\sigma$ level at the $30''$ resolution, despite the fact that IXPE observed near 1~Ms for each pointing of the SNRs.  Moreover, the polarization detections are biased toward brighter regions or have been obtained by summing azimuthal regions, assuming a circular symmetry. The latter results in a depolarization if the polarization pattern deviates from pure circular symmetry. A case in point is Cas A for which the PD is $<5$\%, but assuming a circular symmetric polarization pattern, while a more recent work provides 3--4$\sigma$ polarization detections showing that near the fainter forward shock region the PDs are 10\%--26\% \citep{mercuri25}.

Therefore, limited by the small effective area or insufficient exposure time, IXPE has not fully utilized its imaging capability for SNRs. This is a significant limitation, as good resolution is important for preventing depolarization and constraining magnetic field properties. With a 5-fold larger effective area, eXTP can collect polarization signals more quickly, allowing more efficient measurements of the X-ray polarization distribution.

\begin{figure*}
\center
\includegraphics[width=0.8\textwidth]{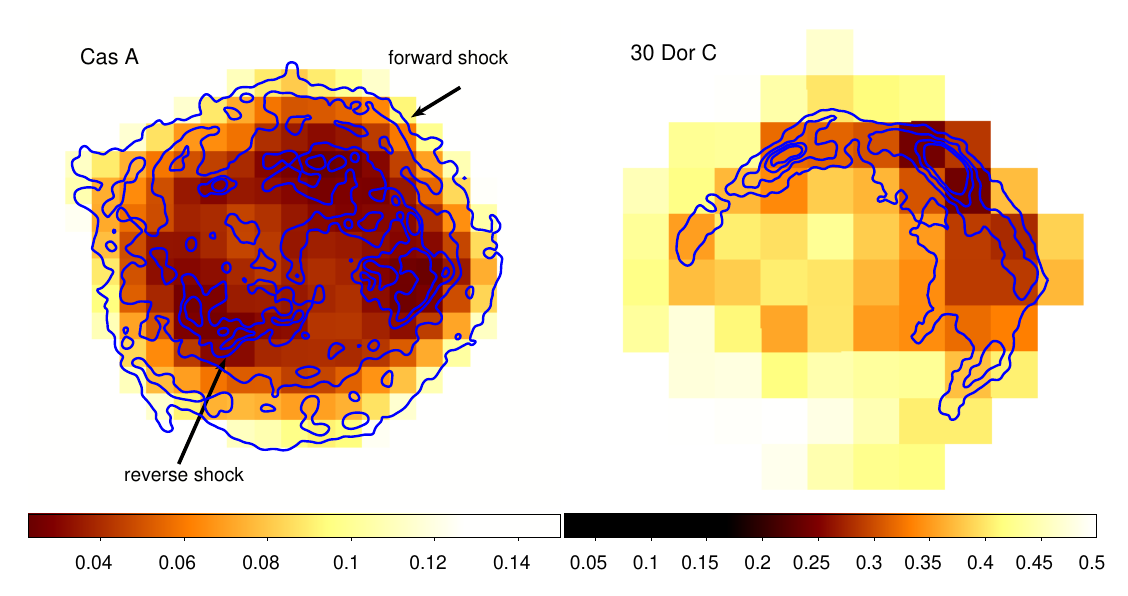}
\caption{Simulation of minimum detectable polarization at the 99\% confidence level (MDP99) maps for SNR Cas~A (left) in the 3--6 keV band and superbubble 30 Dor~C (right) in the 2--5 keV band using 1~Ms PFA observations. The pixel size is $30''$ for Cas~A and $60''$ for 30~Dor~C. The Chandra and XMM-Newton X-ray contours are overlaid, respectively.
}
\label{fig:snr-mdp}
\end{figure*}

eXTP will answer how magnetic turbulence and orientation vary across SNRs by measuring the distribution of X-ray polarization in young SNRs with a resolution $\lesssim 30''$. Cas A is an exceptionally suitable target for eXTP observations, as it represents not only one of the youngest and brightest SNRs in the X-ray band but also serves as a prominent prototype for comprehensive studies of SNR astrophysics. 
With 1~Ms observation with eXTP, we will obtain an X-ray polarization degree and angle distribution in Cas~A with unprecedented detail. 
The left panel of Figure~\ref{fig:snr-mdp} shows a simulated PFA map of the minimum detectable polarization at the 99\% confidence level (MDP99) for SNR Cas~A  in the 3--6 keV band.
With the PFA we can measure PDs without a brightness bias down to the PD=10\% level at the rim and 2.5\% at the reverse shock, using 30\arcsec\ pixel sizes.
eXTP will show whether the PD and PA change between the forward shock and reverse shock and between different azimuthal angles, where shock velocities are different. eXTP will also tell whether the low PD in Cas~A (significantly lower than other young SNRs with PD$>10\%$) is due to a depolarization effect.
We also note that 1~Ms observation of a single object is long. However, Cas~A will be observed with eXTP in multiple epochs as a calibration source. The observation time of Cas~A will be accumulated over the years to increase the measurement statistics.

eXTP can also expand the X-ray imaging polarization to more young SNRs, such as SN 1006, RX~J1713$-$3946, to advance our understanding of magnetic turbulence and orientation for SNRs in different stages of evolution and environments. 

In addition, IXPE lacks the sensitivity to study two very interesting objects: G1.9 + 0.3 ---- the youngest known Galactic SNR \citep{reynolds08}---and 
30 Doradus C (30 Dor C, see a simulation in the right panel of Figure~\ref{fig:snr-mdp}), a large shell in the Large Magellanic Cloud associated with a stellar superbubble, and emitting nonthermal X-rays, probably as a result of a $\sim$5000~yr old SNR expanding rapidly in the tenuous interior of the superbubble \citep{kavanagh19}.
eXTP should be able to measure the polarization properties of these youngest and oldest SNRs to emit nonthermal X-ray emission.

$\bullet$ {\it What are the magnetic configurations and turbulence in PWN structures, and what are the physical processes of particle acceleration?}

PWNe are believed to be among the most efficient particle accelerators in the universe \cite{2021Sci...373..425L}.
The operating acceleration process is an open question that needs a deep study in the compact region of PWNe. With eXTP-PFA, we can study the magnetic fields of the nebulae, which play an important role in particle accelerations.

IXPE observed seven brightest PWNe, including Crab \cite{2023NatAs...7..602B}, Vela \cite{2022Natur.612..658X}, MSH 15-52 \cite{2023ApJ...957...23R}, B0540-69 \cite{2024ApJ...962...92X}, G21.5-0.9, Kes 75, and 3C 58, the last three objects are still under study. These nebulae present a large diversity in polarization properties and magnetic field structures, likely associated with the acceleration mechanism, their ambient environment, and evolution stage. In general, these nebulae have a local high polarization degree that could reach up to 50\%-60\%, approaching the synchrotron limit. A spatially averaged PD is around 20\% for Crab and B0540-69, and an unexpectedly high 45\% for Vela.

\begin{figure*}
\centering
\includegraphics[width=\textwidth]{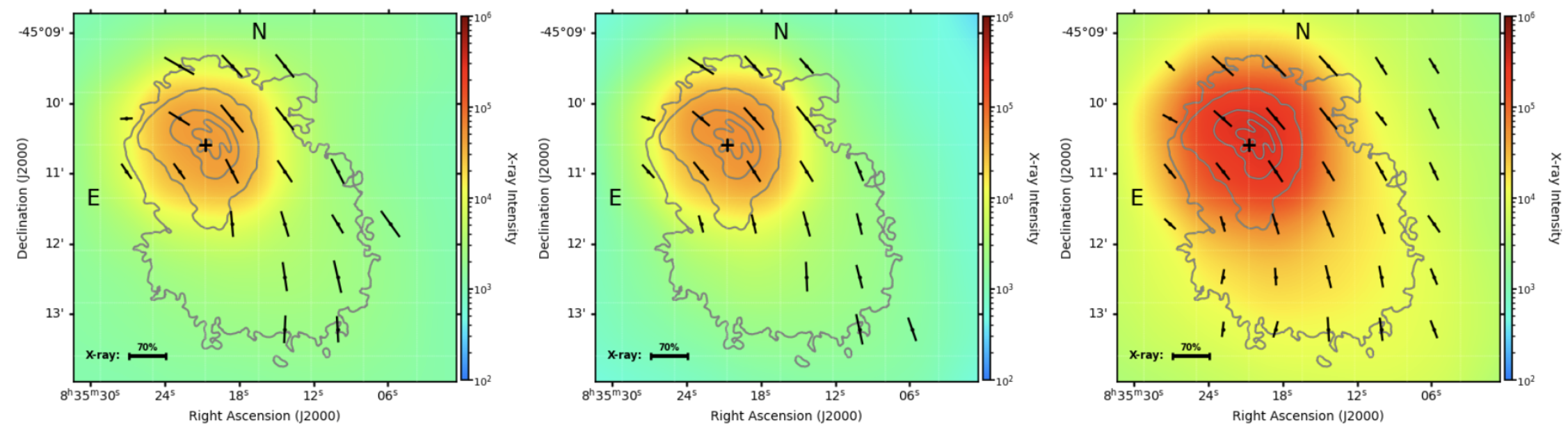} 
\caption{
The 2--8 keV count images of the Vela PWN from IXPE observation (left), IXPE simulation (middle), and eXTP simulation (right), all binned to a pixel size of $45''$ and overlaid with the Chandra X-ray contours. The simulations are based on the polarization map observed with IXPE and the Chandra image. Polarization vectors are overlaid only for pixels with PD $>$ MDP$_{99}$. The length of the black vectors indicate the PD, while their orientation indicates the projected magnetic field direction.}
\label{fig:Vela}
\end{figure*}

Vela PWN is one of the most polarized and interesting nebula detected by IXPE \citep{2022Natur.612..658X}. The nebula mainly radiates in the X-ray and radio bands. The X-ray nebula is bright and highly structured in the central 1 arcmin square region, but rather weak and diffuse in the outskirt (2--4 arcmin). For now, IXPE only maps the X-ray polarization in the bright center. Due to the faintness of the diffuse emission, IXPE could not obtain a robust result with 860 ks. With the same exposure performed by IXPE, eXTP can observe deeper and complete the missing part of the magnetic field to a larger area, as shown in Figure \ref{fig:Vela}.
This will provide a key to understanding the transport and diffusion of high-energy particles downstream. The result will have important implications for the efficiency of cosmic ray production.

Boomerang G106.6+2.9 is a middle-aged PWN that has a highly polarized radio nebula similar to the Vela PWN. The nebula structure is believed to result from the interaction with the supernova reverse shock. \cite{2023ApJ...959L...2L} proposed that it may be highly polarized with a PD$\sim$ 45\% in the X-ray band. However, the nebula is rather faint, making it not suboptimal target for IXPE. With a larger effective area and higher sensitivity, eXTP could obtain a meaningful 5$\sigma$ PD detection using an exposure of 370 ks (assuming an integrated PD of 45\%), which can help us better understand the nebula interacting with the reverse shock. 

Many young pulsars have strong X-ray emission that could contaminate the
measurements of the magnetic fields in the surrounding PWN. A good time resolution of the eXTP is critical to help mitigate this issue, such that we can select only the off-pulse time period to perform phase-resolved analysis.

The large collecting area of eXTP will allow one to map the magnetic fields of PWN
jets in young systems and misaligned outflows in bow-shock systems
\citep{2024ApJ...976....4D}. Since many of these structures are only detected in the X-ray band, eXTP observations are critical for revealing their formation and
confinement mechanisms.

$\bullet$ {\it What are the temporal properties of PWNe and what can we learn about the coevolution between PWNe and their pulsars?} 


The young pulsar PSR B0540-69 is the first to display an increase in brightness in its PWN following a change in its spin-down rate \cite{2019NatAs...3.1122G}. This discovery offers valuable insights into the correlation between the central pulsar engine and the fluctuations of the nebula brightness. Consequently, understanding the relationship between young pulsars and their surrounding PWNe is essential for examining the braking mechanisms of pulsars, as discussed by \cite{2019NatAs...3.1122G}. If the enhanced energy is converted to more particles, these particles could be injected into the surrounding environment and potentially affect the structure of PWNe. This process might gradually change the polarimetry properties of PWNe, which could be observed and studied utilizing the observations of eXTP-PFA.  X-ray polarization measurements could help us to understand the temporal evolution and formation of the PWNe structure affected by the enhanced particle flow. IXPE has detected a PD of 24.5\% $\pm$ 5.3\% at 4-6 keV in the off-pulse phase bin, which covers 35\% of the entire phase \cite{2024ApJ...962...92X}. Using the same exposure of 850 ks, eXTP would provide a $\sim$ 15 $\sigma$ detection with a smaller PD error of 1.6\%. Assuming that the fluence increases by 30\% and the increased portion is not polarized, the PD X-ray would decrease $\sim$ 5\%. 
In this case, the decrease is difficult to detect, requiring 7 Ms to achieve a 3 $\sigma$ detection.  If the outburst disturbs the magnetic fields, it may lead to a larger drop in the PD. Assuming a decrease of 10\%, we could achieve a 3 $\sigma$ detection with an exposure of 195 ks. For a 15\% decrease, a 5 $\sigma$ detection would be achieved with only 260 ks observation. 
The same could be applied to other PWNe the central pulsars, which show flux increase or outburst behavior. For example, the high magnetic field pulsar
PSR J1846$-$0258 in the young SNR Kes 75 exhibited a magnetar-like outburst
\cite{2008Sci...319.1802G} and the pulsar flux increased by 6 times with an emerging thermal component \cite{2008ApJ...686..508N}.
Rapid follow-up of similar events with eXTP could reveal the corresponding
changes in the PWN magnetic field.

\begin{figure*}
\center
\includegraphics[width=\textwidth]{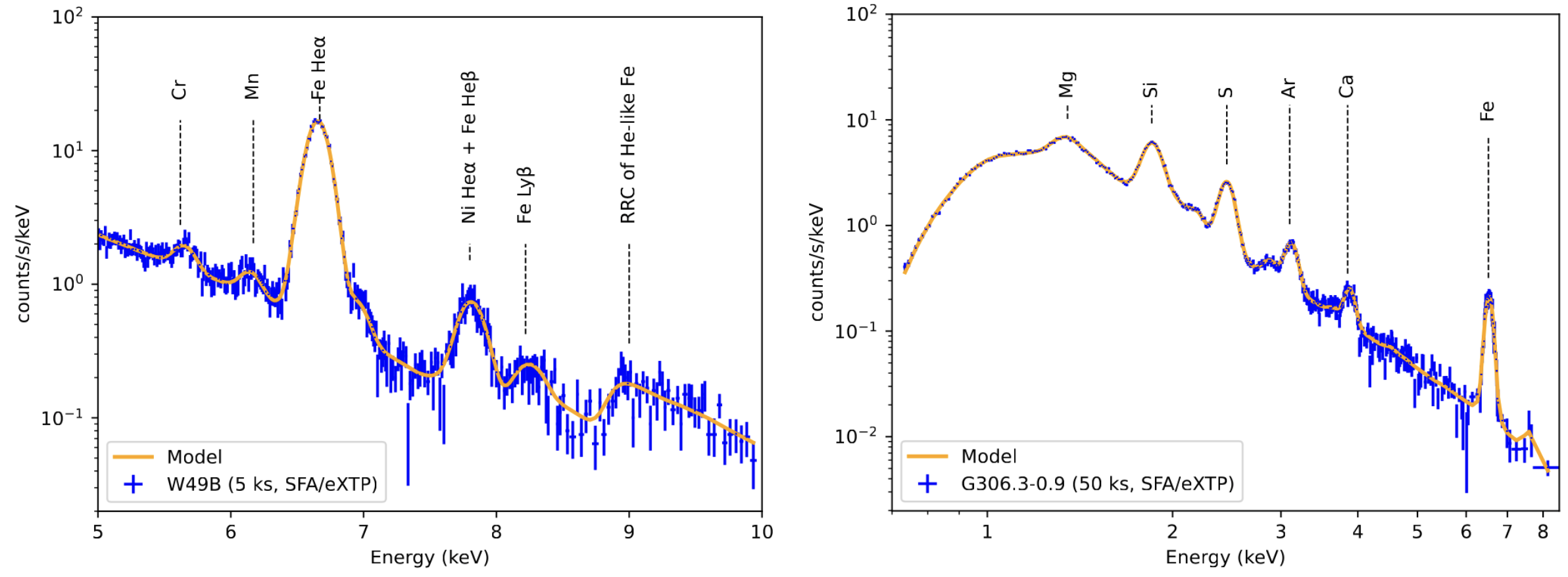}
\caption{Simulated SFA/eXTP spectra of SNRs W49B (left; 5~ks) and G306.3$-0.9$ (right; 50~ks) with main line features labeled. The instrumental background is subtracted from the spectrum.
}
\label{fig:snr-spec}
\end{figure*}

$\bullet$ {\it What are the metal composition and progenitor systems of SNRs?} 

Supernovae regulate the chemical evolution of galaxies via the ejection of heavy elements into the interstellar medium \citep{nomoto13}. 
However, the total chemical output of supernovae remains highly uncertain. These uncertainties come from ongoing debates regarding which stellar systems ultimately result in successful supernova explosions and the identification of viable explosion mechanisms. 
Metal measurements of supernovae and their remnants are thus of great importance in addressing the above debates, as they provide direct observational constraints. The metal yields of an SN depend on the progenitor and explosion mechanism, and thus the metal measurements provide a viable way to constrain which systems can produce supernovae \citep{nomoto13}.

Metal measurements of SNRs are mostly conducted using X-ray spectroscopic observations, which are proven to be an efficient way to constrain the physical and chemical parameters of the SN materials. SFA of eXTP works in the  0.5--10 keV range, covering the lines of elements from Nitrogen to Nickel.   By comparing the measured abundances with those predicted by nucleosynthesis models, we can build an SNR-progenitor connection.

With a large effective area, SFA offers distinct advantages for collecting X-ray photons, allowing for the efficient observation of SNRs in a shorter time than other X-ray telescopes. 
In addition, the SFA has its effective area optimized at 6~keV with a spectral resolution of 150~eV. 
This helps in searching for and studying rare metal lines in the hard X-ray band, such as Mn and Cr, which are critical for distinguishing the explosion mechanisms of thermonuclear SNRs \citep{seithenzahl13,fink14,leung18,leung20}.

The left panel of Figure~\ref{fig:snr-spec} shows a simulated spectrum of SNR~W49B with only 5~ks SFA observation, based on a model obtained by fitting 270~ks Chandra spectra \citep{zhou18}.
W49B has been extensively studied in X-rays, but its origin remains a subject of debate over whether it is a thermonuclear or a core-collapse SNR.
With 5~ks observation, SFA can clearly detect faint Cr and Mn lines, Fe lines at different ionization states, and the radiative recombination continuum (RRC) of Fe. By fitting the spectrum with a recombining plasma model $vvrnei$, we found that the metal abundances of Cr, Mn, and Fe can be constrained with a precision of $\lesssim 10\%$ ($\sim 20\%$ for Ni, 1-$\sigma$).  
This verifies the ability of the SFA to measure the metal composition and physical properties of SNRs. Note that SFA-Ts is not designed for spatially resolved studies, given its large pixel size, so the spectrum is simulated for the whole SNR. SFA-I can provide an angular resolution of $30''$, which will greatly benefit the observations of SNRs by resolving structures and detecting faint features.

eXTP can establish an SNR-progenitor connection for a sample of SNRs in our Galaxy and Magellanic Clouds, and provide valuable insight into the SN progenitor and chemical outputs. Its large effective area and moderate FOV ($18'$) makes it well-suited for observing small and medium-sized SNRs, although mosaic observations are also possible for some large SNRs. Some of these SNRs exhibit elevated metal abundances as revealed with XMM-Newton and/or Chandra observations, but their spectra, particularly in the hard X-ray regime, suffer from insufficient statistical quality. SFA will be able to conduct a survey of these SNRs and provide spectra with high statistical quality within a manageable time. For example, the right panel of Figure~\ref{fig:snr-spec} shows the simulated spectrum of SNR G306.3$-0.9$ with a 50-ks SFA observation, based on an absorbed $apec+vrnei+vrnei$ model described by \cite{weng22}. This SNR is proposed as a remnant of a Ca-rich transient, which is a subgroup of peculiar thermonuclear SNe. Our simulation shows that the abundance of Ca and Fe will be determined with an accuracy of 20\% (90\% confidence level), and other elemental abundances will have an accuracy better than 10\%. This precision will help distinguish the explosion mechanisms and properties of the progenitor system.

$\bullet$ {\it What are the properties of synchrotron X-rays from secondary electrons of proton-proton interaction in SNR-MC systems?} 

What is the gamma-ray production mechanism of the SNR-MC system? Is it produced by direct or indirect interaction? Several dozen SNR-MC systems have been detected so far in the gamma-ray band, and their radiation is generally believed to be produced by hadrons accelerated by shock waves bombarding molecular clouds (MCs) \citep[e.g.,][]{Ackermann2013,Tang2019.SNR}. However, the bombardment process remains a topic of debate. There are currently two scenarios. One is that the particles accelerated by the shock wave are confined near the shock. As the shock wave expands and reaches the MC, energetic hadrons collide with the MCs to produce hadronic gamma rays, namely direct collision \citep[e.g.,][]{Uchiyama2010,Tang2011.model,Tang2014.C}. The other is that the accelerated particles escape the acceleration site, and impact the MCs before the shock wave arrives. Gamma rays come from the accumulation of hadrons that escape from the shock wave, that is, indirect collision or escaping collision\citep[e.g.,][]{gabici2007,Fujita2009,Li2010.W28,Ohira2011,Li2012.9SNRs,Mitchel2024}. Both types of models can explain some of the observed gamma-ray spectra, but the predicted spatial distributions are different. The gamma rays of the direct collision model are mainly concentrated near the shock, while the gamma rays of the escape model can be in the far upstream area of the shock. Due to the poor angular resolution of current gamma-ray telescopes ($> 0.1$ degree), it is difficult to distinguish the two models. X-ray telescopes often have much better angular resolution  and may distinguish the two models by searching for synchrotron X-rays of secondary particles produced from hadronic processes. For an MC with a mass of 10,000 $M_{\odot}$ and a distance of 30 pc from the center of the SNR, the synchrotron X-ray flux of secondary particles from hadronic processes can reach a few $10^{-13}$ erg $\rm s^{-1}$ if the SNR is within a few thousand years old and within a few kpc from the Earth. This emission may be detected by the SFA because of its large collection area.

With the discovery of ultra-high-energy (UHE) gamma-ray sources by LHAASO, determining whether this radiation is leptonic or hadronic still requires further multi-wavelength observations. If this gamma-ray emission originates from the decay of $\pi^0$ mesons produced in SNR-MC systems, secondary electrons can potentially emit synchrotron photons in the X-ray band \citep{Tsuji2024}.
Detecting this emission with the eXTP SFA would test this scenario and provide strong constraints on the primary CR spectrum.

\section{Cataclysmic variables} \label{sec:cv}

Cataclysmic variables  \citep[CVs; see reviews in][]{2003cvs..book.....W, 2011ASPC..447....3K} are semi-detached binary systems in which a white dwarf (WD) accretes material from a low-mass companion. These systems possess diverse accretion modes, leading to a wide range of variability time scales (from seconds to minutes, up to months or even years), and thus serve as ideal laboratories for studying binary star accretion. CVs are also among the most common end products of binary star evolution, granting them a prominent position in the field of binary star evolution studies. Various issues are involved, including mass transfer, orbital evolution, thermonuclear reaction on the white dwarf surface, and the origin of magnetic fields in binary systems. In addition, CVs, especially magnetic systems, have been proposed to account for the Galactic Ridge X-ray emission \citep[GRXE; e.g.][]{Muno2004, Revnivtsev2006,Revnivtsev2008,Nobukawa16}. The contribution of these WD accreting systems to the GRXE should be quantified to help us better understand the origin of the GRXE, one of the most puzzling aspects in X-ray astrophysics \citep{2014MNRAS.445...66W,2015Natur.520..646P, 2016ApJ...826..160H,2016MNRAS.457.4507H, 2018PASJ...70R...1K}.

Thermonuclear runaway events in the accreted envelope of the WD lead to nova explosions \citep[e.g., see review ][]{2016stex.book.....J}. The ejecting events in an accreting system represent an excellent opportunity to examine the mechanisms behind accretion, ignition, outflows, and the subsequent chemical enrichment of the interstellar medium. These events also raise the important question of whether accretion dominates over outflow, potentially causing the WD to exceed the Chandrasekhar limit and resulting in type Ia supernovae.

Although the X-ray domain is essential for understanding the physical conditions near compact stars, progress in addressing many unresolved questions is hindered by the low quiescent luminosities ($\rm L_{X}\sim 10^{29-33}\,{erg}\,{s}^{-1}$) and the unpredictable large-scale variations, which can range from luminosities as low as $\rm \sim 10^{32}-10^{34}\,{erg}\,{s}^{-1}$ during dwarf nova (DN) outbursts to as high as $\sim 10^{38}\,\mathrm{erg}\,\mathrm{s}^{-1}$ in nova explosions. Most persistent, steadily nuclear-burning accreting WDs believed to exist in the Milky Way, its satellites, and M31 have yet to be detected \citep{1994ApJ...437..733D, 2016MNRAS.455.1770W}. These persistent ``supersoft X-ray sources'' are considered potential progenitors of Type Ia supernovae and may exhibit oscillations on very short timescales ($\lesssim$ 100 s) for reasons that remain unclear \citep{2015A&A...578A..39N}.

Thanks to the high sensitivity, excellent time resolution, and reasonable energy resolution of the SFA payload aboard eXTP, we are able to characterize rapid X-ray variability and monitor X-ray spectra for CVs. The PFA payload, specifically designed for polarimetry, is expected to provide polarization information of X-rays emitted by magnetic CVs. eXTP will be powerful in addressing the following questions about CVs:


$\bullet$ {\it How is matter accreted onto WDs and mixed with the CO/CNe substrate?} 

X-ray variability, both periodic and aperiodic, enables the tracing of the accretion flow near the surface of the WD. In magnetic CVs, X-ray emission is powered by magnetically funneled accretion, modulated by the orbit of the system and the spin of the WD. The X-ray spectral analysis allows for the determination of the physical conditions in the post-shock region, the hot spot on the WD surface, as well as complex absorption effects from pre-shock material. Furthermore, quasiperiodic oscillations (QPOs) with a mean period of a few seconds are predicted to result from shock oscillations \citep{2000SSRv...93..611W}, although they have been detected so far only in a few systems at optical wavelengths. The high shock temperatures ($\sim$20 -- 30 keV) suggest that QPO searches should be extended into the X-ray regime.

X-ray polarimetry offers a promising avenue for probing the geometry and physical conditions of the accretion flow in accreting WD systems. Theoretical studies have shown that Compton scattering within the accretion column of magnetic CVs can induce X-ray-polarized radiation \citep{Matt2004,McNamara2008}. The expected polarization properties of magnetic CVs could reflect the geometrical parameters and the Thomson optical depth of the accretion column. Theoretical simulations show that the degree of X-ray polarization can reach levels of up to $\sim$ 4-8 per cent \citep{Matt2004,McNamara2008}. Bright magnetic CVs should therefore be added to the lists of targets for polarimetric missions.

In non-magnetic CVs, when accreting at high rates, DN oscillations with periods of a few seconds have been detected in optical and soft X-rays, with frequency changing during outbursts. Slower, lower-coherence QPOs (on a time scale of minutes) appear at higher energies during the decline phase \citep{2004PASP..116..115W}. Their appearance is linked to the transition of the boundary layer from optically thick to optically thin, offering significant potential for probing the details of boundary layer dynamics. The frequencies of these QPOs exhibit a consistent ratio, similar to those of the low- and high-frequency QPOs observed in low-mass X-ray binaries. Moreover, the presence of a break in the power spectra during quiescence could allow the inference of disk truncation \citep{2012A&A...546A.112B}. The striking similarities between CVs and X-ray binaries make these systems particularly interesting for developing a unified model of accretion.

The maximum temperatures achieved during a nova explosion are limited by the chemical abundance pattern observed in the ejecta and do not appear to exceed $4 \times 10^8$ K. As a result, it is unlikely that the high metallicities detected in the ejecta are the result of thermonuclear processes. A more plausible explanation involves mixing at the core-envelope boundary. The exact nature of this mixing mechanism has remained unclear for decades. However, multidimensional models of the explosion suggest that mixing might be triggered by Kelvin-Helmholtz instabilities at the core-envelope interface \citep[e.g.,][]{2011Natur.478..490C}.

eXTP can address the above issues by studying selected samples of CVs. The SFA will be capable of detecting low-amplitude fast periodic and aperiodic variabilities in both magnetic and nonmagnetic systems, as well as performing time-resolved spectroscopy. For example, an exposure time of 30 ks or longer using the SFA enables the detection and analysis of rapid oscillations in nearby outbursting DNe (e.g., VW Hyi, U Gem, and OY Car). The PFA will be able to detect potential X-ray polarization signatures in bright magnetic CVs during their high states. 
For instance, the 2--10 keV flux of the magnetic CV AM Herculis in its high state is about 10$^{-10}$ erg cm$^{-2}$ s$^{-1}$. An exposure time of 60 ksec with the PFA is expected to provide an MDP$_{99}$ of $\sim 4\%$ for this source.

\begin{figure*}[ht!]
\center
\includegraphics[width=\textwidth,angle=0]{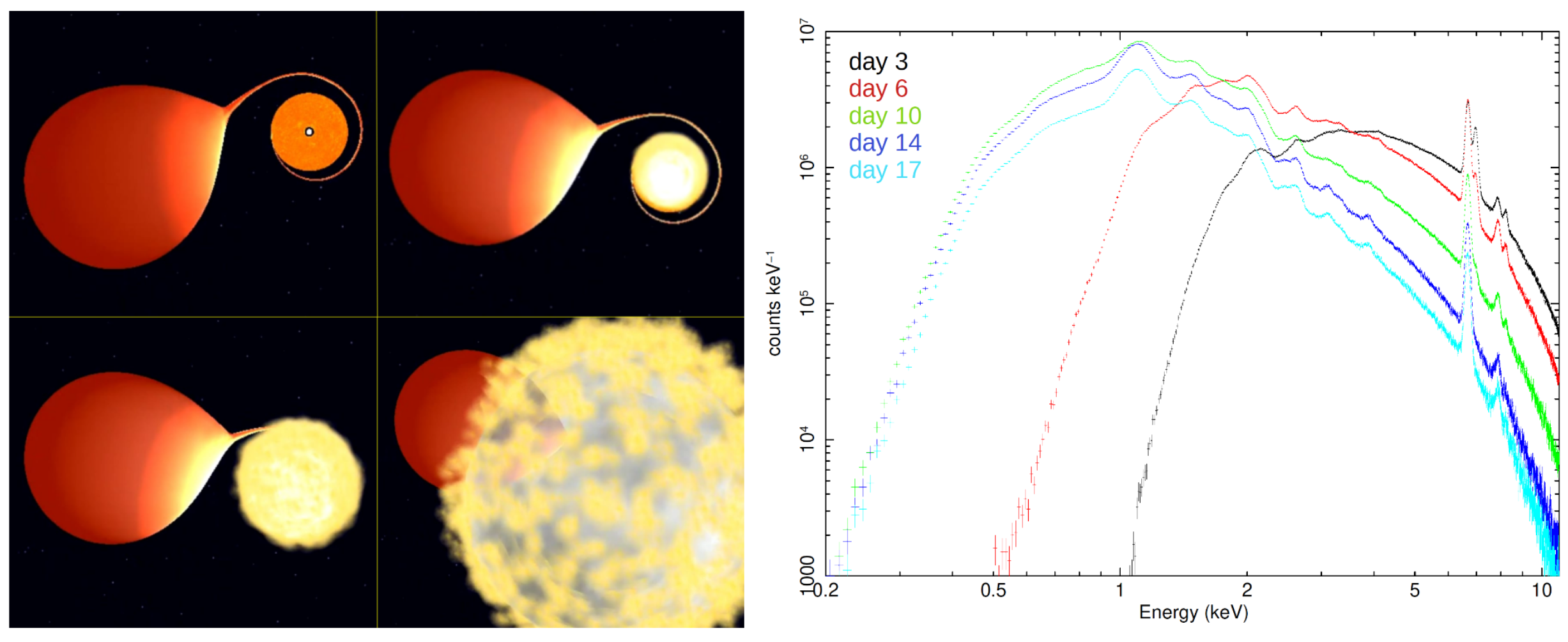}
\caption{Artistic picture of a nova explosion on a cataclysmic variable (left; credit: Josep Guerrero, IEEC, and Gloria Sala) and simulations of the spectral evolution of the symbiotic recurrent nova RS Oph during the expansion of the ejecta (right). SFA simulations for RS Oph assume 10 ks exposure times. The spectra include a thermal plasma model ({\it mekal}) with solar abundances, evolving from $kT= 8.2$ keV and a flux of $\rm 3.2\times10^{-9}$~\cgsflux\, (0.5--20 keV) on day 3, down to $kT= 2.5$ keV and a flux of $\rm 4.3\times10^{-10}$~\cgsflux\, (0.5--20 keV) on day 17 \citep[after][]{2006Natur.442..276S}.}
\label{figRsOph}
\end{figure*}

Novae are, after thermonuclear supernovae, the second most luminous X-ray sources among accreting WDs, with emission spanning from very soft to hard X-rays. The soft X-rays probe continuous nuclear burning on the WD surface, while the hard X-rays are thought to trace the initial shocks within the ejecta and between the ejecta and the surrounding circumstellar material, as well as the onset of resumed accretion. However, the details of the mass outflow and the shaping and evolution of the ejected material are poorly understood. Detection of high-energy gamma rays ($E>100$ MeV) from an increasing number of novae by the {\it Fermi}-LAT \citep{2014Sci...345..554A, 2016ApJ...826..142C} suggests that these systems are sites of particle acceleration. This was previously predicted for RS Oph \citep{2007ApJ...663L.101T}, a recurrent symbiotic nova, where strong shocks between the nova ejecta and the red giant wind were detected in hard X-rays by RXTE \citep{2006Natur.442..276S}. The recent detection of this nova at even much higher energies up to TeV \citep{Acciari22,HESS22} indicates that symbiotic systems are crucial testbeds for studying particle acceleration. In classical novae, however, there is no red giant wind interacting with the nova mass outflow, as the companion star is a main-sequence star. Therefore, particle acceleration in these systems is more difficult to explain \citep{2015MNRAS.450.2739M}. Prompt observations of nova explosions by eXTP would be crucial for a better understanding of these processes in both types of novae (see Fig.~\ref{figRsOph}).

It is crucial to observe the spectral evolution of nova explosions across the broadest X-ray range, from the initial hard X-ray onset to the later super-soft X-ray phase \citep{2006Natur.442..276S,  2011ApJS..197...31S, 2011ApJ...727..124O}. Furthermore, X-ray variability and periodicities on a variety of timescales, from days to sub-minute oscillations, remain poorly understood. Short oscillations in soft X-rays may originate from H burning instabilities \citep{2015A&A...578A..39N}, and the short-time variability in the harder component of the ejecta and/or the accretion disc is still to be investigated.
The wide-field W2C with an exposure time of one or several days is expected to detect the early X-ray emissions from bright novae, such as the recurrent nova RS Oph, which has a hard X-ray flux (20 -- 100 keV) of $\sim$10$^{-10}$ erg cm$^{-2}$ s$^{-1}$ during the initial days of its outburst \citep{Ferrigno2021}. 
The SFA will then track the X-ray spectral evolution from the initial hard X-ray phase to the later super-soft X-ray phase and investigate the characteristics of the fast variability in these bright novae. The large effective area and superb time resolution of the SFA, combined with the high orbit of eXTP, make it well-suited for exploring the variability of novae in outburst and in particular the short-time oscillations.

$\bullet$ {\it What causes the diversity of DN outbursts and what are the conditions for the launch of the diskjet?} 

Dwarf nova (DN) outbursts are thought to be the result of disk instabilities, but the changes that occur at the inner boundary layer are not yet fully understood. The boundary layer is optically thin during quiescence and becomes optically thick during outburst \citep{2003MNRAS.345...49W}. A suppression of hard X-rays during optical outbursts was previously believed to be a universal feature. However, the few DNs observed so far have shown considerable diversity, and not all exhibit the expected X-ray/optical anti-correlation \citep{2011PASP..123.1054F}. Understanding the fundamental parameters that distinguish different regimes of the boundary layer is essential. Another challenge is the detection of radio emission from DNs and CVs of high mass accretion rate \citep{2008Sci...320.1318K, 2011MNRAS.418L.129K, 2016MNRAS.463.2229C}, which has been interpreted as evidence of a radio jet. CVs were once thought to be incapable of launching jets, but recent studies have shown that nova-like CVs can have optically thin and inefficient accretion disk boundary layers \citep{2014ApJ...794...84B}, allowing the retention of enough energy to power jets. If this is the case, a radio/X-ray correlation, similar to those observed in neutron star and black hole binaries \citep{2011MNRAS.414..677C}, should be present in these systems.

\extp\ can tackle these complex issues by monitoring a selected sample of DNe with comprehensive multi-outburst coverage, using time-resolved spectroscopy to explore these systems.

$\bullet$ {\it Where are the periodic bouncers?}

Period bouncers are CVs that have evolved past the orbital period minimum ($\sim$80 min), where the donor star becomes degenerate (brown dwarf-like) and the system ``bounces back'' to longer orbital periods. Theoretical models predict that 40–80\% of CVs should be period bouncers \citep[e.g.,][]{belloni2020}.
Their cumulative X-ray emission could significantly contribute to the Galactic Ridge X-ray Emission (GRXE), but their low individual luminosities ($10^{29}$ erg s$^{-1}$) require deep surveys for detection \citep{rev2009,morihana2022}.
However, observations reveal that only $\sim$ 7–14\% of the observed CVs are period bouncers \citep{pala2020, munoz2024}, creating a significant discrepancy. 
The main challenges include their intrinsic faintness and low mass transfer rates \citep{munoz2024}. Recent works have proposed approximately 200 candidates, with over 80 located within 500 pc \citep{Giraldo2024}. Among these, several candidates have been confirmed by X-ray surveys (e.g., eROSITA) by detecting accretion signatures \citep{munoz2023}. However, whether the majority of the candidates are truly period bouncers remains an open question because of the lack of X-ray detection and/or X-ray spectroscopy.  

The confirmation of a period bouncer depends on the detection of X-ray emission, which signals ongoing accretion. Although H$\alpha$ emission is commonly used to identify CVs, this method is unreliable for period bouncers due to contamination of chromospheric activity in their low-mass donors \citep{munoz2023}. Dedicated pointing observations on strong period-bounce candidates with eXTP would be crucial for probing their properties and population. With adequate X-ray exposure (e.g. $\sim$50 ks exposure time for sources with a distance of $\sim$500 pc), spectroscopic and timing analyses will provide critical constraints on period bouncer physics, particularly their accretion mechanisms and WD properties. X-ray spectra can constrain the maximum emission temperature, while timing analysis reveals the accretion geometry (e.g., disk-fed for nonmagnetic WDs versus column accretion in magnetic systems). The WD mass can then be derived via the $T$-$M$ relation \citep{xu2019}. These observations would also constrain magnetic field strengths, key parameters for testing binary evolution models, particularly the crystallization-driven dynamo scenario \citep{schreiber2023}. 
Considering SFA's large effective area and its good sensitivity to point sources, eXTP is capable of detecting X-ray emission of candidate period bouncers within a distance range of 500-1000 parsecs. 

$\bullet$ {\it Are there any correlations between X-ray emission, accretion of planetary material, and outer companions?}

The Sun will eventually evolve into a WD after its main-sequence phase. However, the evolution of planetary systems around solar-type stars remains poorly understood because of the limited detection of such planets around WDs. Similarly to WDs in CVs that accrete material from visible stars, single WDs may also accrete metal-rich material from their planets. These ``metal-polluted'' WDs are potential targets for detecting X-ray emissions caused by planetary accretion events \citep{cunningham22}.

With SFA-I's high angular resolution and sensitivity to point sources with a flux of 8 $\times$ 10$^{-15}$ erg cm$^{-2}$ s$^{-1}$ for a 100 ks exposure, eXTP is expected to detect X-ray emission from metal-polluted WDs, such as WD 2226-210, which has an X-ray flux of about 1 $\times$ 10$^{-13}$ erg cm$^{-2}$ s$^{-1}$ caused by the accretion of planetary material \cite{Estrada-Dorado25}. This detection would require an exposure time of less than 1 ks.  
Another example is KPD 0005+5106, which has an X-ray flux of $\sim$10$^{-14}$ erg cm$^{-2}$ s$^{-1}$ \cite{chu21}. Among the 139 WDs within 20 parsecs of the Sun \citep{hollands18}, at least 27\% are metal-polluted by planetary materials \citep{koester14}, and thus are potential targets for detecting X-ray emission. 

The delivery of planetary material to WDs is likely driven by secular perturbations from outer companions \citep{veras16}, making astrometry and direct imaging crucial for detecting these potential perturbers \citep{mullally24}. By combining these techniques with X-ray observations, eXTP will probe the relationship between ongoing accretion (revealed by X-ray emission), the presence of metal pollution, and the existence of outer companions, thereby providing critical insights into the dynamical evolution of planetary systems around WDs.

\section{X-ray binaries} \label{sec:xrb}

\subsection{Low-mass X-ray binaries}

Low-mass X-ray binaries (LMXBs) are interacting binary systems in which a compact object like a neutron star or a black hole, accretes matter from a low-mass companion star (typically $<1 M_\odot$). The accreted material, transferred via Roche lobe overflow, forms an accretion disk that radiates strongly in X-rays as gravitational potential energy is converted into heat near the compact object. These systems exhibit a broad range of temporal and spectral phenomena, including QPOs, relativistic jets, thermonuclear bursts, and pulsations in neutron star accretors. LMXBs serve as laboratories for studying accretion physics, strong gravity and magnetic fields, and the evolution of compact binaries, and they include diverse sub-classes such as black hole X-ray binaries (BHXRBs), accreting millisecond X-ray pulsars (AMXPs), transitional millisecond pulsars (tMSPs), and nuclear--powered bursting neutron stars.

\subsubsection{Accreting Millisecond X-ray Pulsars}
AMXPs represent a distinct subclass of neutron stars in LMXBs that exhibit coherent pulsations in the X-ray band due to the misalignment of their magnetic and rotational axes \citep{pw21}. These pulsars, with spin periods typically in the $\sim$1--10 millisecond range, provide a crucial observational link between the accretion phase in LMXBs and the formation of rotation-powered millisecond pulsars. Observations of AMXPs have confirmed that accretion-induced spin-up can recycle neutron stars into millisecond pulsars, but have also revealed a richer phenomenology than initially expected, including pulse amplitude variability, mode switching, and complex interactions between accretion flow and magnetosphere.

Despite significant progress, key questions remain. What determines the long-term spin behavior of AMXPs, and how is it influenced by magnetospheric torques or disk instabilities? What triggers transitions between accretion and propeller-dominated regimes? How does the geometry of the emission region vary with luminosity and spectral state? Are outflows a common feature in these systems, potentially influencing angular momentum loss?

The eXTP mission is suited to address these questions. For example, the high time resolution and large effective area of the SFA-T will allow precise tracking of the pulse frequency evolution and phase-resolved spectroscopy to investigate the accretion geometry and the coupling of the disk-magnetosphere. In parallel, the PFA will provide direct measurements of X-ray polarization, giving access to emission geometry and magnetic field topology. The variation of polarization properties over the pulse phase will further constrain models of beam formation and magnetospheric scattering.

Transitions between accretion and propeller states can be studied in detail by monitoring spectral and timing variability with the SFA. At the same time, changes in the angle or degree of polarization, measurable with the PFA, can reveal whether the emission geometry or the flow pattern have changed fundamentally \citep{pap25}. Combined, these diagnostics provide a multi-dimensional view of state transitions.

\begin{figure*}[ht!]
\centering
    \includegraphics[width=0.9\textwidth]{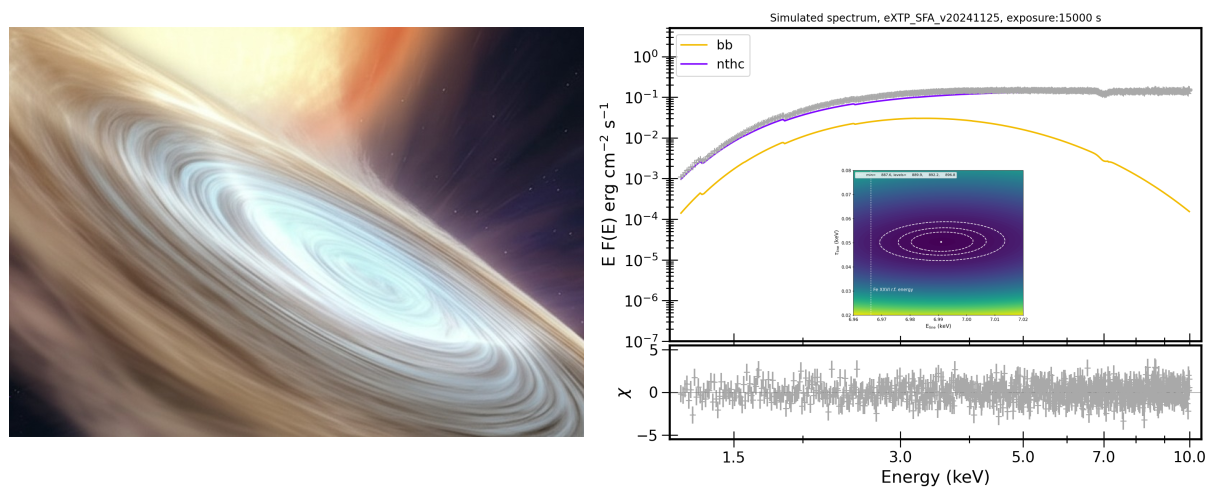}%
\hfill
\caption{Illustration of the inner accretion disk regions of an AMXP with a potential outflow (left), modified from original work by Gabriel P\'{e}rez D\'{i}az, SMM (IAC) and simulated eXTP-SFA spectrum of Swift J1749.4--2807 for a 15 ks exposure using the best-fit NICER model from \cite{mar22} (right). An absorption line at $\sim$7 keV is clearly detected and significantly blue-shifted. The inset shows the corresponding contours at 1, 2 and $3~\sigma$ confidence levels.}
\label{fig:absorption}
\end{figure*}

An intriguing recent findings in AMXP research is the possible presence of disk-driven outflows, observed via absorption features in their X-ray spectra \citep{mar22}. These features, including blue-shifted Fe XXV and Fe XXVI lines near 7 keV, have traditionally been detected with long exposures of high-resolution spectrometers such as those on board \textit{XMM-Newton} and \textit{Chandra}. However, simulations using eXTP’s response matrices show that the SFA can also detect these lines in moderately deep exposures. In particular, simulated spectra for the eclipsing AMXP Swift J1749.4--2807, based on NICER observations and including a Gaussian absorption feature, demonstrate that with 15--25 ks exposures and six SFA modules, eXTP can detect absorption lines a factor of three fainter than that observed in this source, and determine whether they are significantly blue-shifted. A representative fit is shown in Figure~\ref{fig:absorption}, where the contour plot and spectral model illustrate both detectability and constraints on blueshift. The PFA may also reveal complementary evidence for outflows by detecting scattering-induced polarization changes associated with extended structures, such as disk winds.

From a spectral-timing perspective, the detection and characterization of iron K-shell transitions—especially Fe XXV and Fe XXVI near 6.7--7 keV, offer crucial diagnostics of plasma conditions near the neutron star. The SFA's energy resolution ($\sim$180 eV at 6 keV) and large effective area allow for the detection of such features even in short exposures ($\sim$10–20 ks), enable phase-resolved spectroscopy to study how the line parameters vary across the pulsar spin phase. In sources with outflows, this can reveal changes in the ionization state and dynamics of the absorber on short timescales. Simultaneously, the high timing resolution (10 $\mu$s) permits detailed modeling of pulse profiles and their energy dependence, allowing separation of Comptonized and thermal components and tracking of phase lags or frequency drifts. These joint capabilities will be especially powerful in bright AMXPs undergoing outbursts, where variations in the iron line strength or width may correlate with pulse amplitude, providing direct observational evidence of inner disk dynamics modulated by the star’s magnetic field.

\subsubsection{Transitional Millisecond pulsars}

Millisecond pulsars (MSPs, defined with spin period $P\le30$ ms) are believed to be formed through the ``recycling'' scenario by accreting mass and angular momentum from a companion star in LMXB systems in a Gyr-long phase \citep{Alpar1982,Bhattacharya1991}. Currently, 552 MSPs have been detected in the Galactic field \footnote{\url{https://www.astro.umd.edu/~eferrara/pulsars/GalacticMSPs.txt}}, representing $10\%$ of the known Galactic pulsar population. In comparison, 80\% of the 344 radio pulsars found in 45 globular clusters (GCs) are MSPs\footnote{\url{https://www3.mpifr-bonn.mpg.de/staff/pfreire/GCpsr.html}}. This phenomenon has been attributed to the higher stellar density thus higher stellar encounter rate in GCs which resulted in more abundant LMXBs in GCs than in the field \citep{Pooley2003,Gendre2003,Menezes2023}. While the majority of MSPs are detected in radio with an increasing fraction in gamma rays\footnote{\url{https://confluence.slac.stanford.edu/display/GLAMCOG/Public+List+of+LAT-Detected+Gamma-Ray+Pulsars}}, most of them being rotation powered \citep[see the ATNF pulsar catalog\footnote{\url{https://www.atnf.csiro.au/research/pulsar/psrcat/}},][]{Manchester2005}, 
two dozen accreting MSPs (AMXPs) in X-ray transients have been found \citep{Patruno2021,Campana2018}.  The discovery of GC MSPs and AMXPs reasonably provided supports for the ``recycling'' scenario. 

Discovered just a decade ago, transitional MSPs (tMSPs) are defined by the switch between the accretion- and rotation-powered states back and forth over a few years. They are considered as the ``missing link'' between radio MSPs and LMXBs, providing conclusive proof of an
evolutionary link between these two classes. 
So far, there are only three firmly identified tMSPs, PSR J1023+0038 \citep{Archibald2009}, XSS J12270$–$4859 \citep{Martino2010} and IGR J18245$–$2452 \citep{Papitto2013} and a few candidates recognized, representing a very limited sample. During the transitions, the X-ray luminosity changes by at least an order of magnitude, accompanied with multi-wavelengh emissions from radio, optical/UV to gamma rays \citep[for a review, see][]{Papitto2022}. The unique ``two-sidedness'' of tMSPs provides a natural laboratory for understanding neutron star evolution, accretion physics, and extreme magnetic field environments.

Despite the progress made by a decade of observation and theoretical modeling of tMSPs, the underlying physics of tMSP is still not well understood and there are many open questions. With its exceptional large effective area, high time resolution, and polarization measurement capability, eXTP has unique advantages in studying the radiation characteristics and state changes of tMSPs, eventually advancing our knowledge of tMSPs by potentially solving the following key problems.

$\bullet$ {\it What are the exact mechanisms that trigger the state transitions?}
In the rotation-powered state ($L_{\rm x}\lesssim10^{33}$ erg $\rm s^{-1}$), tMSPs behave as ``redbacks'', a sub-catagory of MSPs in tight binaries ($P_{\rm orb}<1$ day) with a companion mass $M_{\rm com}\sim 0.1-0.4 \, M_\odot$. Redbacks are thought to ``devour'' their companions through ablation by the pulsar wind, producing synchrotron radiation which is observable at X-rays and gamma rays. Such sources exhibit significant radio eclipses due to the presence of intra-binary material. In the accretion-powered states, so far, the accretion outburst ($L_{\rm x}\sim10^{36-37}$ erg $\rm s^{-1}$) similar to other AMXPs has only been observed in IGR J18245$–$2452. The tMSPs preferentially show a peculiar intermediate state, termed ``sub-luminous disc state'' 
which is characterized by the presence of an accretion disc and $L_{\rm x}\sim10^{33-34}$ erg $\rm s^{-1}$ \citep[][and references therein]{Papitto2022}. 

The state transitions are generally attributed to changes in the mass accretion rate, which reflects the interplay between the infalling material of the companion and the outward pulsar wind. However, it remains unclear what ultimately causes the accretion rate to vary.

eXTP is expected to perform high cadence and high statistic follow-up observations once the state transition is triggered by X-ray monitors, because of the unprecedented large effective area of eXTP-SFA. With spectral and high-resolution  timing analysis, eXTP will be able to capture the abrupt and dynamic processes associated with the state transition, shedding light on the physical mechanisms that are onset during the transition.

$\bullet$ {\it What powers the multi-band emission in the sub-luminous disc state?} \\
The sub-luminous disc state exhibits complex multi-band emissions. The X-ray 
emission is variable and three intensity modes (\textit{high}, \textit{low} and \textit{flaring} modes) have been identified from the X-ray light curves \cite{linares2014}. The tMSPs show unexpectedly fast mode transitions on a time scale of $\sim$ 10 s with modes duration ranging from a few tens of seconds to a few hours. In this state, X-ray, UV and optical pulsations from two of the tMSPs have been detected in the \textit{high} mode \citep{archibald2015,ambrosino2017,Zanon2022}. Gamma-ray emission is several magnitudes brightened when switching from the rotation-powered state to the sub-luminous disc state, and vice versa, dimmer when turning into the rotation-powered state \citep[][and references therein]{Papitto2022}. 

The sub-luminous disc state has triggered many theoretical efforts to explain the enigmatic behaviour of the tMSPs. However, it is a major challenge to determine whether the multi-wavelength emission observed in the subluminous disc state is accretion- or rotation-powered, or both are in work. The enshrouded radio pulsar models assume that a radio pulsar hides behind the enshrouded intrabinary matter and the accretion disc is truncated far from the pulsar ($d\approx10^9-10^{10}$ cm) by the pulsar wind. The shock-accelerated electrons produce gamma rays by inverse Compton scattering (ICS) of the disc UV photons and synchrotron X-ray radiation by interacting with the magnetic field \citep{Takata2014,Li2014,Zelati2014}. These models were unable to explain the X-ray pulsation detected due to unreasonable X-ray efficiency required to convert the pulsar spin-down power to X-rays. In the propeller model, most of the accreting material would be ejected and emit X-ray synchrotron photons in the magnetized turbulent disc/magnetosphere boundary region, which are subsequently upscattered to gamma rays via ICS \citep{Papitto2014,Papitto2015}. However, this model has difficulty explaining the simultaneous appearance and disappearance of optical and X-ray pulsations, as well as their similarity in pulse profiles and spectra. This strongly points to a common underlying process. Recently, it has been proposed that the pulsed emission seen at optical, UV and X-ray energies in the high mode is powered by synchrotron emission of electrons accelerated from the interaction between the electromagnetic wind of a rotation-powered pulsar and the accretion disk beyond the light cylinder surface ($d\simeq 100$ km) \citep{Papitto2019,Veledina2019,Illiano2023}.

eXTP can provide good energy resolution spectra and good time resolution light curves with a few  times larger effective area than previous telescopes, enabling deeper investigation of the sub-luminous state.
Precise timing solution (ephemeris) of X-ray pulsations will help to finely determine the magnitude of the slight changes of the spin-down rate in the active X-ray state compared to that observed during the rotation-powered radio pulsar state \citep{jaodand2016,Baglio2024}.

\begin{figure*}
\centering
\includegraphics[width=0.8\textwidth]{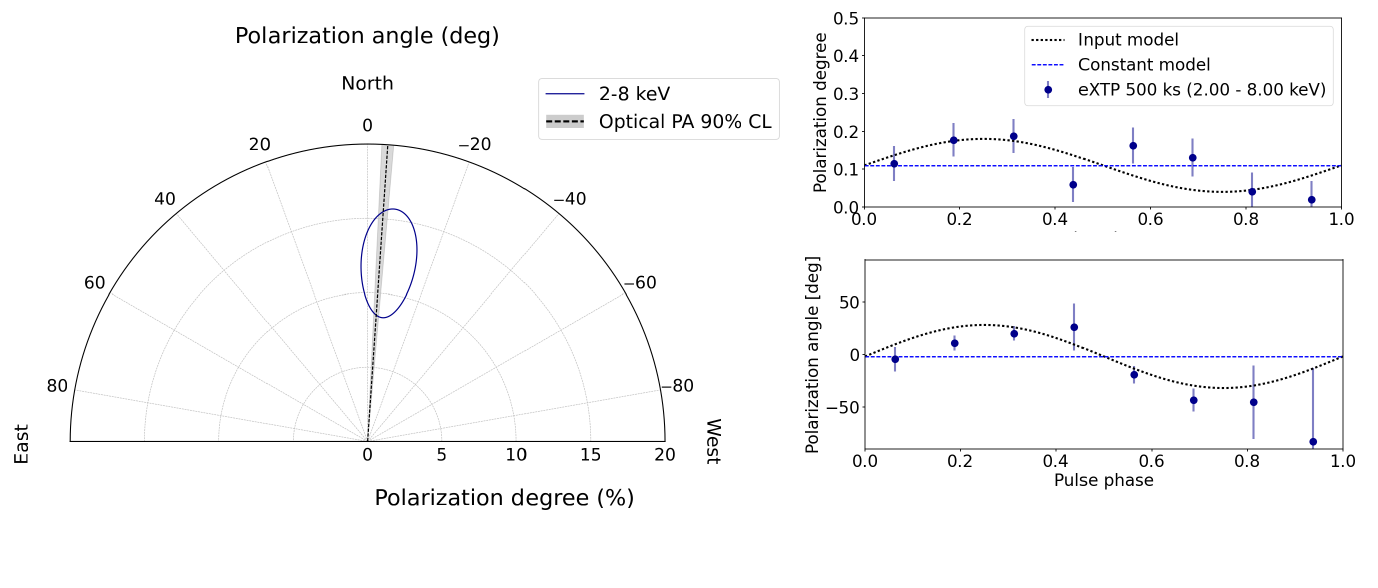}
\caption{
 Protractor plot of the degree and angle of X-ray polarization for the high mode of the transitional MSP PSR J1023+0038, observed for 500~ks by the PFA on board the eXTP assuming the same parameters observed by \citep{Baglio2024}, compared to the optical polarization angle (left panel). Phase resolved variability of the degree and angle of X-ray polarization in the same exposure time, assuming sinusoidal variations of 5\% and $40^{\circ}$, respectively (right panel; courtesy: A.~Di Marco, A.~Papitto).
}
\label{tmsp}
\end{figure*}

Polarimetric measurements will also be crucial to determine the geometry of the accretion flow and the magnetic field in the ``sub-luminous disc state''. Recently, the detection of polarized X-ray emission from the brightest tMSP currently in such a state at an average degree of $(12\pm3)\%$ was reported \citep{Baglio2024}. The compatible angle and energy distribution of the polarized emission seen from optical to X-ray energies suggest a common origin as synchrotron radiation at the shock formed where the pulsar wind interacts with the inner regions of the accretion disk. Given the low X-ray flux of tMSP in the ``sub-luminous disc state'' ($\simeq 10^{-11}$ erg cm$^{-2}$ s$^{-1}$), only crude average estimates  are currently possible with weeks-long integration time. The increased sensitivity of eXTP to polarized emission will allow to detect polarized X-ray in the  \textit{high} mode at a higher significance and to study the dependence of the degree and angle of X-ray polarization on the pulsar spin phase (see Fig.~\ref{tmsp}) for the whole sample of tMSPs.

$\bullet$ {\it Are all MSPs in tight binaries transitional?} \\
The current tMSP sample size is very limited, prohibiting a full understanding of the complex and mysterious phenomena observed. Discovering more tMSPs to enlarge the sample is not only very important to answer the previous two questions but also to explore whether the state transition can occur in all MSPs in tight binaries. So far, the three tMSPs are all redbacks which are mostly found through dedicated multi-wavelength observations of Fermi-LAT unidentified (UnID) sources \citep[over 2200 UnID in the latest 4FGL-DR4 catalog,][]{4FGL-DR4}. On the one hand, the eXTP can perform follow-up observations of current tMSPs whose state transitions are triggered by wide-field X-ray monitors.
However, the high eXTP sensitivity will allow targeting roughly twice as faint X-ray counterparts of UnID gamma-ray sources to look for the distinctive features of tMSPs in the ``subluminous disc state'' such as the \textit{high} and \textit{low} modes, to date the most efficient way to identify candidate tMSPs \citep[e.g.][and references therein]{Papitto2022}.

\subsubsection{Weakly magnetized neutron star systems} 

Another important sub-class of LMXBs is that hosting a weakly magnetized neutron star (NS) with $B \sim 10^{8-9}$ G.
These systems have historically been divided into two further subclasses, the Z and atoll sources
\citep{hasinger89}, on the basis of their temporal pattern in the color-color  or hardness-intensity diagram (CCD and HID, respectively).
In their brightest, high-accretion regimes (High Soft State, HSS), the spectra of LMXBs are usually described by a two-component model consisting of a strong dominating Comptonization spectrum at higher energies plus soft emission attributed either to the accretion disk \citep{lavagetto2004, farinelli2008}, or a direct blackbody (BB)
at temperature $kT_{\rm bb} \sim 1.5$ keV coming from the NS \citep{white1988}. 
The two scenarios were labeled Eastern Model (EM) and Western Model (WM), respectively.
In many cases, in addition to the persistent two-component continuum, in both Z and atoll sources clear evidence of a reflection feature attributed to the accretion disk, which can be fully or partially ionized \citep{piraino1999, mondal2019, iaria2020, ludlam2022}.
All known Z sources (Cyg~X-2, GX~5-1, GX~340+0, Sco~X-1, GX~17+2 and GX~349+2)  always remain the high soft state (HSS), where the Comptonization component has low temperature ($kT_{\rm e} \sim 3$ keV) and high optical depth $\tau \sim$ 5--10
depending on the assumption of plasma geometry, slab-like or spherical. Atoll sources, on the other hand, can be divided into two further subclasses: sources of the first class (GX 13+1, GX 3+1, GX 9+9, GX 9+1) 
have only been observed in the HSS,  while the others experience state transition from HSS to low hard state (LHS), characterized by a dimmer flux in the X-ray band, and a significantly harder spectrum where the electron temperature can achieve about 30 keV, in low optical depth environment \citep[e.g.,][]{guainazzi1998, paizis2006, falanga2006, cocchi2010}.

The fact that both WM and EM have been used for a long time in spectral analysis shows that the latter has not been able to resolve the controversy over the actual geometry of accretion, although studies based on Fourier frequency resolved spectroscopy clearly point in favor of the EM (e.g. \cite{gilfanov2003}).
 
The decisive tool appears to be unequivocally polarimetry, as demonstrated since 2021 with the launch of
the IXPE \citep{weisskopf2016}.
What IXPE has told us so far is that polarization increases with energy, which is only possible if the soft and hard components do not share the same polarization angle \cite{ursini2023b, farinelli2023, dimarco2023, saade2024, gnarini2024b}.
Then a boundary layer with high optical depth and coplanar to the accretion disk seems to be ruled out. However, because of IXPE's statistics, the positive correlation of the PD with energy is visible qualitatively, but it is not possible to define a truly parameterizable trend for individual sources, and in most of the cases the results are presented for the 2-4 keV and 4-8 keV energy bands only.

\begin{figure*}[ht!]
\centering
\centering
\includegraphics[width=\textwidth]{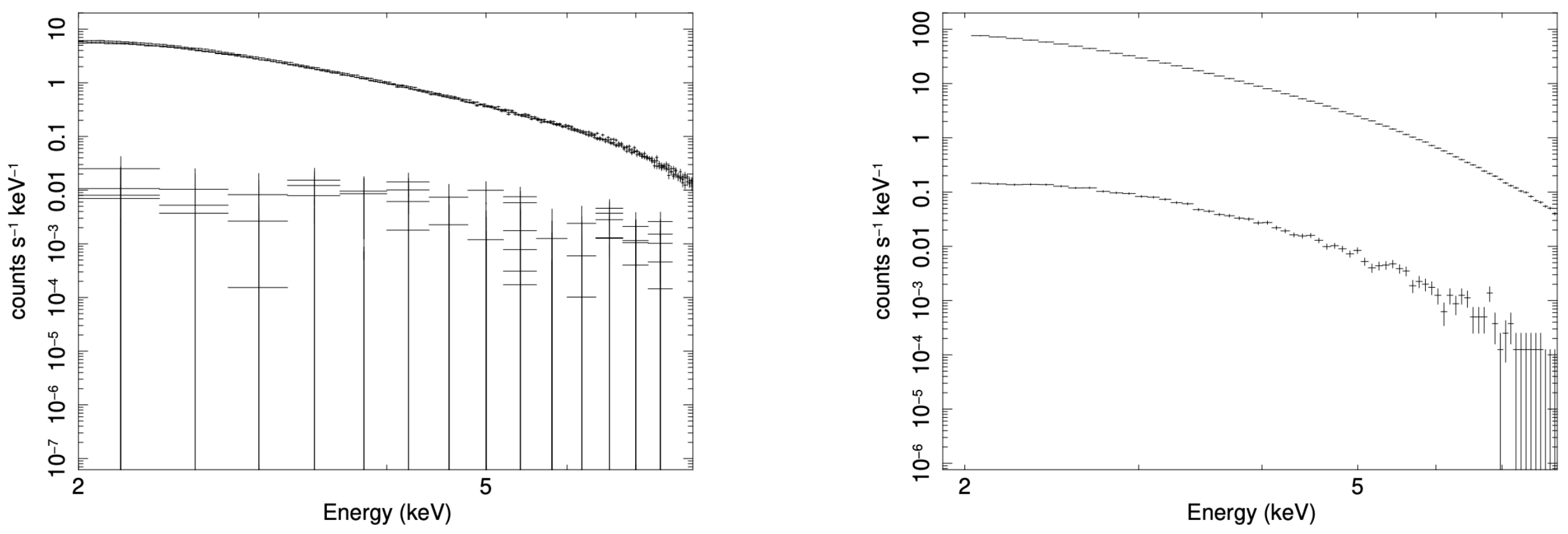}
\caption{Left panel: IXPE spectra (I,Q,U) of the NS-LMXB GX~9+9, for which a polarization degree of 1.4\% was measured \citep{ursini2023b}. The exposure time was $9.3 \times 10^4$ s with a source flux of $4 \times 10^{-9}$ erg cm$^{-2}$ s$^{-1}$ in the 2--8 keV band. Right panel: simulated eXTP I/Q spectra for a source with a polarization degree of 1.2\%, and the same time exposure and flux. Results have been obtained using a Montecarlo ray-tracing code with azimuthal symmetry for which U=0. The statistics of the Q spectrum in the latest case is likely higher than actual because background spectra are not yet available, but the strong improvement in the polarization data quality is evident.}
\label{fig:ixpe_extp}
\end{figure*}

Additionally, spectropolarimetric analysis has provided evidence that the PAs of the accretion disk and those of
the Comptonization plus reflection component are sometimes less than $90^{\circ}$ apart, thus pointing towards the very intriguing possibility of the presence of a warped disk \citep{cocchi2023, fabiani2023}.

Finally, no less importantly, the theory of accretion and radiative transfer tells us that in NS-LMXBs the net polarization layout is the combined result of at least three components, i.e. the Comptonization zone around the NS, the reflection of this radiation off the disk atmosphere and accretion disk itself, but currently, due to limited statistics considering all of them lead to an over modelization when fitting the IXPE Q/U spectra as well. It is therefore standard practice to fix the PD of one of these components to a value expected from theory, in an attempt to constrain at least the other two. Even that might not be enough: A telling example is the component associated with the accretion disk, for which so far only upper limits have been established for all observed sources. Therefore, we are still not sure whether the accretion disk is polarized according to Chandrasekhar’s law, as is often assumed, perhaps with an excessive degree of arbitrariness.

Exactly as for IXPE, NS-LMXBs will be a main target for eXTP as well. The key difference lies in the significantly larger effective area of the PFA, approximately five times larger, which will enable tighter observational constraints, not only on the energy dependence of the polarimetric parameters, but also on the contribution of each emission component to the formation of the net polarization signal. This will provide complementary information to spectroscopy and timing that is essential to disentangling the parametric degeneracy often observed in this class of sources, which has been partially demonstrated by IXPE  (see Fig. \ref{fig:ixpe_extp}).

\subsubsection{Thermonuclear X-ray bursts}


Thermonuclear X-ray bursts, also known as type I X-ray bursts \citep[e.g.][]{Lewin93}, occur as a result of thermonuclear runaways on the surface of a neutron star in a binary system. These bursts occur in the layers above the crust, where freshly accreted fuel, primarily composed of hydrogen and helium, can burn in an unsteady way when compressed to a density of around $10^6$ g cm$^{-3}$. Unstable burning takes place when the cooling cannot balance the heating due to the nuclear reactions. Key factors that affect the burning and its stability are the accreted composition, accretion rate and the energy exchanged with the deeper layers, including the crust. Very importantly, the observed light curves are affected by relativistic effects, especially those due to the gravitational field of the star that determines the path and spectrum of the light emitted from the surface \citep[e.g.][]{Strohmayer2006}.

Despite a general understanding of their origin, type I bursts present several unresolved questions, many of which are fundamental to astrophysics. Also, because of their nature and relation to the neutron star and its environment, the bursts can be used to gain insight into physics of particular relevance. First and foremost, because of the effects of the gravitational field of the star, which depends on its mass and radius, their light curves can be used to put constraints on the internal structure of neutron stars and the exotic nuclear processes occurring within. Other key areas of interest include the geometry of accretion flows, magnetic field interactions, and unusual stellar abundances \citep[see e.g.][]{2016RvMP...88b1001W}. Furthermore, the very phenomenology of the type I bursts still presents challenges to our theoretical modelling. For example, the critical mass accretion rate at which the bursts are quenched seems to be almost an order of magnitude smaller than predicted. This is probably related to the fluid dynamics on the surface, with spin and magnetic fields playing a major role, and to the nuclear reactions involved \citep[e.g.][]{Strohmayer2006,2020MNRAS.499.2148C}.

\begin{figure*}[ht!]
\center
\includegraphics[width=\textwidth]{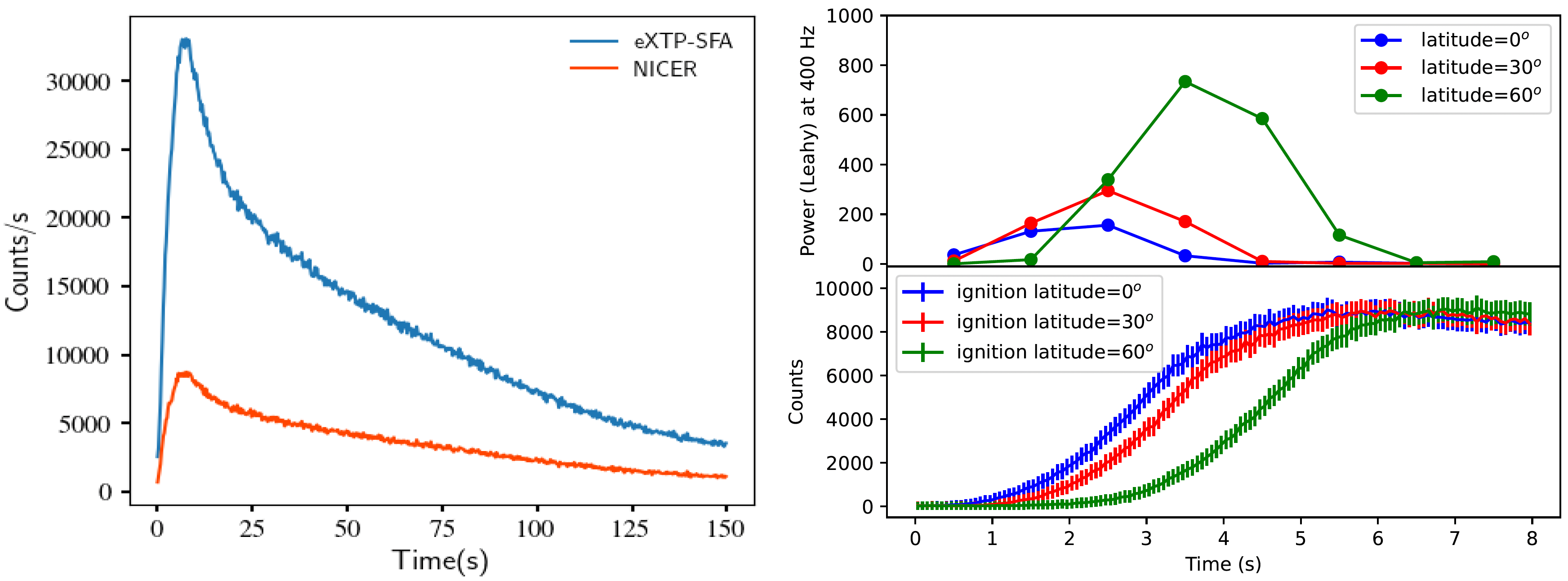}
\caption{Left panel: light curves of simulated type I X-ray bursts as observed by NICER and eXTP-SFA. Right panel: simulations of light curves during the burst rise for a typical NS of 1.4 solar masses spinning at 400 Hz. The three light curves correspond to ignition at different latitudes, as indicated, with the flame speed depending linearly on the sinus of the latitude. We also show the temporal evolution of the expected intensity of the burst oscillation at 400 Hz for each case. 
}
\label{X_ray_burst_lc}
\end{figure*}

In order to extract as much information as possible from the short and bright burst light curves, we need high-precision observations such as those provided by advanced X-ray telescopes and eXTP is particularly well-suited for this task. 
As an example, in the left panel of Figure \ref{X_ray_burst_lc}, we compare the simulated light curves of type I X-ray bursts as detected by NICER and SFA.
With its exceptional sensitivity and timing resolution, eXTP will be able to capture the rapid and dynamic processes associated with thermonuclear bursts, shedding light on the underlying physics and helping to answer many of the open questions in this field.

$\bullet$ {\it What is the nature of burst oscillations?} 

Burst oscillations are a fascinating phenomenon observed in approximately 10\% of all type I X-ray bursts \citep{Galloway08}. These oscillations occur at frequencies closely matching the neutron star's spin frequency, ranging between 11 and 620 Hz. 
They appear as strong signals in the power spectra of the bursts, sometimes at constant frequency and sometimes slowly drifting by a few Hz during the burst evolution. They are thought to be due to asymmetries in the surface emission modulated by the rotation of the star and its gravitational field effects. Thus, they provide a valuable diagnostic tool for studying neutron star properties (since the pulse profiles are affected by its gravitational field) and the dynamics of the burst process (which determines the surface emission pattern) \citep[for a review, see][]{2012ARA&A..50..609W}.

During the rise phase of a burst, oscillations are believed to be caused by a hot spot expanding from the initial point of ignition. This theory, supported by research from \cite{Strohmayer97, Spitkovsky02, Bhattacharyya06a},  and others, suggests that the asymmetry created by the growing hot region produces the oscillatory signal. However, this explanation still requires further investigation to be fully confirmed.

The origin of burst oscillations during the cooling (or tail) phase remains a subject of debate. Theories that explain non-uniform emission include varying fuel depths across the neutron star surface and cooling wakes \citep[e.g.][]{Zhang2013}. An alternative and intriguing hypothesis involves the excitation of global surface modes triggered by instabilities in the burning ocean both during the propagation of the deflagration front and in the cooling phase. This idea was initially proposed by \cite{Heyl2004} and expanded upon by subsequent theoretical work \citep[e.g.][]{Cumming05, Lee05, Piro05, Berkhout2008}. 

eXTP-SFA will enable high-precision tracking of burst oscillation frequency drift, which is crucial for understanding the burning front propagation and oscillation mechanism.
eXTP will also allow for the first high-statistics studies of time-dependent phase-resolved spectroscopy during bursts.
By tracking the spectral evolution of the oscillation signal, we can distinguish between flame spreading,  wave patterns, and other potential burst asymmetry mechanisms.

$\bullet$  {\it Is there a preferred ignition latitude on the neutron star?} 

The ignition of bursts is highly sensitive to the local conditions on the NS surface, such as temperature, pressure, and composition of the accreted material. One critical factor influencing the burst behavior is the latitude of ignition. The shape and duration of the burst rise can vary depending on whether ignition occurs near the equator or at higher latitudes. For example, ignition at the equator could lead to faster spreading of the burning front due to the initial weaker fuel confinement by the NS rotation \cite{Spitkovsky02,2015MNRAS.448..445C}, resulting in shorter and more intense bursts. In contrast, ignition at higher latitudes might produce slower-spreading burning fronts and longer-lasting bursts.

According to theory, the burst rate depends mainly on the accretion rate. However, theory predicts a steady increase in the burst rate, while observationally it has been noted that many sources display a phase when the burst ignition rate decreases with the accretion rate. The reasons for this behavior are still unknown, but it has been suggested that it could be related to some varying conditions on the surface \citep{1998mfns.conf..419B,2007ApJ...657L..29C,2020MNRAS.499.2148C}, or to additional sources of heat \citep{1999AstL...25..269I,keek09,2010AstL...36..848I}, including mysterious shallow heating \citep{Brown_2009_nucurca}.  Since the problem is related to the transition of the bursts to stable burning (see also below), another possibility is errors in the estimates for the reaction rates \citep[e.g.][]{keek14,2016ApJ...830...55C}. A combination of these effects should change the ignition latitude of the bursts, which could also have an effect on the shape of the rise \citep{2008MNRAS.383..387M} and the presence and shape of the burst oscillations. Timing studies with eXTP have the potential to resolve burst oscillations during the rapid rise phase, and thus provide the clues necessary to identifying the ignition latitude and the underlying mechanism behind this striking discrepancy between observations and theory. 
We performed simulations of type I X-ray bursts ignited at three latitudes on the surface of a neutron star spinning at 400 Hz.
In the right panel of Figure \ref{X_ray_burst_lc} we show the resulting light curves of the burst rise phase, as well as the temporal evolution of the 400 Hz burst oscillation intensity, demonstrating how eXTP-SFA could distinguish the three cases.

$\bullet$  {\it Transition from unstable to stable nuclear burning  }

Marginally stable burning occurs in the transition between unstable and stable burning. In this regime, the heating rate is alternatively greater and smaller than the cooling rate, while remaining always very close \citep{heger07}, thus generating QPOs at a typical frequency of 5--10 mHz (mHz QPOs) in the soft X-ray band \citep[e.g.,][]{revni01,heger07,linaries12}. The mHz QPOs occur within a narrow range of non-burst luminosities, $L_{2–20 \rm keV}$ $\simeq$ (5–11) $\times$ 10$^{36}$ erg~s$^{-1}$, and should be detected by the SFA with strong flux variations below 5 keV. The mHz QPOs usually, but not always, disappear right before a type I X-ray burst \citep[e.g.,][]{revni01,diego08}.

Currently, a crucial question is how the mHz QPOs come about. It has been suggested that they are due to nonuniform, area changing, surface emission \citep{stiele16,hsieh20} or to a global variation of the surface blackbody temperature \citep{strohmayer18}. Which scenario is correct?

\begin{figure}[H]
\center
\includegraphics[width=0.52\textwidth]{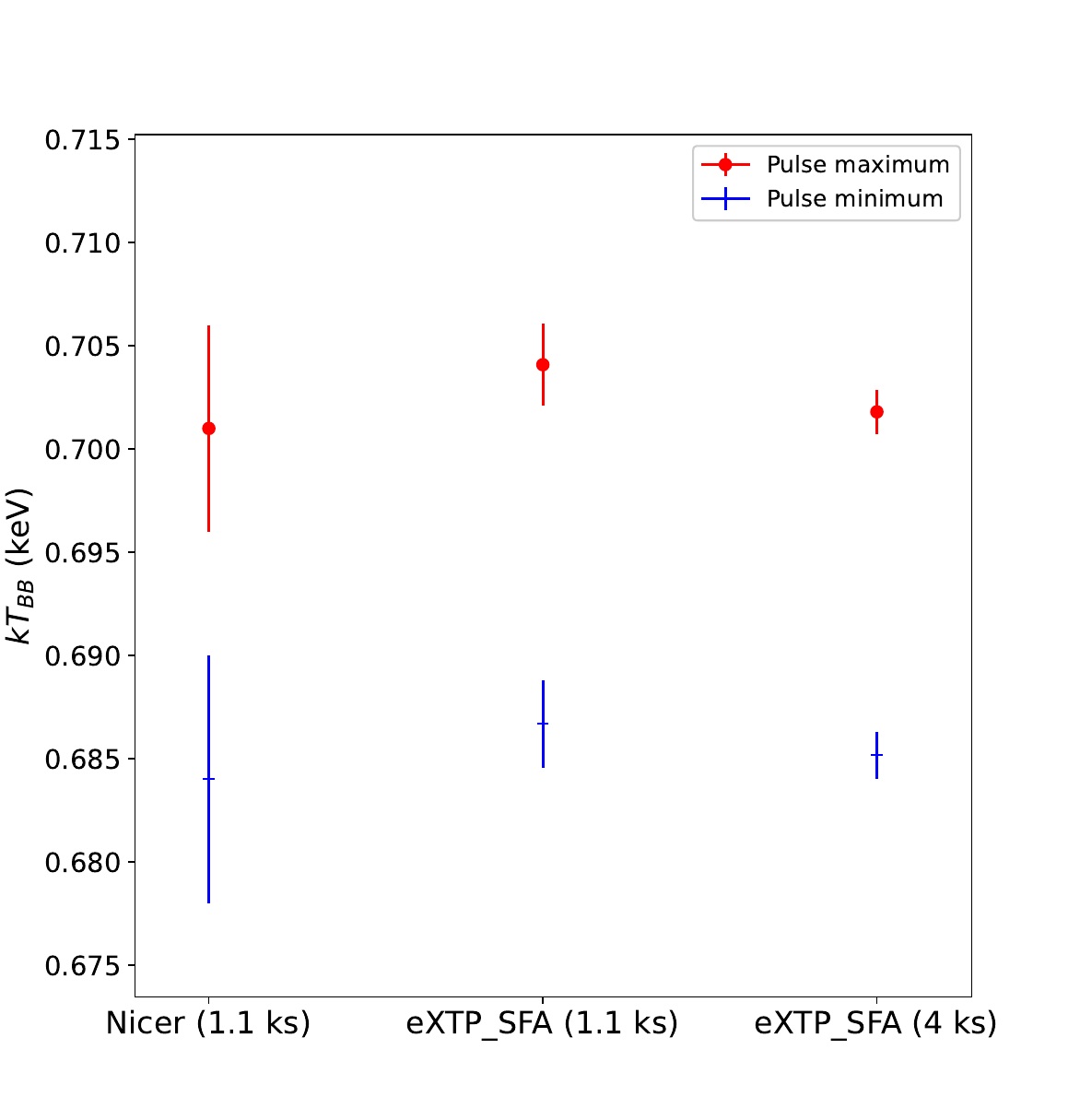}
\caption{Blackbody temperatures and their 1$\sigma$ errorbars constrained by NICER and eXTP at different exposures. The NICER results are taken from \citep{strohmayer18}, where a $\sim$3$\sigma$ level temperature difference is obtained from a joint fit of the spectra at the minimum and maximum QPO phases. The eXTP results are derived from simulations with the same model and the parameter settings as in \citep{strohmayer18}. The eXTP data demonstrate an enhanced capability for precise temperature constraints, thus enabling a robust identification of the possible temperature variations on the neutron star surface.}
\label{kTbb_compare}
\end{figure}

The results supporting the former scenario are based on data from XMM-Newton, while the second interpretation is based on NICER data. The different conclusions probably stem from the fit parameter degeneracy between the blackbody temperature and its normalization. The degeneracy is strong when the photon counts are low.
Leveraging its large effective area, eXTP-SFA is anticipated to provide substantial photon statistics, thereby enabling us to put stringent constraints on the relevant parameters. For example, Figure~\ref{kTbb_compare} compares the fit results from simulated observations with 5 SFA-T modules to those obtained by NICER. The 1-$\sigma$ error bars are reduced by a factor of $\sim$ 2 to 3.

Another key issue is how to trigger marginally stable burning at the observed low accretion rate (10\%--50\% Eddington) \citep[e.g.,][]{revni01,diego08,lyu19} rather than the $\sim$ 100 \% of Eddington rate predicted by simulations \citep{heger07,keek14}. Two scenarios have been proposed to bridge this accretion rate gap: (a) it is the local accretion rate of the burning region, instead of the global average rate of the entire neutron star, that triggers the marginally stable burning \citep{heger07,lyu16}; (b) there is some extra heat flux from the crust and/or a turbulent mixing of helium in the deep layer, which could increase the stability of the burning and hence lower the critical rate for the stable burning \citep[e.g.][]{keek09}. The validity of either scenario is still debated, and it is challenging to directly study the local accretion rate or turbulent mixing from observations. In the transitional spectral state, the frequency of the millihertz QPOs is observed to systematically decrease with time, until the QPOs disappear and an X-ray burst occurs. The turbulent mixing scenario predicts an increase in the oscillation amplitude along with this drift. Thus, precisely measuring the evolution of the amplitude is a possible way to test this scenario. With its large effective area in the soft X-ray band, the SFA is promising in identifying these possible subtle amplitude variations and hence establishing the validity of turbulent mixing theory.

 A further conundrum is the coexistence of mHz QPOs and type I bursts in some sources \citep[e.g.,][]{revni01,diego08,lyu15}, since the two phenomena should appear sequentially \citep{heger07}. A possible explanation can again be the different local conditions on the surface of the star \citep{2007ApJ...657L..29C,lyu16,2020MNRAS.499.2148C}. The spectral-timing capabilities of eXTP-SFA can allow us to study in greater detail both the persistent and the burst/mHz QPOs emission properties, giving clues to identify which solution best explains the observations.

$\bullet$  {\it Interactions between burst radiation and accretion flows} 

The sudden intense X-ray emission from a thermonuclear burst will interact with the surrounding accretion environment \citep[for a review, see][]{2018SSRv..214...15D}, more so if the flux or fluence are higher such as in superbursts or intermediate-duration bursts \citep[]{stroh2002,cumming2002,cumming2006,zand2011}. In contrast, the accretion flow will affect the burst radiation. This interaction is expressed through multiple effects. Bursts can cool the corona (causing a hard X-ray deficit, \citealp{2012ApJ...752L..34C}), alter the accretion flow (leading to an increase of the accretion spectrum, \citealp{2013ApJ...772...94W} or a decrease \citealp[e.g.,][]{intzand11}), and can produce reflection features\citep{2004ApJ...602L.105B} as the burst photons bounce off the accretion disk. Although these interaction mechanisms likely coexist during thermonuclear bursts, their relative dominance remains unclear and is likely to depend on system-specific conditions. 
The hard X-ray deficit has been detected in 10 bursters to date, out of $\approx$120 known, often in systems with low accretion rates. Furthermore, the observed saturation of the deficit fraction at 50\% in several bursters implies the coexistence of two distinct hard X-ray emission mechanisms \citep{2024MNRAS.531.1756C}: a quasi-spherical corona undergoing burst-induced cooling and a persistent jet structure near the polar caps that remains unaffected during bursts. 
Accretion spectrum enhancement (the ‘$f_{rm a}$ effect’, \cite{2013ApJ...772...94W,Worpel2015}) is nearly ubiquitous in bright bursts (e.g., for peak luminosities higher than 10\% of the Eddington limit). However, the correlation between burst-induced accretion spectrum enhancement and burst emission varies significantly between burst sources \citep{2015ApJ...806...89J}, indicating a source-dependent coupling mechanism between these phenomena, likely depending on the geometry and perspective of the accretion flow.
There exists a degeneracy among the interaction mechanisms described above. For example, both the increased accretion rate model and the disk reflection scenario could explain the observed enhancement in the soft X-ray band \citep{2018ApJ...855L...4K}. 
The accretion circumstance can also have an impact on the burst emission. For instance, for the most luminous bursts with photospheric expansion (PRE), there is a peculiarity that the bolometric flux seems to keep increasing through the Eddington-limited phase. This is observed in bursts at a high mass accretion rate, but not for bursts with a faint persistent emission, which has been predicted theoretically but only observed in a single burst \citep{2023JHEAp..40...76C}. If this effect is due to disk obscuration, the phenomenon indicates that the anisotropy of the burst emission is accretion-rate dependent, which could be evidence of the truncated disk in the low/hard state, and the phenomenon could be used to estimate the inner disk radius.
An enhancement in hard X-ray emission has been observed during short-duration bursts in two bursts. This phenomenon is thought to result from photon upscattering by either the accretion disk corona or the boundary layer between the accretion disk and neutron star. Notably, burst sources that exhibit elevated persistent luminosities, such as Cyg X-2 and Cir X-1, frequently demonstrate atypical characteristics, including the absence of spectral softening. These deviations from standard burst behavior are similarly thought to originate from coronal-scattering processes, where the persistent emission's hot electron population interacts with the burst photons.

With its unmatched effective area below a few keV, the SFA will uniquely probe dominant burst-accretion interactions and probe the accretion physics for burst sources with relatively low interstellar absorption.
The SFA observations of bursts will enable a more accurate determination of the enhanced accretion rate and disk reflection parameters, providing a more detailed description of the accretion conditions throughout the bursts.

\subsubsection{Compact and transient jets in BHXRBs} 

Theoretical models and numerical simulations suggest that the accretion flows surrounding black holes generate the magnetic fields and energy necessary to launch relativistic jets \citep[e.g.][]{BZ1977,BP1982,Tchekhovskoy2011,Liska2020}. In transient BHXRBs, two types of jet are observed: a compact jet during the hard-state transition and a transient jet during the hard-to-soft-state transition. Significant progress has been achieved through multi-wavelength observations, offering valuable insights into jet activities and their couplings with accretion processes \citep[see reviews such as][]{Fender2016}. Nevertheless, a comprehensive understanding of the mechanisms of jet formation and acceleration, as well as its large-scale propagation, is still lacking. Likewise, the physical processes that govern individual jet components and their connection with the rapid changes of accretion flow continue to present unresolved questions in both stellar-mass and super-massive accreting BHs.

Multi-wavelength spectral, timing, and polarization observations are critical for constraining the physical properties of relativistic jets in accreting black hole systems. The eXTP, with its unique capability to simultaneously acquire X-ray spectral, timing, and polarization data, will play a pivotal role in future coordinated multi-wavelength campaigns. Its extended observational exposure times, compared to low-orbit X-ray satellites, significantly improve the feasibility of achieving truly simultaneous observations with instruments operating at other wavelengths, such as optical and radio facilities. Spectral analysis across gamma ray, X-ray, optical, and radio bands reveals energy distributions and particle acceleration mechanisms, while multi-wavelength timing studies establish causal connections between variability in distinct emission regions of the corona and jet. Polarimetric measurements, particularly in X-ray, optical, and radio regimes, provide critical insights into the geometry and orientation of magnetic fields within both the jet and corona, constraining jet-launching mechanisms and collimation processes. Together, these techniques form a unified framework for determining key properties of jet and/or corona, such as magnetic-field topology, electron energy distributions, and coronal geometry. This approach not only resolves degeneracies in spectral modeling but also advances our understanding of jet dynamics and the energetics in accreting BHs.

High-cadence observations of eXTP will enable us to capture X-ray spectral, timing, and polarimetric variations associated with changes of the corona, while radio imaging will track the onset and propagation of downstream transient jet components. The prolonged exposure capacity of eXTP is particularly well suited to capture the precise moment of the launch of the transient jet during the state transition \citep[e.g.][]{Homan2020}. When combined with multi-epoch radio interferometry observations, eXTP promises to unravel the causal relationship between corona evolution and transient jet ejection. This synergy will advance our understanding of the physical process underlying state transitions, such as geometric and/or magnetic reconfiguration of the corona, and energy drainage into jet launching.

\subsection{High-mass X-ray binaries}

\subsubsection{Gamma-ray binaries} 

Gamma-ray binaries are a rare subclass of high-mass X-ray binaries (HMXBs) comprising a compact object (neutron star or black hole) and a massive O- or Be-type companion star. These systems derive their name from their characteristic spectral energy distributions (SEDs), which peak at high energies (HE; $\geq1$~MeV) and extend to very high energies (VHE; $\geq100$~GeV) \citep{Dubus2013}. To date, only 7 confirmed gamma-ray binaries have been identified within the Milky Way, with one additional system discovered in the Large Magellanic Cloud \cite{Chernyakova2020}. Their broadband emission exhibits strong orbital modulation, attributed to relativistic particle acceleration in colliding stellar winds or jets, synchrotron radiation at lower energies (radio to X-rays), and  ICS of stellar photons at the HE and VHE bands \citep[see, e.g., ][ and references therein]{Dubus2013}.

The detection of radio pulsations in PSR~B1259$-$63 \citep{Johnston1992}, PSR~J2032+4127 \citep{Camilo2009}, and LS~I~+61$^\circ$~303 \citep{Weng2022}, provides compelling evidence that at least some gamma-ray binaries are powered by the spin-down energy of a young, rotation-powered neutron star. In these three binary systems, gamma-ray emission is thought to originate from shocked pulsar winds interacting with the companion's stellar wind or circumstellar disk (see the left panel of Figure \ref{polar_B1259}).
Apart from these three cases, the compact object in most gamma-ray binaries remains unconfirmed. For example, other gamma-ray binaries lack solid pulsar signatures, leaving their energy mechanisms debated (in particular, for LS~5039, see, \citep{Yoneda2020, Volkov2021}.)

Therefore, the microquasar model, where relativistic jets from accretion onto a black hole or neutron star produce nonthermal emission, remains a viable alternative to the colliding wind model \citep{Paredes2019}. Disentangling these models requires multiwavelength studies of orbital-phase-resolved spectra, including the X-ray band and its polarization, as well as deeper pulsation searches.

In the 0.5--10 keV energy band, gamma-ray binaries typically show nonthermal spectra characterized by hard power-law emission with photon indices in the range $1.5-2$, with no measured Fe line component and flux modulation over orbital timescales \citep{Dubus2013}.
The usual assumption is that this X-ray emission is attributed to synchrotron radiation from relativistic electrons accelerated within the system, possibly in the shock formed by the interaction of a pulsar wind and the wind of the massive companion star, or in jets. Other models have
been proposed where the X-ray emission is due to inverse Compton emission
\citep{Chernyakova2009, Zdziarski2010, Chernyakova2020}.

The sensitivity of the telescopes on board eXTP will provide crucial key information and address the following two fundamental questions:

\begin{figure*}
\centering
\includegraphics[width=0.8\textwidth]{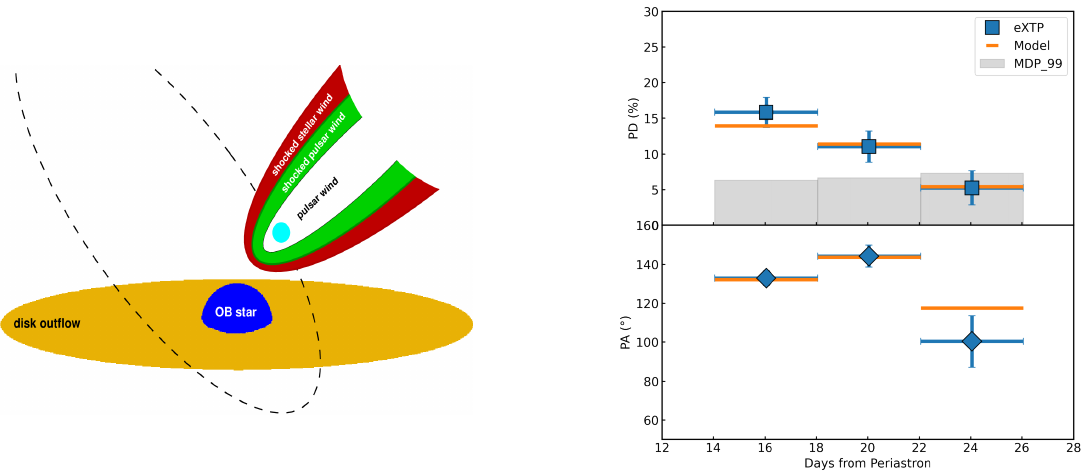}
\caption{Left panel: the relativistic outflow from a rotation-powered pulsar collides with the stellar wind (and, if present, the circumstellar disk) of its OB type star companion. The green and red areas represent the shocked pulsar wind and the shocked stellar wind, respectively. Right panel: simulating 14 days of eXTP-PFA observations based on IXPE data of PSR~B1259-63 yields higher significance, allowing clear detection of orbital-phase-dependent polarization variations. The PFA simulation results are represented by blue data points, while the corresponding model predictions are marked in orange.}
\label{polar_B1259}
\end{figure*}

$\bullet$ {\it What drives high-energy emissions?}

Recently, {\it IXPE} detected significant X-ray polarization from the gamma-ray binary PSR~B1259$-$63 \citep{Kaaret2024}.
The measurement, performed during an X-ray bright phase following the passage of the periastron in 2024 June, revealed a degree of polarization of 8. 3\% $\pm$ 1. 5\% with a significance of 5.3$\sigma$. The X-ray polarization angle was found to be aligned with the axis of the shock cone at the time of the observation. For synchrotron emission, the polarization angle is perpendicular to the magnetic field,  which indicates that the dominant component of the magnetic field within the particle acceleration region of PSR~B1259$-$63 is oriented perpendicular to the axis of the shock cone. This result is consistent with the scenario of a toroidal magnetic field structure around the shock, previously proposed by \citep{Xingxing2021} along with other possible magnetic field configurations. In this model, the PA is expected to change with the orbital phase, which can be tested with the PFA data.
Based on the IXPE polarization measurements, our simulation of a 14-day observation with eXTP-PFA not only achieves higher significance (7.3$\sigma$, $PD = 8.8\% \pm 1.2\%$) but also enables us to clearly detect variations in the degree of polarization with orbital phase (see the right panel of Figure \ref{polar_B1259}).

Alternatively, the microquasars SS~433 \citep{Kaaret2024a} and Cyg~X-3 \citep{Veledina2024}, exhibit distinct X-ray polarization properties: their higher polarization degrees (PD) and magnetic field orientations aligned with their relativistic jets differ markedly from the predictions of the collision wind scenario which fit the case of PSR~B1259$-$63. Therefore, these observational signatures provide a critical diagnostic to distinguish between emission mechanisms. eXTP will not only probe particle acceleration physics in these extreme environments but also refine orbital parameters, in particular for systems with moderate eccentricity (e.g. LS~I~+61~303 \cite{Aragona2009, Kravtsov2020}).

$\bullet$ {\it What is the nature of compact objects?}  

The detection of pulsed emission would prove the pulsar scenario. Although the X-ray flux of gamma-ray binaries is dominated by the unpulsed synchrotron emissions in the acceleration region, rotation-powered pulsars themselves can also emit a significant amount of X-rays due to rotational energy loss ($L_{\rm X}/\dot{E} \sim 10^{-6}-10^{-1}$ \cite{Xu2025}). Furthermore, the detection of short bursts will be evidence of magnetars or young energetic pulsars in binary systems, e.g. LS~I~+61~303 \citep{Papitto2012, Torres2012}. With its high sensitivity and excellent time resolution, eXTP is well suited for these types of studies.

Finally, we note that the recent literature reports few marginal detections of transient gamma-ray sources spatially coinciding with accreting pulsars in HMXBs that are neither microquasars nor colliding-wind binaries \citep{Li2012, Xing2019, Harvey2022}. These outliers might represent only a small fraction of the total population of HMXBs, which, on the basis of X-ray observations, is known to be significantly larger than those of conventional gamma-ray binaries. Although numerous hypotheses have been proposed, the mechanism driving gamma-ray emission from accreting X-ray pulsars remains poorly understood \citep[][ and references therein]{Ducci2023}.
In this context, the superb photon collecting area, excellent timing resolution, and good spectral capabilities of the SFA and PFA will be critical for obtaining deeper insights into the accretion flow dynamics and magnetic field geometry of these sources.

\begin{figure*}[ht!]
\centering
\includegraphics[width=1.0\textwidth]{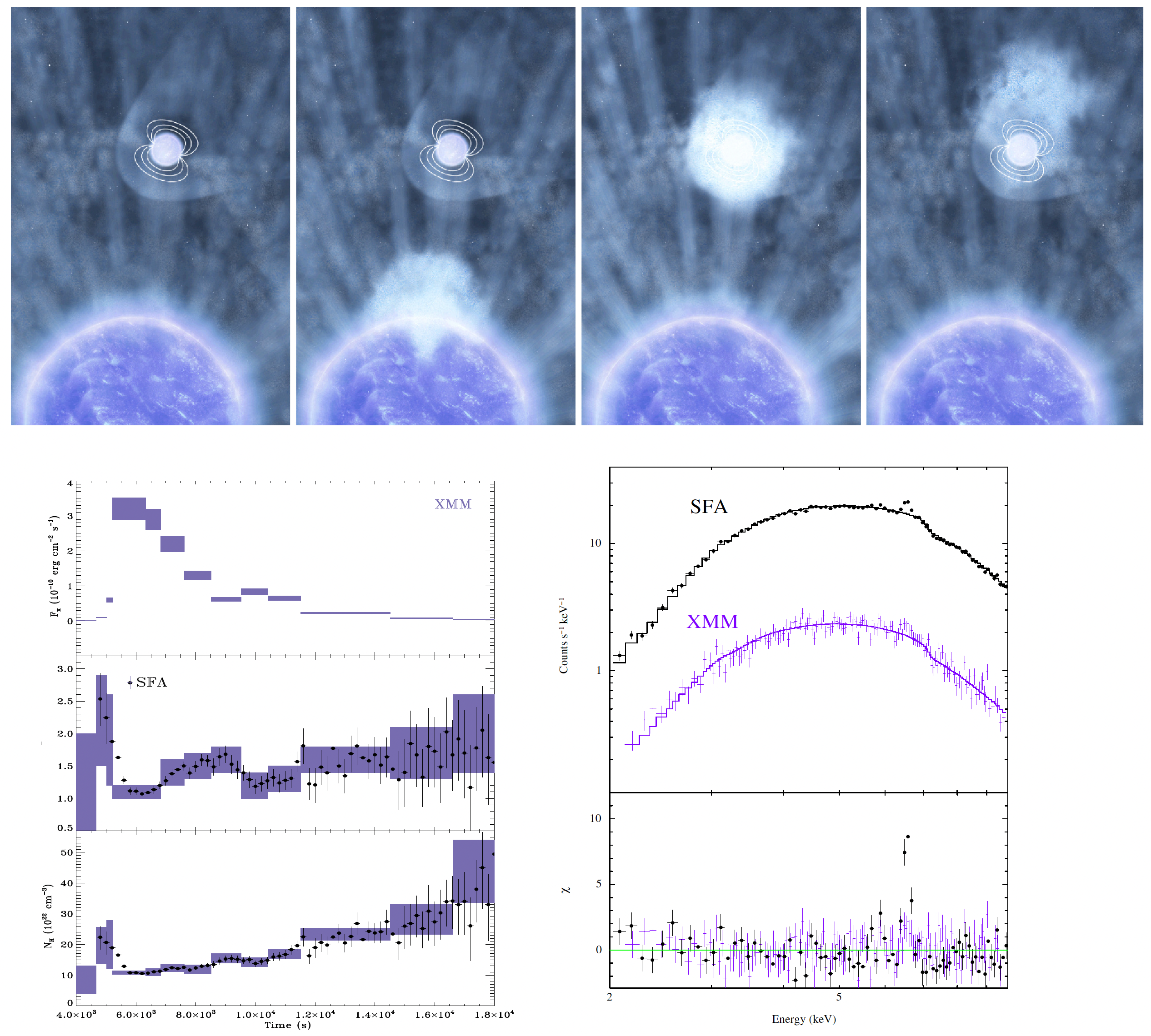}
\caption{{\it Top}: 
  Sketch of the upper neutron star accreting 
  a clump ejected by the lower supergiant star (credits: ESA). {\it
    Bottom left}: Changes in the flux and spectral parameters during
  the flare recorded by \xmm, from the SFXT IGR~J18410-0535
  \citep{Bozzo2011}. Violet boxes represent the measurements obtained
  with \xmm, while black points represent values obtained from the
  simulated \lfa spectra with exposure times as short as 200\,s.
  Compared to \xmm, the dynamic process of the clump accretion can be
  studied in much more detail and the fast spectral changes can be
  revealed to an unprecedented accuracy (we remark also that the
  pile-up and dead-time free data provided by the  \lfa
  greatly reduce the uncertainties affecting the \xmm, data obtained
  so far from HMXBs during bright flares). {\it Bottom right:} Shown,
  for comparison, is an example of an \xmm\ spectrum extracted in a
  1\,ks-long interval (source flux 3$\times$10$^{-10}$~\cgsflux) during
  the decay from the flare shown on the left and the corresponding
  simulated \lfa spectrum.  The Fe-K line at 6.5~keV, used to
  probe the clump material ionized by the high X-ray flux, is barely
  visible in \xmm, but very prominently detected in the \lfa spectrum.}
\label{fig:flare_hmxb} 
\end{figure*}

\subsubsection{Supergiant High Mass X-ray Binaries} 

Massive supergiant, hypergiant, and Wolf-Rayet stars
($M\gtrsim$10~$M_{\odot}$) have the densest, fastest, and most
structured winds. The radiatively accelerated outflows trigger star
formation and drive the chemical enrichment and evolution of Galaxies \citep{kudritzki2002}. 
The amount of mass lost through these winds has a
large impact on the evolution of the star. The winds can give rise to
an extremely variable X-ray flux when accreted onto an orbiting
compact object in a high-mass X-ray binary (HMXB).  The understanding
of the relation between the X-ray variability and stellar wind
properties has been limited so far by the lack of simultaneous large
collecting area and good spectral and timing resolution.

In the past decades, observational evidence has been growing that
winds of massive stars are populated by dense clumps.  The presence
of these structures affects the mass loss rates derived from the
optical spectroscopy of stellar wind features, leading to
uncertainties in our understanding of their evolutionary paths
\citep{Puls2008}.  HMXBs were long considered an interesting possibility
to probe clumpiness \citep{Sako2003,bozzo2008,nunez17,kre19}. As X-rays released by accretion
trace the mass inflow rate to the compact object, an HMXB provides a
natural in situ probe of the physical properties of the massive star
wind, including its clumpiness \citep{zand2005,Walter2015:HMXBs_IGR}. The
so-called Supergiant Fast X-ray Transients (SFXTs) form the most
convincing evidence for the presence of large clumps. In X-rays, the
imprint of clumps is two-fold: 1) clumps passing through the line of
sight to the compact object cause (partial) obscuration of the X-ray
source and display photo-electric absorption and photo-ionization; 2)
clumps lead to temporarily increased accretion and X-ray flares.  Several hour-long flares displayed by the SFXTs could be
convincingly associated with the accretion of dense clumps
\citep{rampy09, Bozzo2011, bozzo17, ferrigno22}. 

The instruments on-board eXTP will dramatically open up perspectives
for research in the above mentioned fields.

$\bullet$ {\it What are the physical properties of massive star wind structures and how do these affect the mass loss rates from these objects?}  

With current instruments, integration times of several hundreds to thousands of seconds are needed to get a rough estimate of clump properties and only an average picture of the clump accretion process can be obtained.  Similar observations performed with the SFA-T on-board eXTP will dramatically improve our present understanding of clumpy wind accretion and winds in massive
stars in general by studying (spectral) variability on time scales as short as a few to tens of seconds (see Figure~\ref{fig:flare_hmxb}).
  This will permit a detailed investigation of the dynamics of the clump accretion process and obtain more reliable estimates of the
  clump mass, radius, density, velocity, and photoionization state. In turn, it will improve our understanding on the mass loss rates from massive stars. Because structured winds are not spherically symmetric, they are also predicted to emit polarized X-ray
  radiation. Measurements with \gpd can yield additional information to possibly constrain the stellar mass loss rate \citep{Kallman2015}.  The improved capabilities of the SFA-T
  compared to the current facilities will permit conducting also similar studies on the SyXBs \citep{Enoto2014}, thus exploring the structure and composition of the still poorly known winds of red giant stars (which are accelerated through the absorption of the stellar radiation by dust grains rather than heavy ions). 


$\bullet$ {\it What role does X-ray reprocessing play in diagnosing accretion flow structures?}

\begin{figure}[H]
    \centering
   \includegraphics[width=0.5\textwidth]{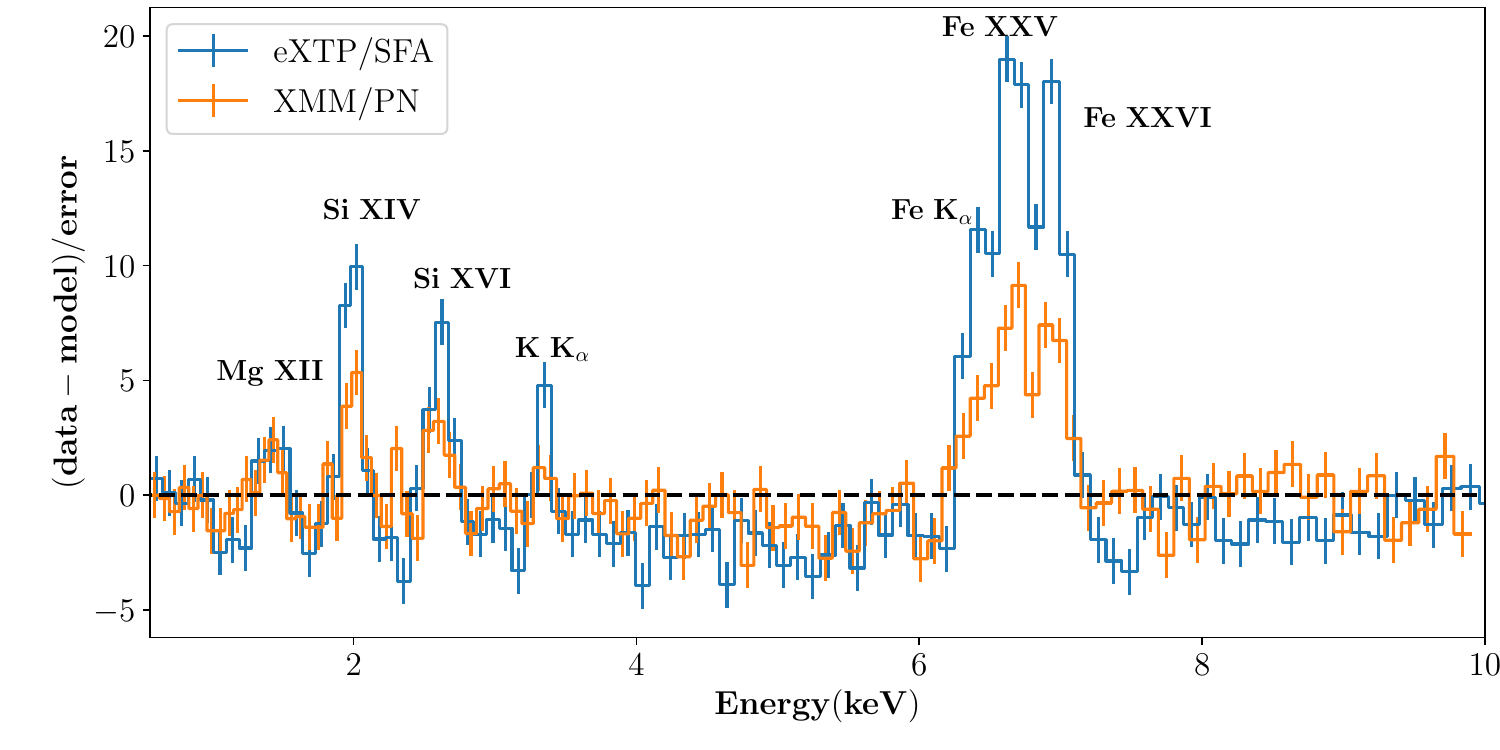}
    \caption{Residuals resulting from fitting a continuum model to simulated SFA and XMM-Newton EPIC-PN spectra, which are produced by using a model including both the continuum and additional narrow emission lines. We assumed an exposure of 10\,ks and a flux of $10^{-11}$\,$\rm erg/s/cm^2$, and input spectral parameters were adopted from a HMXB Cen X-3 \citep{Aftab2019}. 
    }
    \label{fig:compare_XMM_eXTP_residuals}
\end{figure}

X-rays originating near compact objects are subject to absorption by surrounding matter, followed by subsequent re-emission.
This phenomenon, known as ``X-ray reprocessing", manifests primarily through characteristic fluorescence lines, such as Fe K$_\alpha$ at 6.4 \,keV, Fe XXV He triplets around 6.7\,keV, and FeXXVI Lya line at 6.97\,keV.
These features are related to the distribution of the accreting matter, which could therefore serve as a powerful diagnostic probe for the environment in binary systems \citep{Torrejo2010,Aftab2019,Ji2021}.
eXTP, with a large effective area of the SFA, is quite suitable for studying line emissions, especially in faint sources and low states of variable sources (see the comparison between eXTP-SFA and XMM-Newton-PN in Figure~\ref{fig:compare_XMM_eXTP_residuals}).
For example, Cen X-3 shows enhanced Fe lines in its low states, suggesting that the suppressed flux is due to obscuration by dense matter rather than a reduction of the accretion rate \citep[e.g.][]{LiuJR25}. 
In addition, in a scenario where primary X-ray emission is heavily obscured and the observed flux is dominated by reprocessing, a high degree of polarization is expected, which may reach 25\% as suggested by IXPE in thick AGNs Cyg X-3 and Compton \citep{Veledina2024,Ursini2023a}.
In this case, PFA polarization studies will play a pivotal role in understanding the structure and geometry of accretion flows.

\section{Ultraluminous X-ray sources} \label{sec:ulx}

Ultraluminous X-ray sources (ULXs) are typically luminous, off-nuclear point sources with X-ray luminosities $L_\mathrm{X} \gtrsim 10^{39}\,\mathrm{erg\,s}^{-1}$, which exceed the Eddington luminosity
\(
L_\mathrm{Edd} \simeq 1.3\times 10^{38} \, (M/M_\odot)\,\mathrm{erg\,s}^{-1}
\)
of a $10\,M_\odot$ black hole accretor \citep{Kaaret2017,Fabrika2021,King2023,Pinto2023}. 
Several explanations have been proposed for ULXs, including accretion onto intermediate-mass black holes \citep{Colbert1999}, photon-bubble instabilities that alter the accretion disc structure \citep{Begelman2002}, beaming (anisotropic emission) leading to reduced accretion rate estimates \citep{King2001}, and strong magnetic fields modifying photon opacities in pulsating ULXs \citep{Mushtukov2015}. 
The significance of ULXs has been recognized across a broad range of topics, including the growth of early supermassive black holes, formation channels of black holes and neutron stars, binary evolution pathways such as mass transfer and accretion efficiency, and estimates of neutron star magnetic fields, particularly in pulsating ULXs \citep{Tsygankov2017,Misra2020,Suh2025}.

$\bullet$ {\it What is the true proportion of pulsating ultraluminous X-ray sources?}

So far, nearly 2{,}000 ULXs and candidate ULXs have been discovered in approximately 1{,}300 galaxies \citep{2024Tranin,2022MNRASWalton}. However, only a subset of ULX with pulsations (PULX) has been identified \citep{2025AN....34640102I}. Among all ULXs observed by \textit{XMM-Newton}, the known PULXs comprise only about 2\%, and most lack long-term monitoring data. ULXs are primarily discovered via \textit{XMM-Newton} and \textit{Chandra}, but only \textit{XMM-Newton} has adequate time resolution to detect periodic pulse signals from PULXs. Given that ULXs are located at significant distances from Earth (spanning a few to hundreds of Mpc), they are often confused with foreground or background X-ray sources (especially AGNs) \citep{2011MNRAS.416.1844W}.
Moreover, pulsation search algorithms such as power spectral density or $Z^2$ searches require high-quality data with sufficient count rates to detect typically weak PULX pulsations \citep{2020ApJ...895...60R}. When only \textit{XMM-Newton} data with sufficiently high count rates are considered, the fraction of PULXs within the ULX population can increase to approximately 30\% \citep{2025AN....34640102I}, strongly suggesting that the currently known PULXs are likely a small fraction of the actual population. Consequently, we anticipate identifying additional PULXs by reobserving known ULXs with the \textit{eXTP-SFA} instrument and targeting those ULXs for which the existing time resolution or statistical quality of data is insufficient. Furthermore, by conducting long-term rotational monitoring of PULXs and applying accretion/torque models, we can constrain the magnetic field, thereby advancing our understanding of how PULXs achieve super-Eddington luminosities.

$\bullet$ {\it Probing intermediate‑mass black holes in ULXs}


\begin{figure*}
    \centering
   \includegraphics[width=\textwidth]{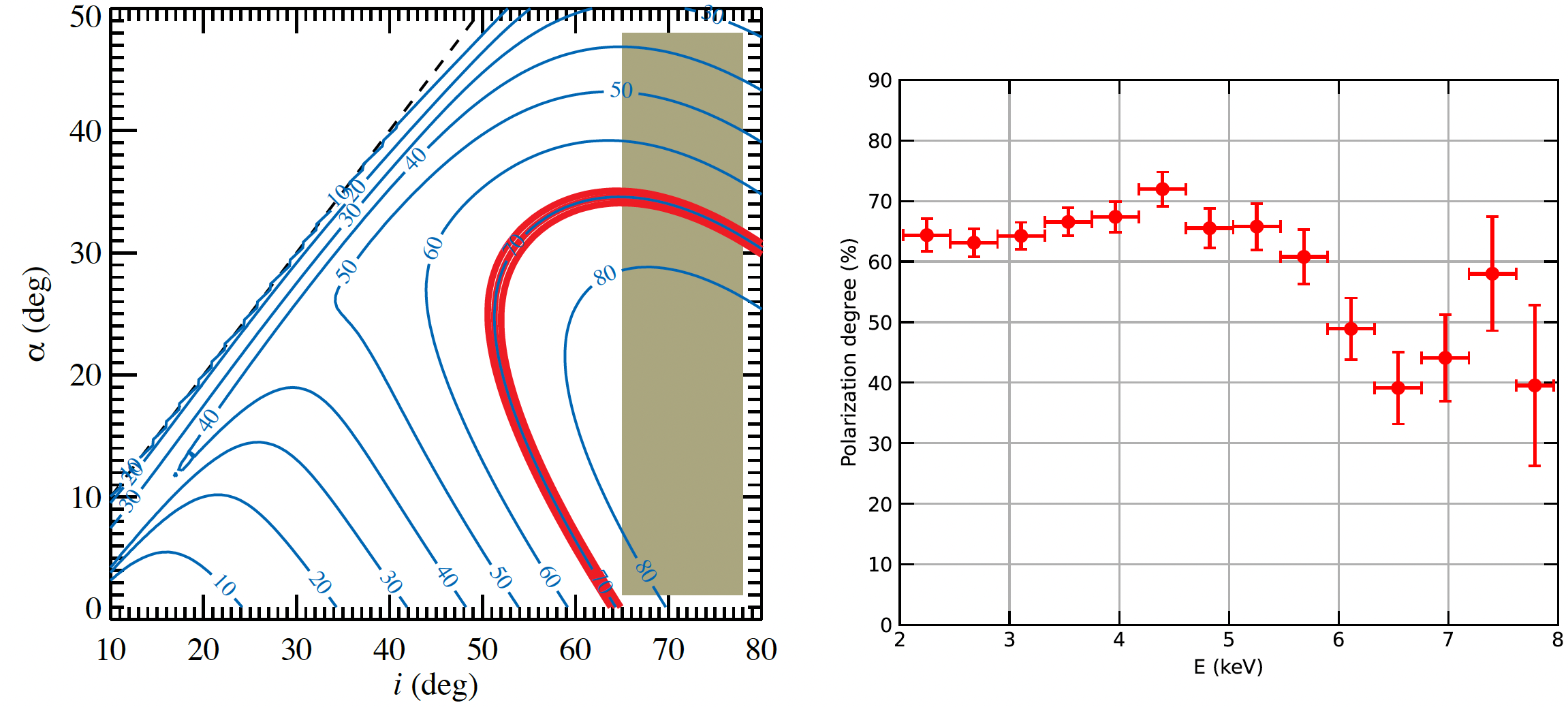}    
\caption{Left panel: Dependence of the funnel opening angle on the inclination of the binary system for a ULX seen from side. Contour plots correspond to expected polarization levels (in per cent). The shaded area corresponds to the inclinations of potential Galactic ULXs. Using 100~ksec exposure time with the PFA on a source of 5 mCrab, we will be able to constrain the polarization degree with the accuracy of $\sim1$\%, which translates to the estimates of the opening angle better than $5^{\circ}$ (red stripe in the figure). Based on Figure~4 of \citep{Veledina2024}. 
   Right panel: simulated PFA data with reflected continuum polarized at 70\% and an unpolarized iron line at 6.5 keV.}
   \label{fig:funnel_alpha-i}
\end{figure*}

Intermediate-mass black holes (IMBHs) have been proposed as possible accretors in some ULXs. The large effective area and superior timing resolution of the SFA enable detailed spectral and timing analyses, offering a pathway to test this scenario. In particular, QPOs serve as a potential diagnostic tool. Although QPOs have been well established in stellar-mass black hole binaries (BHBs; $M \sim 10\,\mathrm{M}_\odot$), they have also been detected in AGN hosting supermassive black holes ($M \sim 10^6$--$10^{10}\,\mathrm{M}_\odot$), such as RE~J1034$+$396~\citep{2008Natur.455..369G} and 1ES~1927$+$654~\citep{Masterson2025}, albeit at much lower frequencies due to the larger black hole masses.
Using SFA' enhanced timing capabilities, ULXs can be probed for QPOs on a wide range of frequencies, allowing mass-scale comparisons with BHBs and AGNs. Such analyses could provide critical evidence for the presence of IMBHs in ULX systems.

$\bullet$ {\it What is the innermost accretion geometry of super-critical sources?}

The accretion mode of an accretion disk is fundamentally influenced by the accretion rate. In regimes characterized by low accretion rates, the disk can be accurately modeled as a geometrically thick and optically thin configuration \citep{Shakura1973}, governed mainly by radial migration. In contrast, under super-Eddington accretion conditions, strong radiation pressure inflates the disk, launching outflows loaded with matter and forming funnel-like structures \citep{Abramowicz1978,Abramowicz1988,Poutanen2007}. These funnels enhance emission along the outflow axis while blocking radiation at high inclinations, making ULXs appear dimmer to observers at large viewing angles. The opening angle of the funnel depends on the mass accretion rate and determines the emission amplification factor in the direction of its axis \citep{Takeuchi2013}.

The innermost accretion geometry of super-critical sources leaves distinct imprints on the observed spectral and timing properties. A high-inclination observer predominantly sees reflection-dominated spectra formed at the outer parts of the funnel. These spectra are often accompanied by strong emission lines (with high equivalent widths). Galactic sources such as Cyg~X-3, V404~Cyg, V4641~Sgr, and GRS~1915$+$105 display these characteristics \citep{Koljonen2020}.

An alternative and robust approach to probing the geometry of an emitting system is through polarization.
Polarization signatures of a purely reflected signal can be quite distinctive: typically, the polarization degree is high (over $\sim$ 10\%) and largely energy independent, reaching above 80\% in systems viewed edge-on, as illustrated in Figure~\ref{fig:funnel_alpha-i}. 
Moreover, the polarization angle is orthogonal to the funnel axis, which is often determined by radio imaging of the system’s jets or ejections.
The prominent iron lines around 6.5\,keV are expected to be unpolarized, thus reducing the overall polarization. 

Cyg~X-3 is the first super-critical accretor for which X-ray polarization has been measured \citep{Veledina2024}. 
IXPE observations provided robust evidence for a funnel-like accretion geometry, revealing an unexpectedly small opening angle for the funnel in this source, thus supporting beam-driven ULX models \citep{King2001}. 
Studies of ULXs are further complemented by X-ray polarimetry of the first Galactic PULX, Swift~J0243.6+6124 \citep{Poutanen2024}, which, however, shows no evidence of a funnel, favoring a strong magnetic field explanation \citep{Mushtukov2015}.

These studies were performed for sources at moderately bright fluxes: Cyg X-3 has a relatively low inclination, $i\sim30^{\circ}$ \citep{Antokhin2022}, thereby enabling significant X-ray flux reflection from the funnel walls; the transient Swift J0243.6+6124 did not reach a significant super-Eddington luminosity phase during the past observations \citep{Poutanen2024}.
Studies of both dimmer sources, which are located at higher inclinations, and brighter transients are currently challenging due to the IXPE effective area and telemetry limitations.

The large effective area and good telemetry constraints of the eXTP-PFA will enable detailed studies of both highly-obscured (high inclination) sources and extreme super-Eddington transients.
Polarization constraints will refine funnel opening angles, enabling proper comparison between different sources and enabling direct constraints on the mass accretion rate estimates.
Furthermore, these constraints will help distinguish between funnel-like (high polarization degree orthogonal to the jet) versus corona-like geometries (polarization degree smaller than about 10\% and direction along the jet), which represents an important diagnostic for the possibility of testing the strong gravity effects (see \citep{WP-WG2}).

In the left panel of Figure~\ref{fig:funnel_alpha-i} we show the constraints on the opening angle of the funnel ($\alpha$), as a function of the inclination of the system ($i$), for different observed polarization levels.
For the simulated point-like source with flux 5~mCrab, located at an inclination between $65^{\circ}<i<78^{\circ}$ (motivated by the existing constraints on the obscured sources) with the degree of polarization of the reflection continuum 70\% and 100~ksec exposure time, provides constraints on the opening angle with accuracy better than $5^{\circ}$.
The right panel gives the simulated dependence of the degree of polarization on energy. 
The drop around 6.5~keV owes to the presence of the unpolarized iron line.
These observations will help answer the long-standing questions of the accretion engine in binaries, but also have direct implications for the torus-like geometry in Seyfert 2 galaxies \citep{Ursini2023}.

\section{Active Galactic Nuclei} \label{sec:agn}

\subsection{Radio-Quiet AGNs}


~

Active galactic nuclei (AGN) are powerful X-ray emitters located at the center of the host galaxy and fueled by accretion onto a supermassive black hole (SMBH). According to the standard model of AGN, the accreting material forms an optically thick and geometrically thin accretion disk \cite{Shakura1973} that emits in the optical / UV waveband. The detection of significant amounts of X-ray flux released as a power law and extending up to hundreds of keV, requires the presence of a corona of hot (with energy up to $\sim 100-200$ keV, e.g. \citenum{Tortosa2018}) and optically thin plasma capable of ICS the optical/UV seed photons received from the disk into the X-ray band \citep[e.g.][]{Galeev1979,Haardt1991,Haardt1993}. A fraction of the primary X-ray radiation is scattered back towards the disk, where it interacts with the relatively colder accreting gas producing a reflection spectrum \citep[e.g.][]{Ross&Fabian1993,Bambi+2021}. The fraction of hard X-ray flux that irradiates the disk depends on the exact coronal geometry, which is currently unknown. Recently studies have shown that there are non-thermal electrons in the AGN hot corona \citep{2018ApJ...869..114I,2019ApJ...880...40I,2024MNRAS.527.5627L}, and could lead to gamma-ray emissions \citep{2025NatAs.tmp..141L}. A much larger extended corona is proposed to explain the observed gamma-ray emission. One of the strongest features in the X-ray reflection spectrum is the fluorescence Fe K$\alpha$ line. The profile of this spectral line is highly distorted by relativistic effects if the emission is from the innermost radii of an accretion disk extending close to the last stable orbit \citep[e.g.][]{Matt2006}. The reflection spectrum and the relativistic Fe K line are, therefore, powerful diagnostics of the strong gravity field close to the SMBH. One of the defining properties of AGN is their extreme X-ray variability. The timescales of the fastest registered variations suggest that the hot corona is very compact and is located in the innermost regions of the accretion flow.\\
The phenomenology of AGN is quite similar to that observed in stellar-mass BH X-ray binary systems. However, since the size of the innermost orbit scales with the BH mass ($\propto GM_{BH}/c^2$), the timescales that can probe the variability of such regions are longer in AGN than in stellar BH systems by a factor $10^{5-6}$, determined by the difference in mass. As a consequence, the number of collected photons on these characteristic timescales is significantly larger in AGN, making such systems ideal for time-dependent studies of the dynamics of the innermost accretion flow.
In addition, 
studying the dependence of the UV and X-ray bolometric corrections ($L_{\rm bol}/L_{\rm UV}$ and $L_{\rm bol}/L_{\rm X}$), the UV-to-X-ray luminosity ratio ($L_{\rm UV}/L_{X}$), and the spectral index between the optical/UV and X-ray continuum $\alpha_{\rm OX}$ on the SMBH 
Parameters such as accretion rate and SMBH mass provide critical insights into the coupling mechanisms of matter and energy between the disk and coronae in the vicinity of SMBHs \citep[e.g.,][]{vasudevan2009,lusso2012,2017A&A...602A..79L,2020A&A...636A..73D,2024A&A...691A.203G}. 

The outer regions of AGN are populated by clouds of gas, the so-called broad line region (BLR, e.g. \citenum{Czerny2011}), moving in roughly Keplerian motion around the BH and producing broad optical/UV emission lines. The BLR is delimited by an equatorial, dusty, and probably clumpy pc-scale torus \citep[e.g.][]{Buchner2014,Markowitz2014}. All these components are enveloped by a narrow line region (NLR) where the clouds orbit at lower speeds. Although the bulk of their emission is at longer wavelengths, these outer regions may contribute to the X-ray spectrum with reflection features due to reprocessing, and with absorption from ionized and cold material. 

The large effective area and the CCD-class spectral resolution of the SFA onboard eXTP coupled with the complementary simultaneous high-quality X-ray polarization data from the PFA will allow significant steps forward in our understanding of the AGN structure. 
The unprecedented effective area of the SFA will allow hundreds of AGN to be observed down to a flux of $10^{-13}\rm{erg~s^{-1} cm^{-2}}$ in the local universe with a very good accuracy (i.e., a signal-to-noise ratio in excess of 100). 
In particular, eXTP capabilities will be fundamental to answer the following questions.

$\bullet$ {\it What is the nature of the soft X-ray excess?} 

Below $\sim 2\ \text{keV}$ a {\it soft excess} above the extrapolation of the hard X-ray power-law continuum is ubiquitously observed in unobscured type 1 AGN, 
with an occurrence rate $\gtrsim 50\%$ \citep[e.g.][]{Bianchi+2009}. 
This feature is particularly prominent in narrow-line Seyfert 1 (NLSy1) galaxies, which are also characterized by high Eddington ratios and strong X-ray variability. Despite its prevalence, the physical origin of soft excess remains an open question.
The proposed ``warm corona'' models suggest that the soft excess represents the hard tail of a broad-band (from optical/UV to soft X-rays) continuum component, produced via Compton up-scattering of the disk photons in a warm ($kT\sim 1$ keV), optically thick ($\tau\simeq10-20$) plasma \citep[e.g.,][]{Magdziarz+1998, Done+2012, Petrucci+2018}, possibly covering the inner disk. On the other hand, reflection models explain the soft excess as relativistic blurred ionized reflection \citep[e.g.,][]{Ross&Fabian1993,Crummy_etal_2006,Bambi+2021}.
Recent theoretical \citep[e.g.,][]{Xiang+2022, Ballantyne2024} and observational studies \citep[e.g.,][]{Chen+2025} suggest that both components may coexist in AGN. However, the exact contribution of each component to the soft X-ray excess and the way in which they vary as a function of the physical parameters of the AGN (e.g., BH mass, and accretion rate) remain poorly understood. Disentangling these two components is crucial for understanding the origin of the soft excess, precisely constraining the geometry of the innermost accretion flow, determining the BH spin, and providing us with a more comprehensive understanding of the energy dissipation processes in the vicinity of the BH. 
In addition, resolving the nature of the soft excess will be key to deciphering the physics behind the extreme properties of some sources, such as NLSy1 galaxies.

\begin{figure*}[ht!]
\centering
\includegraphics[width=0.9\textwidth]{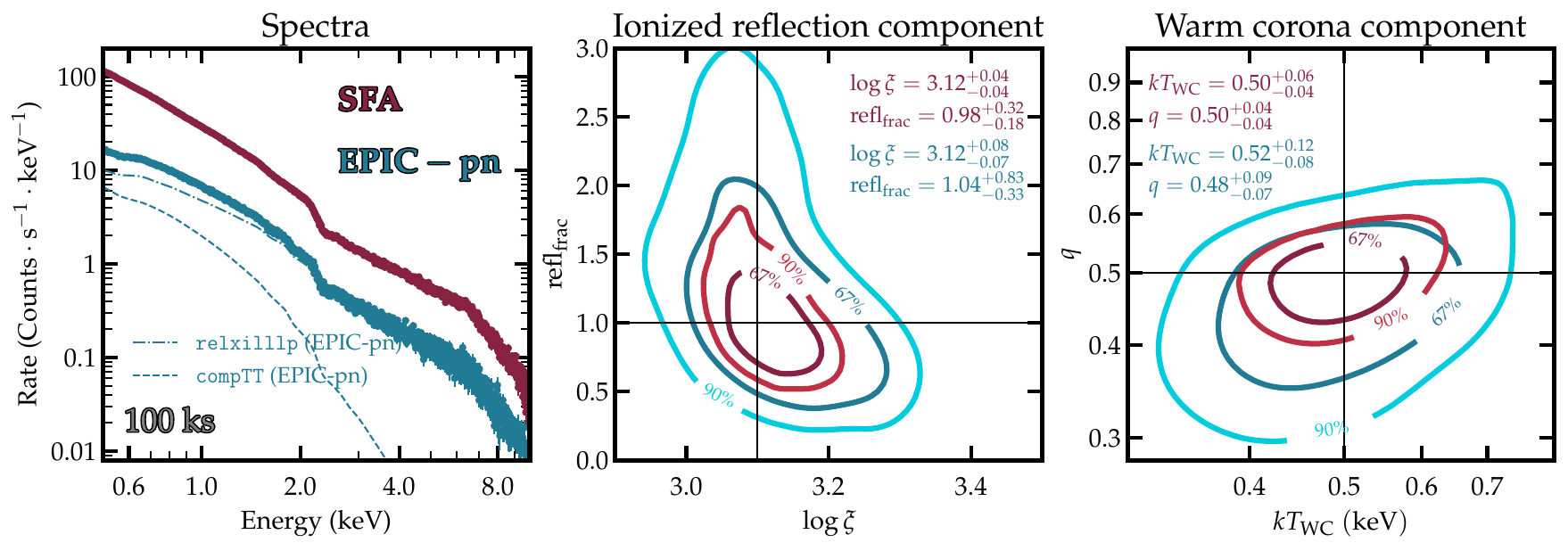}
\caption{ Left: A typical type 1 AGN spectrum as observed by eXTP-SFA (red) and XMM-Newton EPIC-pn (blue) for 100 ks, assuming input model: \texttt{TBabs*zTBabs*(relxilllp+compTT+zgauss)}. For the ionized reflection (along with primary continuum) component, the reflection fraction ($\mathrm{refl}_\mathrm{frac}$) is set to 1, $\log\xi$ to 3.1, photon index $\Gamma$ to 2.0, 0.5-2 keV flux to $10^{-11}\mathrm{erg}\ \mathrm{s}^{-1}\ \mathrm{cm}^{-2}$ (the corresponding 2-10 keV flux is $9.9\times 10^{-12} \mathrm{erg}\ \mathrm{s}^{-1}\ \mathrm{cm}^{-2}$), while other parameters kept at their default values. For the warm corona component we set $kT_\mathrm{WC}$ to 0.5 keV, optical depth to 10, and 0.5-2 keV \texttt{comptTT}-to-\texttt{relxilllp} luminosity ratio ($q$) to 0.5. 
Middle and Right: Using the same parameter values as input, we simulate 1000 fake spectra and re-fit with the same model. The middle panel shows the probability contour for $\mathrm{refl}_\mathrm{frac}$ and $\log\xi$, while the right panel shows the contour for $kT_\mathrm{WC}$ and $q$. The numbers on the contour indicate the portion of cases enclosed, defining the corresponding confidence region. Notably the error bars from eXTP SFA are generally half as small as those of XMM-Newton EPIC-pn.
}
\label{fig:SE_simulation} 
\end{figure*}

With its unprecedented effective area, the eXTP-SFA offers a unique opportunity to address this problem. The left panel of Figure~\ref{fig:SE_simulation} compares the simulated X-ray spectrum of 
an AGN resembling the NLSy1 galaxy Mrk 335, with a soft X-ray flux of $\sim 10^{-11} {\rm erg\ s^{-1} cm^{-2}}$, observed over 100 ks with XMM-Newton EPIC-pn and eXTP SFA. Both simulated spectra are obtained by assuming the same hybrid model. This model consists of a linear combination of emission from the warm corona (\texttt{compTT}) and ionized reflection (\texttt{relxilllp}). While the SFA spectrum clearly exhibits a substantial improvement in signal-to-noise, the middle and right panels of Figure~\ref{fig:SE_simulation} further illustrate the significant enhancement in parameter constraints enabled by the SFA. These panels present confidence contours for key parameters of the ionized reflection component (i.e. the reflection strength $\mathrm{refl}_\mathrm{frac}$ and the ionization parameter $\log\xi$) as well as for the warm corona component (i.e. the soft X-ray, $0.5–2$ keV, warm corona-to-ionized reflection flux ratio $q$ and the warm corona temperature $kT_\mathrm{WC}$). With an effective area four times larger than that of XMM-Newton EPIC-pn, eXTP-SFA reduces parameter uncertainties by approximately half, demonstrating its superior capability for constraining AGN X-ray emission mechanisms.
SFA spectra can advance our understanding of the heating and dissipation processes in warm coronae \citep{Palit2024}
and the geometric changes in the warm corona as a function of
accretion rate \citep{Palit2024,Hagen2024}. 
In addition, the reflected emission is expected to be intrinsically polarized up to $\sim30\%$ \citep[e.g.][]{Dovciak2004,Marin2018,Ursini2022}. Under certain conditions (relatively high inclination and/or high reflection fraction), the reflection component may contribute a significant fraction of the measured polarization degree in the $2-10$ keV. 
For instance, some NLSy1 galaxies exhibit sharp drops in X-ray flux that coincide with the emergence of complex spectral features, which have been attributed to a dominant reflection component \citep[e.g.][]{Gallo2006}. 
Detailed spectro-polarimetric analysis with simultaneous eXTP-PFA and SFA data will allow the contribution of reflection to be more precisely estimated. For example, spectropolarimetric simulations based on a simple toy model (including the same spectral components as those shown in Figure~\ref{fig:SE_simulation}, but incorporating polarization in the reflection component using \verb|KYNSTOKES| \citep{Dovciak2004,Podgorny+2023}, while assuming that the hot / warm corona emission is not polarized) suggest that simultaneous eXTP-PFA and SFA observations could further reduce statistical uncertainty in the reflection strength by 10–20\%.

$\bullet$ {\it What is the geometry of the hot corona?}

The dominant contribution to the X-ray spectrum of radio-quiet AGN originates from the hot corona, whose physical nature is currently unknown. The proposed models mainly invoke magnetic processes, suggesting that the corona might be a compact region powered by magnetic reconnection events \citep[e.g.][]{Beloborodov1999}, the base of a (failed) jet \cite{Ghisellini2004}, or a slab-like structure sandwiching the accretion disk. Each of these models predicts distinct coronal geometries, making it crucial to obtain strong observational constraints on its structure to uncover its physical origin.

Constraints on coronal geometry can be obtained through various methods. However, previous attempts based on X-ray spectral and timing analysis have encountered significant degeneracies \citep[e.g.][]{Bambi+2021}, which limits the ability to derive robust conclusions about the corona's size, shape, and location relative to the accretion disk. Although larger effective area detectors, such as the eXTP-SFA, and advances in spectral-timing models offer promising improvements, overcoming these degeneracies may remain a challenge. A breakthrough in this regard has been the advent of X-ray polarimetric analysis with IXPE \cite{weisskopf22}. Since polarization properties are highly sensitive to the geometry of the emitting region \citep[e.g.][]{Ursini2022}, measuring the polarization degree and angle of X-ray emission in radio-quiet AGN provides a powerful tool to break the degeneracies inherent in spectral and timing analyses alone. By combining polarization data with traditional X-ray observations, it is possible to place much stronger constraints on the structure and orientation of the corona. However, current efforts to constrain the geometry of the corona in type 1 radio-quiet AGN through polarimetric analysis of IXPE data have faced several challenges. One major limitation is IXPE’s relatively small effective area, which has allowed for significant constraints on a few of the brightest Seyfert galaxies in the local universe \citep[e.g.][]{Gianolli2024}. 
Additionally, achieving robust constraints on the different spectral components requires a strong synergy with other X-ray telescopes to obtain simultaneous high-quality X-ray spectra.

The advent of eXTP represents a major step forward in spectro-polarimetric analysis. With an effective area approximately five times that of IXPE, eXTP-PFA will significantly enhance the sensitivity and precision of polarimetric measurements, reducing uncertainties and allowing the detection of polarization signals in weaker sources. Furthermore, the simultaneous observations provided by eXTP-PFA and SFA will enable comprehensive spectro-polarimetric fits for a broad range of AGN, eliminating the need to rely on external instruments for complementary spectral information. This unprecedented capability will offer a more complete and self-consistent picture of the coronal geometry and emission mechanisms in AGNs.

$\bullet$  {\it What is the origin of the observed variability in AGN?} 

AGNs are highly dynamic systems where physical components undergo continuous changes, giving rise to extreme variability phenomena. The X-ray flux from AGN exhibits stochastic, large-amplitude variations over a broad range of timescales, from years down to just a few minutes, the latter corresponding to the light-crossing time of one gravitational radius for typical local radio-quiet AGN \citep[e.g.][]{Uttley2002,Vaughan_2003,McHardy2006}.  
This variability is most commonly attributed to perturbations in the mass accretion rate, which modulate shorter-timescale fluctuations \citep[e.g.][]{Uttley2005}. 
However, additional factors, such as changes in coronal properties and sporadic absorption events \citep[e.g.][]{Parker2017,Alston2020,DeMarco2020}, can further contribute to flux variations. These mechanisms may even produce extreme variability events, in which the observed flux changes by one or more orders of magnitude on a timescale of months to years \citep[e.g.][]{Kaastra2014,Kaastra2018}. In this context, time-resolved spectral studies and spectral time analysis techniques \citep[e.g.][]{Uttley2014} are essential to unravel the physical origins of these fluctuations and unravel the various physical contributions to the variability of AGN. The large effective area of eXTP SFA will make such investigations possible with unprecedented accuracy.

Over the past two decades, an increasing number of AGN have been observed to exhibit large-amplitude variability that cannot be easily explained by intrinsic local stochastic perturbations in the accretion rate. In many cases, these events appear to be driven by extrinsic phenomena, such as obscuration by low-ionization clumps (potentially originating from the dusty torus or the outer BLR) crossing our line of sight to the X-ray source \citep[e.g.][]{Kaastra2014,Kaastra2018}. In other cases, known as ``changing-look AGN'' (CLAGN, e.g. \citenum{Ricci2023}), extreme increases or drops in optical and X-ray flux, on timescales of months to years, are accompanied by the appearance or disappearance of the broad optical/UV emission lines. 
It is not yet clear if such transitions are driven by global changes
in accretion rate or locally-operating instabilities such as the
H-ionization \citep{Noda2018} or radiation pressure
\citep{Sniegowska2023} instabilities.  Nonetheless, such events
provide observational constraints on AGN duty cycles, and allow us to
explore accretion flow state changes analogous to those occurring in
BHXRBs \citep{Ruan2019}.  

Because changes in the ionizing X-ray flux can have a significant impact on the BLR, fully understanding the CLAGN phenomenon requires not only constraining the contribution from the standard accretion disk but also tracking variations in the warm and hot Comptonization coronae. As discussed earlier, combined eXTP-PFA and SFA observations will be crucial in constraining the two contributions, thus deciphering the intrinsic physics behind changing-look events. In particular, the large effective area and the ability to perform simultaneous spectro-polarimetric fits will enable precise detection of variations in key physical parameters and coronal geometry. At the same time X-ray spectral-timing analysis will offer critical insights into the mechanisms that drive these dramatic events, 
potentially revealing transitions in the accretion state of the source \citep[e.g.,][]{Lyu2021}. 

Although the Galactic Center black hole (Sgr A$^*$) is currently in a quiescent state, it remains a unique target for the AGN topic \cite{mou2023}. X-ray polarization and Fe K$\alpha$ echo studies reveal that Sgr A$^*$ experienced luminous flares $100-200$ years ago \cite{ponti2010,marin2023}. The abundant molecular clouds near the Galactic Center provide essential conditions for probing historical X-ray outbursts during the past $10^{2-3}$ years. Such accretion rate variability could fill a gap bridging short-term variability (years) and long-term evolutionary phases (millions of years).  

\begin{figure*}[ht!]
\centering
\includegraphics[width=\textwidth]{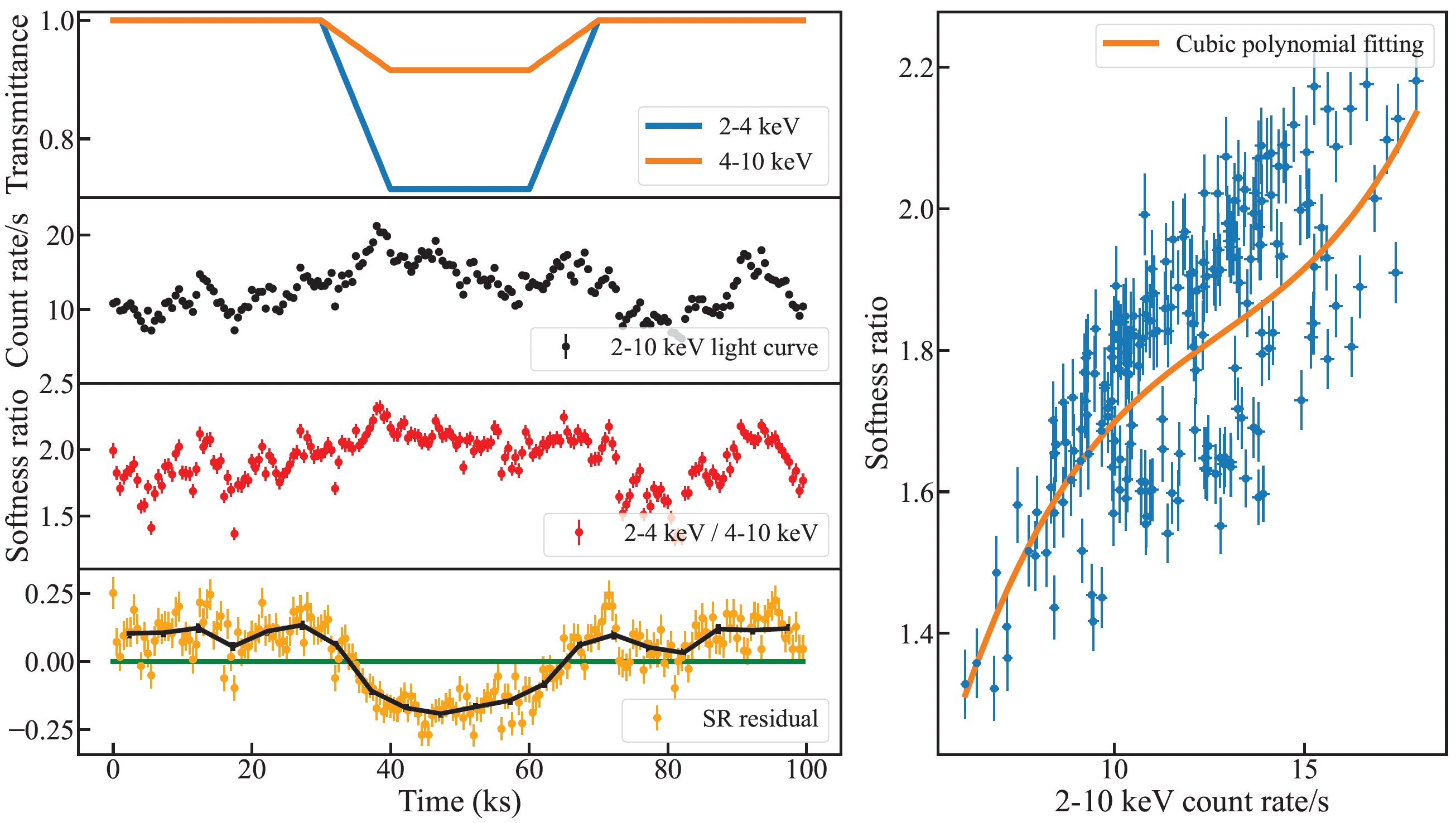}
\caption{\label{fig:eclipse} A method to search for eclipse events with SFA high quality light curves. Left panel, from top to bottom: 1) the transmittance curves of 2--4 keV and 4--10 keV; an eclipse event happens at 30--70 ks. 2) the 2--10 keV count rate curve. 3) the 2--4 keV/4--10 keV softness ratio curve. 4) the softness ratio residual curve, derived by subtracting the data with the empirical relationship at the right panel; black points show the smoothed curve. Right panel: the CR-SR curve and its cubic polynomial fitting result. The simulation is conducted with the method of \cite{Wu_2020} and the response files of SFA. 
}
\end{figure*}

$\bullet$  {\it What causes circumnuclear obscuration in AGN?} 

Rapid variations of the neutral absorption column density ($N_{\rm H}$) on timescales ranging from a few hours to days are commonly observed in the X-ray spectra of bright, local, radio-quiet AGNs \citep[e.g.][]{Torricelli-Ciamponi2014,Maiolino_2010,Sanfrutos_2013,Markowitz2014}. 
The most extreme cases of such variability manifest themselves as eclipses, where an obscuring cloud partially or fully blocks the X-ray source. Tracking these obscuration events provides critical insights into the structure, distance, and kinematics of the intervening material, offering a unique opportunity to probe the circumnuclear environment of AGN. The large effective area of SFA will enable high time-resolution spectral analysis, allowing the detection of these variations on the shortest timescales, which is essential for constraining the properties of clouds on the smallest orbits around the SMBH (\citep[e.g.][]{Nardini2011}.

Particularly valuable are observations that capture a complete eclipse, i.e. both the ingress and egress phases, as these provide measurements of cloud location, density, size, and even morphology \citep{Maiolino_2010, Kang_2023}.
However, detecting such events remains challenging due to the difficulty in distinguishing eclipses from the intrinsic, short-timescale stochastic variability of AGN. As a result, only a limited number of full eclipses have been reported so far \citep{Lamer_2003, Maiolino_2010, Rivers_2011, Sanfrutos_2013, Markowitz2014, Gallo_2021}.
The superior effective area of eXTP-SFA will significantly increase the rate of eclipse detections, even for weak obscuration events. Figure \ref{fig:eclipse} presents simulations of a 100 ks eXTP-SFA light curve for a source like NGC 4051 (with a $2-10$ keV flux of $2.4\times 10^{-11}\rm{erg\ cm^{-2}\ s^{-1}}$), undergoing a weak eclipse event ($N_{\rm H} =10^{23} \rm cm^{-2}$, $\log{\xi} =1.0$ and $f_{\rm cov} = 0.5$ at maximum obscuration) occurring between 30 and 70 ks. The top-left panel of Figure~\ref{fig:eclipse} illustrates the variation in transmittance at different energies for a power-law continuum with photon index $\Gamma=1.9$.
This weak eclipse does not produce a clearly distinguishable dip in either the count rate or the softness ratio time series (middle-left panels). However, following the method of \citep{Wu_2020}, the correlation between the softness ratio and count rate is fit with an empirical polynomial relation (Figure~\ref{fig:eclipse}, right panel), and the best-fit model is subtracted from the softness ratio data. The lower-left panel of Figure~\ref{fig:eclipse} demonstrates that the eclipse signal becomes apparent in the residuals of the softness ratio.

The ability to detect the weakest eclipses strongly depends on the signal-to-noise ratio of the light curves. Thanks to its large effective area, eXTP-SFA will significantly 
improve the precision of this kind of measurements, enabling the detection of subtle eclipsing events that would otherwise remain hidden in the noise. Consequently, we can accumulate statistics on clouds'
locations and densities so we can test models of cloud
distributions. Example model classes include ensembles of BLR clouds
driving X-ray obscuration \citep{Pietrini2024}, or radiatively-driven
winds that originate in the inner accretion flow and contain numerous
partial-covering, X-ray-obscuring clumps
\citep{Kaastra2014,Mehdipour2017,Mao2022}.  
This will allow for a more complete census of obscuration in AGN, ultimately refining our understanding of their circumnuclear environment.

$\bullet$  {\it Which mechanisms contribute to feedback in radio-quiet AGN?} 

Beyond neutral absorption, AGN spectra also show evidence for the presence of powerful outflows which span a wide range of velocities, ionization states, and distances from the central engine. Such outflows carry momentum and energy, thus they might represent an important feedback mechanism in radio-quiet AGN, playing a crucial role in regulating the growth of both the SMBH and the host galaxy \cite{Fabian2012}. X-ray observations, particularly with the large effective area of SFA, will provide a unique window into the physics of ionized gas outflows, probing the innermost regions where the most energetic processes occur. In this context, ultrafast outflows (UFOs) are particularly intriguing, characterized by extreme velocities, typically 10\%-30\% of the speed of light, and detected as highly blueshifted absorption features primarily in the Fe K band \citep{Tombesi2010}. Their high velocities and ionization parameters suggest that UFOs originate very close to the SMBH. The combination of high velocity and substantial column density implies that UFOs can carry significant accretion-driven kinetic energy, potentially playing a critical role in the AGN feedback \citep{Tombesi2013}.
The UFOs from AGN could produce gamma-ray emission \citep{2021ApJ...921..144A}, which is also a proof of carrying significant kinetic energy.
The diagnostic power of eXTP-SFA will be valuable, particularly for faint AGN with typical $0.5–8$ keV X-ray fluxes around $10^{-12}$ erg cm$^{-2}$ s$^{-1}$. Simulations using a baseline model of a typical radio-quiet AGN -- including a power-law continuum, a reflected component from neutral material (pexrav), and absorption by fast outflowing ionized gas (modeled using a grid of XSTAR photoionized absorption models) -- show that with a 100 ks eXTP-SFA exposure, it is possible to place tight constraints on the column density and ionization parameters of the absorber with low statistical uncertainties.
Tracking the time-dependent ionization response of the winds to the
rapidly variable continuum will allow us to constrain the density,
location, 3-D structure, and kinematics of the winds, and thus pinpoint the wind
launch mechanism and the
wind acceleration zone \citep{Fukumura2010,Giustini2019}.

$\bullet$  {\it What is the nature of QPEs?}

Quasi-periodic eruptions (QPEs) are recurring intense soft X-ray bursts from galactic nuclei. These eruptions last under an hour and recur every few hours to hundreds of hours, often alternating between long and short intervals.  \citep{miniutti2019nine,giustini2020x}. Following the first QPE detection in GSN 069, rapidly increasing discoveries have been made, revealing their association with galaxies hosting low-mass SMBHs.\citep{miniutti2019nine,giustini2020x,arcodia2021x,arcodia2024more,guolo2024x,nicholl2024quasi}.Their thermal X-ray spectra show temperatures ranging from hundreds of eV during eruptions to tens of eV in quiescence \citep{wevers2022host,miniutti2019nine}. Furthermore, the observed decaying quiescent flux in two sources supports a possible connection with previous tidal disturbance events (TDEs) \citep{miniutti2023repeating,nicholl2024quasi}.

QPEs are thought to originate from SMBHs, though their physical mechanism remains unclear. The proposed models include disk accretion instabilities \citep{raj2021disk,pan2022disk,pan2023application,kaur2023magnetically}, objects orbiting around the SMBH \citep{king2020gsn,zhao2022quasi,wang2022model,krolik2022quasiperiodic,lu2023quasi,xian2021x,sukova2021stellar,linial2023unstable,linial2023emri+,franchini2023quasi}, and accretion flows precessing \citep{middleton2025quasi}. The leading scenario involves extreme mass ratio inspirals (EMRIs) with two main classes. The first attributes QPEs to periodic mass transfer between a stellar-mass object on an eccentric orbit and an SMBH, where gravitational wave emission shrinks the orbit, triggering QPEs during pericenter passages \citep{king2020gsn,zhao2022quasi,wang2022model,wang2024Orbital}. The second suggests collisions between stellar objects and post-TDE accretion disks, explaining alternating recurrence times and TDE-QPE connections \citep{xian2021x,linial2023emri+,franchini2023quasi}.

Although QPE theories have advanced, challenges persist. Mass transfer models struggle to explain the alternation of long- and short-short recurrence times of QPE and their TDE associations \citep{miniutti2023repeating,wang2022model,king2023angular,wang2024tidal}. Collision models debate whether stellar objects collide directly with gaseous disks or stripped debris interacts with them \citep{linial2023emri+,linial2023unstable,yao2024star}. Furthermore, the diversity and complexity observed in the long-term evolution of QPE challenge current models \citep{giustini2024fragments,pasham2024alive,zhou2024probing2,linial2023emri+,franchini2023quasi,wang2024Orbital,miniutti2025eppur}. eXTP will advance QPE studies through:

(A) Expanded Sample Size. 

To date, eight QPEs sources have been identified, half discovered in the past year \citep{arcodia2024more,guolo2024x,nicholl2024quasi}, indicating a likely substantial undiscovered population. The upcoming eXTP mission will be pivotal in expanding this sample.

While QPEs currently show an exclusive association with TDEs with only two sources \citep{miniutti2019nine,nicholl2024quasi}, confirming the universality of this connection demands larger observational datasets. Increased sample sizes will also clarify QPE diversity, particularly in long-term evolutionary paths and recurrence patterns (e.g., alternating long-short cycles).

QPE host galaxies spectroscopically resemble TDE hosts, frequently showing evidence of past nuclear activity \citep{wevers2022host,wevers2024quasi}. Theoretical models proposing QPE formation via stars in near-circular SMBH orbits similarly depend on historical galactic activity\citep{pan2021wet,zhou2024probing}. Expanded samples will enable statistically rigorous investigations of QPE population characteristics and origins.

(B) Long-term monitoring. Long-term monitoring of known QPEs sources facilitates the study of their temporal evolution, governed by the SMBH environment, orbital dynamics, and associated accretion disk evolution\citep{linial2024coupled,zhou2024probing2}. This observational approach not only provides direct tests of theoretical models, but also advances understanding of both SMBH environments and TDE disk evolutionary processes.

$\bullet$  {\it What are the key X-ray spectral and variability characteristics of IMBH candidates?}

Accretion of matter onto a SMBH powers the central engine, which is known to play a key role in the formation and evolution of galaxies. 
However, the formation and growth of black hole seeds remain an open question in the context of galaxy evolution. Intermediate-mass black holes (IMBHs, $10^2-10^6$ $M_{\odot}$) fill the gap between the stellar mass BHs and SMBHs and are potential analogs of primordial seeds of SMBHs, making AGN in this BH mass regime particularly interesting targets.
AGNs with such low mass are intriguing sources, as they can potentially provide important clues on the nature of primordial SMBHs. Besides that, they represent the link between stellar mass and SMBHs. The high-quality observations obtained from SFA on-board eXTP will be fundamental to explore the X-ray spectral and timing properties of low-mass AGNs.

A fraction of low mass AGNs were previously detected by the ROentgen SATellite (ROSAT) and noted for their soft X-ray luminosity \citep{Ludlam_etal_2015}. Although short snapshot X-ray observations of many of these objects have been performed with Chandra \citep{Greene_Ho_2007, Desroches_etal_2009, DongRB_etal_2012}, detailed X-ray spectral fitting has thus far only been performed for a sample including less than 15 targets \citep{Dewangan_etal_2008, Miniutti_etal_2009, Ludlam_etal_2015}. These studies primarily focus on the X-ray spectral properties, soft excess emission, and variability of low-mass AGNs. 
Key findings include: 1) the X-ray spectra are similar to those of PG quasars; 2) soft excess emission is detected in some sources; 3) rapid and long-term variability is observed, and 4) optically selected low-mass AGNs exhibit a different variability behavior compared to higher-mass AGNs. 
Recently, some further works have been performed by SRG/eROSITA \citep{2024MNRAS.527.1962B, 2024A&A...681A..97A}.

However, IMBHs, typically fainter by a factor of ~10–100 compared to standard Seyfert 1 galaxies, often fall below the detection sensitivity of current telescopes. In addition, the limited photon counts associated with known IMBH candidates make it difficult to disentangle X-ray spectral components. These limitations remain a major obstacle in systematically characterizing the X-ray properties of a large sample of IMBHs \citep{Greene_Ho_2004, Greene_Ho_2007}. The eXTP mission, promises a breakthrough in this field. With an effective area four times larger than XMM-Newton, the SFA will enable more 
efficient detections of IMBHs. Moreover, the large photon collection capabilities will allow for more accurate spectral modeling. This will be crucial for disentangling and characterizing the main X-ray spectral components in individual IMBH candidates, testing disc-corona models \citep{Haardt1993}, and advancing our understanding of accretion physics in this intermediate mass regime.

X-ray variability has become apparent in IMBHs and can be considered to be a defining characteristic of these targets. The timescale of variability shown by IMBH candidates spans from minutes to hours on a short timescale and months to years on a long timescale \citep{Ludlam_etal_2015}. The two well-known prototypical IMBH candidates, NGC 4395 and POX 52, are seen to be rapidly variable in X-rays, with strong amplitude variability \citep{Cameron2012, Kawamuro2024}, consistent with scaled-down type 1 Seyferts. 
The high time resolution of $2 ~\rm \mu$s and large effective area provided by eXTP SFA will allow accurate characterization of the short-term variability of these sources. 
For rapidly variable targets, such as NGC 4395, the SFA's long uninterrupted exposures 
will enable users to construct the PSD (power spectral density) over a broad frequency range, which is crucial for accurately identifying key features, such as characteristic break frequency timescales, and study their correlation 
with fundamental parameters, such as the BH mass and the mass accretion rate. Since this requires high signal-to-noise observations, the large effective area of eXTP-SFA will provide an opportunity to robustly examine the PSD. While single long exposures 
can be used to study the short-term flux variations, 
eXTP multi-epoch observations will enable accurate studies of the long-term X-ray variability, to investigate the underlying physical mechanisms driving these changes, thus shedding light on the structure of the accretion flow.


\subsection{Radio-Loud AGNs}


The production of relativistic jets from AGNs is intricately linked to the properties of the central SMBH and the surrounding accreting material. 
X-ray observations with eXTP, together with observations at other wavelengths will provide crucial information to investigate radiation mechanisms, particle acceleration processes,  hydrodynamical/magneto-hydrodynamical properties, and magnetic field features in radio-loud AGNs. 
The key questions that will be addressed are: 


$\bullet$ 
{\it What are the radiation and particle acceleration mechanisms in jets? 
}

The emission from radio-loud (RL) AGNs spans nearly the entire electromagnetic spectrum and is predominantly attributed to radiation from their relativistic jets. 
The broadband spectral energy distributions (SEDs) of blazars typically exhibit a double-peaked structure. The first (lower frequency) peak, with peak frequency observed at infrared to X-ray wavelengths, and the second (higher frequency) peak, from MeV to GeV. 
The first peak is attributed to synchrotron radiation from relativistic electrons, while the origin of the second peak remains debated. Proposed mechanisms include inverse Compton (IC) scattering by relativistic electrons \citep[e.g.,][]{1992ApJ...397L...5M,1996MNRAS.280...67G,2009ApJ...704...38S,2012ApJ...752..157Z,2014ApJ...788..104Z} or emission from hadronic processes \citep[e.g.,][]{1993A&A...269...67M,2000NewA....5..377A,2003APh....18..593M}.

Blazars are classified into two categories based on their emission line properties: flat spectrum radio quasars showing strong emission lines and BL Lac objects showing only weak emission lines or a featureless continuum \citep{Urry1995PASP}. A different classification into three categories has been proposed, based on the frequency of the synchrotron peak ($\nu_{\rm s}$): low-synchrotron-peaked blazars (LSPs; $\nu_{\rm s}<10^{14}$ Hz), intermediate-synchrotron-peaked blazars (ISPs; $10^{14}<\nu_{\rm s}<10^{15}$ Hz), and high-synchrotron-peaked blazars (HSPs; $\nu_{\rm s}>10^{15}$ Hz) \citep{2010ApJ...710.1271A}, see \citep{Fan2016ApJS,Yang2022ApJS}. 
The origin of X-ray emission varies across blazar classes.
Tight constraints are expected to come from the investigation of multi-band intensity correlations, especially the correlations between the radio and the X-ray band, and the X-ray band and the TeV band.
The detection of $\gamma$-ray emission from the large-scale jets of radio galaxies (RGs) \citep[e.g.,][]{2010Sci...328..725A,2016ApJ...826....1A,2024ApJ...965..163Y} confirms that some 
substructures also serve as sites for high-energy particle acceleration. 
However, the mechanisms governing particle acceleration and radiation, along with the composition of the jets, remain subjects of ongoing debate. Utilizing observations from eXTP and other multiwavelength detectors can provide valuable insights into the fundamental physics of relativistic jets in AGNs.

The synchrotron peak frequency ($P_2$) and the curvature parameter ($p_1$) in the flux-frequency plane are linked by the relation $\log\, \nu f_{\nu} = p_1 (\nu - p_2)^2 + p_3$ \citep{Fan2016ApJS, Yang2022ApJS}. This kind of relations can be used to investigate the particle acceleration mechanism, provided simultaneous radio to X-ray observations are available. Synergic multi-wavelength campaigns with eXTP and both ground-based and space telescopes will be key to answer questions regarding the acceleration mechanism during the flaring and quiescent states \citep{Xiao2024ApJ966}, as well as to better constrain the broad-band contribution from the different physical components (disk, synchrotron, and inverse Compton).

$\bullet$  {\it What can we learn from blazars flaring/variability detected by eXTP?}

Flares represent prominent observational phenomena during active phases of blazars. They are characterized by considerable flux variations over a wide range of timescales, from months to years \citep[e.g.][]{impiombato2011,chen2013}, down to hours \citep[e.g.][]{acciari2011,hayashida2015}, or even minutes \citep[e.g.][]{Fan2005ChJAA,xue2005,albert2007}. Flares represent an important feature of blazar emission across the entire electromagnetic spectrum, which can provide direct insights into the microphysical processes in jets, such as particle acceleration mechanisms and energy dissipation. 
The physical origin of rapid flares remains debated, though these events are presumably due to strong and rapid energy dissipation in a relatively compact region within jet \citep[e.g.][]{wagner1995,rani2013}.
eXTP, with its high time resolution of $2-10$ $\mu$s, will allow the detection and optimal sampling of the most extreme flux variations occurring in blazars. At the same time, eXTP's spectral capabilities will enable dynamic tracking of spectral variations during these extreme flaring events, providing insights into the time-dependent evolution of the electron distribution. Moreover, by correlating the X-ray and $\gamma$-ray 
spectral properties and studying their evolution it will be possible to differentiate between SSC and EC contributions to the high-energy emission. Finally, time-dependent spectral modeling across different energy bands will place constraints on the location and physical properties of the emission regions.

eXTP-SFA will also provide the opportunity to estimate the Doppler factor ($\delta$). The $\delta$-factor is an important parameter for radio loud AGNs, as it determines the amount of intrinsic flux enhancement and the apparent decrease of variability timescales. However, it can only be estimated using indirect methods \citep{Ghisellini1993ApJ,Readhead1994ApJ,1999AJ....117.1168L}. As shown in \citep{Fan2013RAA}, assuming that the X-ray and $\gamma$-ray emission come from the same emitting region, combined measurements of the X-ray spectrum and the corresponding variability timescale, can be used to estimate $\delta$. 

$\bullet$ 
{\it 
Which physical mechanisms can be constrained by eXTP-PFA polarization measurements?}

{\it Magnetic field structure, jet formation, and jet composition:} 
X-ray polarization measurements can provide a new perspective for studying the properties of relativistic jets, including radiation mechanisms, particle acceleration, macro/micro dynamics, magnetic field topology, and geometry of emission region.
For instance, the Imaging X-ray Polarimetry Explorer (IXPE) observed a progressive decrease in the degree of polarization from X-rays to optical and radio wavelengths in Mrk 501 \citep{Liodakis2022Natur611}, 
which has been associated with a shock front as the origin of particle acceleration. 
This result highlights the importance of X-ray observations for our understanding of jet physics and structure. 
The magnetic-driven model posits that the rotational energy of the black hole accretion system is converted into Poynting flux, which subsequently accelerates the jet \citep{Blandford1977, Blandford1982, Chen&Zhang2021ApJ906}. According to this model, a helical magnetic field plays a crucial role in the jet's formation, collimation, and acceleration \citep{Chen&Zhang2021ApJ906}. However, observational evidence supporting this theory remains limited.
IXPE observations revealed polarization angle rotation in Mkn 421 by $\sim$360 degrees, hinting at the presence of a helical magnetic field in this source \citep{DiGesu2023NatAs7}. However, this evidence is still inconclusive, emphasizing the need for further investigations with enhanced sensitivity and high-quality data. 
Targeted studies of High-Synchrotron-Peaked (HSP) blazars using the 
eXTP-PFA will be crucial in providing deep insights into the magnetic field configuration of AGN jets, thereby advancing our understanding of the mechanisms that drive jet formation, collimation, and acceleration.
In addition, thanks to its larger effective area, eXTP-SFA, together with other multi-wavelength facilities, will allow the SED to be constrained, in order to properly infer important physical parameters, such as the electron energy ($\gamma_p$), the magnetic field strength ($B$), and the $\gamma$-ray emission location \citep{2023ApJS..268...23F}.

\begin{figure}[H]
    \centering
\includegraphics[width=0.49\textwidth]{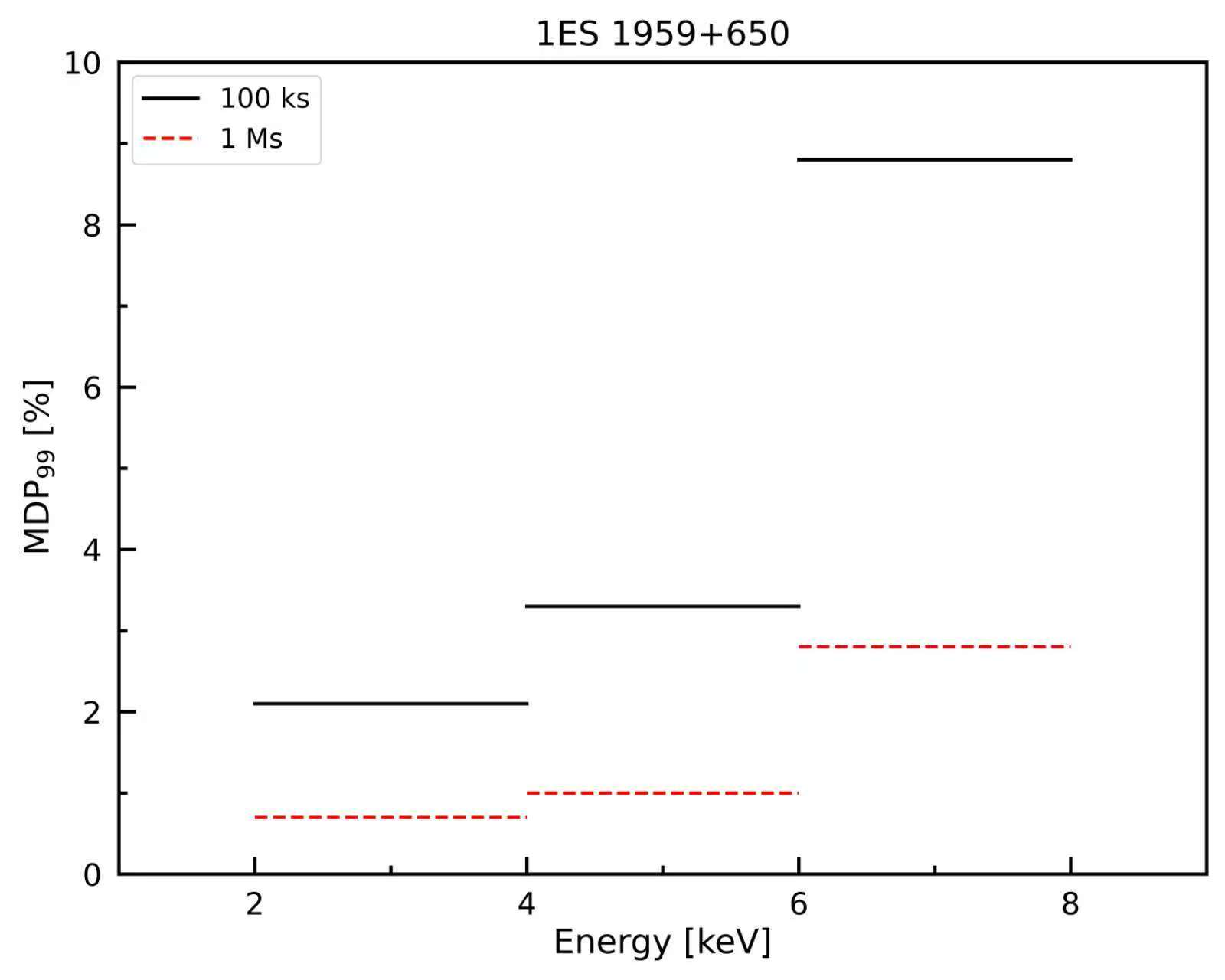}
    \caption{The simulated MDP$_{99}$ of eXTP for a bright object 1ES 1959+650 
    in different energy intervals with different exposures.
    The spectrum of 1ES 1959 is fitted by an absorbed log-parabolic blazar model with the parameters of $a=1.73\pm0.05$, $b=0.48\pm0.05$, and $pivotE=1 $~keV(frozen). 
    }
    \label{fig:polar-blazar}
\end{figure}

With a large effective area, eXTP will provide high-quality X-ray spectral and polarization measurements for RLAGNs, which is key to determining whether hadronic models should be invoked to explain the observed emission. In particular, in the case of the IC process, X-ray PD is expected to be significantly lower than that of the seed photons \citep{Bonometto1973A&A23, Nagirner1993A&A275, Poutanen1994ApJS92, Liodakis2019ApJ880, Peirson2019ApJ885}. Conversely, for hadronic interactions, X-ray PD should be similar to or even higher than that observed in radio or optical wavelengths (e.g., in pure proton-synchrotron models) \citep{Zhang2013ApJ774, Paliya2018ApJ863, Zhang2019ApJ876}. 

eXTP-PFA will enable unprecedented energy-dependent polarimetric studies. To demonstrate this, we consider the BL Lac object, 1ES 1959+650, with X-ray flux of about $10^{-11}~\rm{erg~cm^{-2}s^{-1}}$ \citep{2014styd.confE.142K}, and linear polarization degree of $8.0\%\pm2.3\%$ as constrained by IXPE \citep{2024ApJ...963....5E}. Simulations of the expected $\rm{MDP}_{99}$ as a function of energy for 1ES 1959+650 are shown in Figure~\ref{fig:polar-blazar} for different exposure times, demonstrating PFA's unprecedented sensitivity to polarization. Here, MDP means the minimum detectable polarization.

The X-ray emission 
of some substructures in large-scale jets 
is thought to originate from synchrotron radiation \citep[e.g.,][]{1987ApJ...314...70R,1999AJ....117.1168L}. However, this emission is apparently not a simple extrapolation of the radio-to-optical synchrotron spectrum and requires an additional component for explanation. Such component could either arise from ICS \citep{2006ARA&A..44..463H} or synchrotron radiation from a distinct population of high-energy particles \citep[e.g.,][]{2015ApJ...805..154M,2020ApJ...903..109T}. 
High sensitivity X-ray polarization observations will be a powerful tool to elucidate the X-ray radiation mechanism, particularly for the substructures with highly ordered magnetic fields, such as the western hotspot (WHS) in RG Pictor A \citep[e.g.,][]{1987ApJ...314...70R,1995ApJ...446L..93T}.
The WHS in Pictor A is a faint emission region with X-ray flux of $\sim 10^{-13}~\rm{erg~cm^{-2}~s^{-1}}$. According to eXTP-PFA simulations, the $\rm{MDP}_{99}$ will be about 30\% and 60\% in the energy bands of $2-4$ keV and $4-6$ keV, respectively. 
Although the polarization measurement for the detailed structure in a global jet is important to investigate the magnetic field and dynamical properties in a non-uniform jet, it is a challenge for the eXTP polarization to detect such weak emission.

Rapid changes in polarization degree during a blazar flare may indicate localized energy release and particle acceleration, possibly triggered by the radiative magnetic reconnection or turbulence \citep{2015ApJ...815..101N, 2023MNRAS.523.3812S, 2024ApJ...964...14D, 2014ApJ...780...87M, 2022ApJ...936L..27C}. 
In particular, it is interesting to investigate the critical Synchrotron/SSC transition regions of ISPs. During flaring states, the synchrotron peak may shift to higher energy bands, leading to changes in X-ray emission polarization as flux varies, transitioning from unpolarized to significantly polarized states. Energy-dependent polarization analysis during these transitions may reveal the Synchrotron/SSC transition regions within the broadband SED of ISPs \citep{2023ApJ...948L..25P}.
According to simulations shown in Figure~\ref{fig:polar-blazar}, eXTP-PFA will be able to perform such polarization measurements at least for the brightest blazars (with X-ray flux of $\sim10^{-11}~\rm{erg~cm^{-2}~s^{-1}}$). At the same time, a comparative analysis of polarization variations across different 
wavebands can 
shed light on the energy dissipation across different timescales. 

\section{Pulsar-based positioning and time-keeping} \label{sec:pos}

The advent of interplanetary exploration has exposed critical limitations in conventional deep-space navigation and systems. Traditional radio-based tracking networks, while effective for near-Earth missions, suffer from escalating positional errors, signal attenuation near the Sun, and latency exceeding hours for distant spacecraft. These constraints hinder real-time maneuverability and mission reliability, particularly in scenarios requiring rapid decision-making. In deep space applications, where GPS/BEIDOU signals are unavailable, spacecraft must rely on on-board atomic clocks for precise timekeeping. However, atomic clocks that cannot be calibrated will gradually accumulate timing errors, and these deviations grow quickly over time.  This underscores the necessity of autonomous calibration technologies or multi-clock redundancy systems for long-duration missions beyond Earth’s orbit.

Against this backdrop, X-ray Pulsar Navigation (XPNAV) has emerged as a promising solution. Leveraging millisecond pulsars (MSPs)—neutron stars emitting highly stable X-ray pulses with timing stability rivaling atomic clocks \citep{1991IEEEP..79.1054T, 1997A&A...326..924M, 2024Univ...10..174Z}—this technology enables autonomous positioning and time-keeping. Unlike earth-dependent systems, pulsar navigation provides subkilometer-level 3D positioning and submicrosecond timing synchronization across the solar system, with precision independent of distance. Early demonstrations, such as China’s Tiangong-2 space station \citep{2017SSPMA..47i9505Z}, the Insight-HXMT telescope \citep{2019ApJS..244....1Z}, the XPNAV-1 telescope \citep{2019JATIS...5a8003H} and the NICER onboard ISS \citep{2020AcAau.176..531Y}, demonstrated feasibility in low Earth orbit, achieving positional errors less than 5 kilometers. In addition, atomic clock timekeeping innovations now enable drift correction via pulsar timing residuals \citep{2020MNRAS.491.5951H}, eliminating reliance on GPS/BDS signals or long-time drift of the atomic.

The eXTP, with its unprecedented effective area and microsecond-level timing resolution, establishes a robust experimental platform for X-ray pulsar navigation and timekeeping. It will integrate embedded algorithms for autonomous orbit determination and time-keeping, laying the foundation for its future in-orbit applications.

To achieve pulsar-based positioning and timekeeping, we propose and evaluate an in-orbit observation strategy. Six MSPs with exceptional timing stability \citep{2024Univ...10..174Z} are selected, including PSRs B1937+21, B1821$-$24, J0437$-$4715, J0030+0451, J0218+4232, and J2124$-$3358. A sequential observation scheme is implemented, in which these pulsars are observed in a predefined cyclic order to accumulate X-ray photon time series. Positioning and time-keeping solutions are subsequently derived through time-of-arrival (TOA) measurements of their pulsed signals.

The simulation incorporates the SFA telescopes with their response matrix and employs an elliptical orbital configuration. Precise astrometric and timing parameters for the six MSPs (positions, spin frequencies, first- and second-order frequency derivatives) are derived from NICER observational data \citep{2024Univ...10..174Z}. The eXTP spacecraft’s onboard atomic clock is modeled with the following specifications:  frequency accuracy: $\pm 5\times 10^{-12}$, frequency stability: $1\times 10^{-12}~\rm s^{-1}$, and frequency drift rate: $2\times 10^{-12}~\rm day^{-1}$.

While operational constraints occasionally render specific pulsars temporarily unobservable, these observational gaps are strategically neglected in our analysis as they do not significantly compromise the overall solution quality.

\vspace{0.2in}
$\bullet$ {\it pulsar-based Positioning results}

In this simulation, the six MSPs are observed sequentially, each with 1000 s without clock errors considered. Using the measured times of arrival (ToAs) of the pulsars and the spacecraft precision orbit dynamics model, the states of the spacecraft (i.e. position and velocity) are estimated by the Unscented Kalman Filter (UKF)\cite{WANG202344}\cite{WANG201427}. The simulation lasts 70 days, and the initial state error of the spacecraft is set as (100km, 100km, 100km, 10m/s, 10m/s, 10m/s).

\begin{figure*}
    \centering 
   \includegraphics[width=0.9\textwidth]{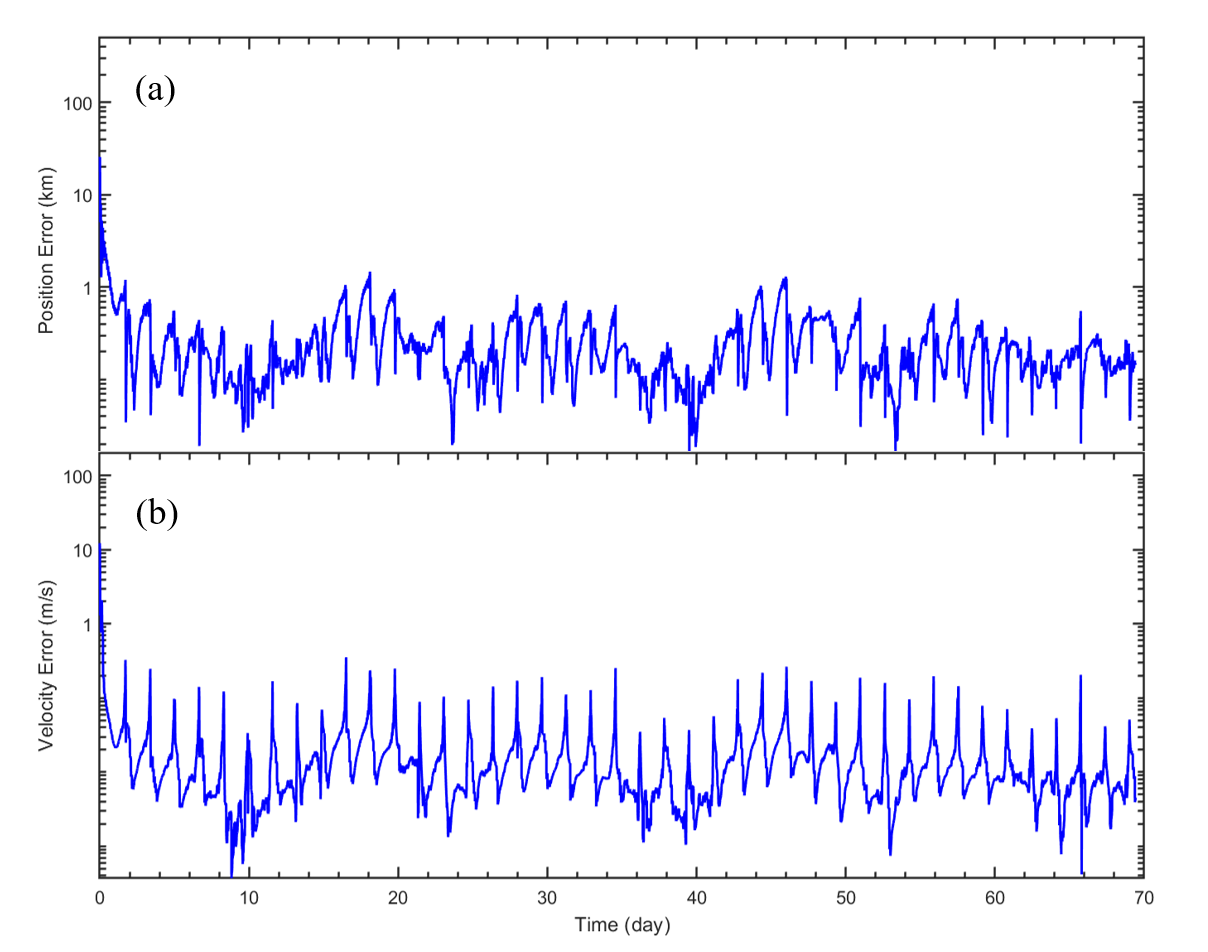}
    \caption{The 70-day simulation results of pulsar-based positioning: (a) the error of estimated position and (b) the error of estimated velocity.}
    \label{fig:pulsar2}
\end{figure*}

\begin{figure*}
    \centering
   \includegraphics[width=0.95\textwidth]{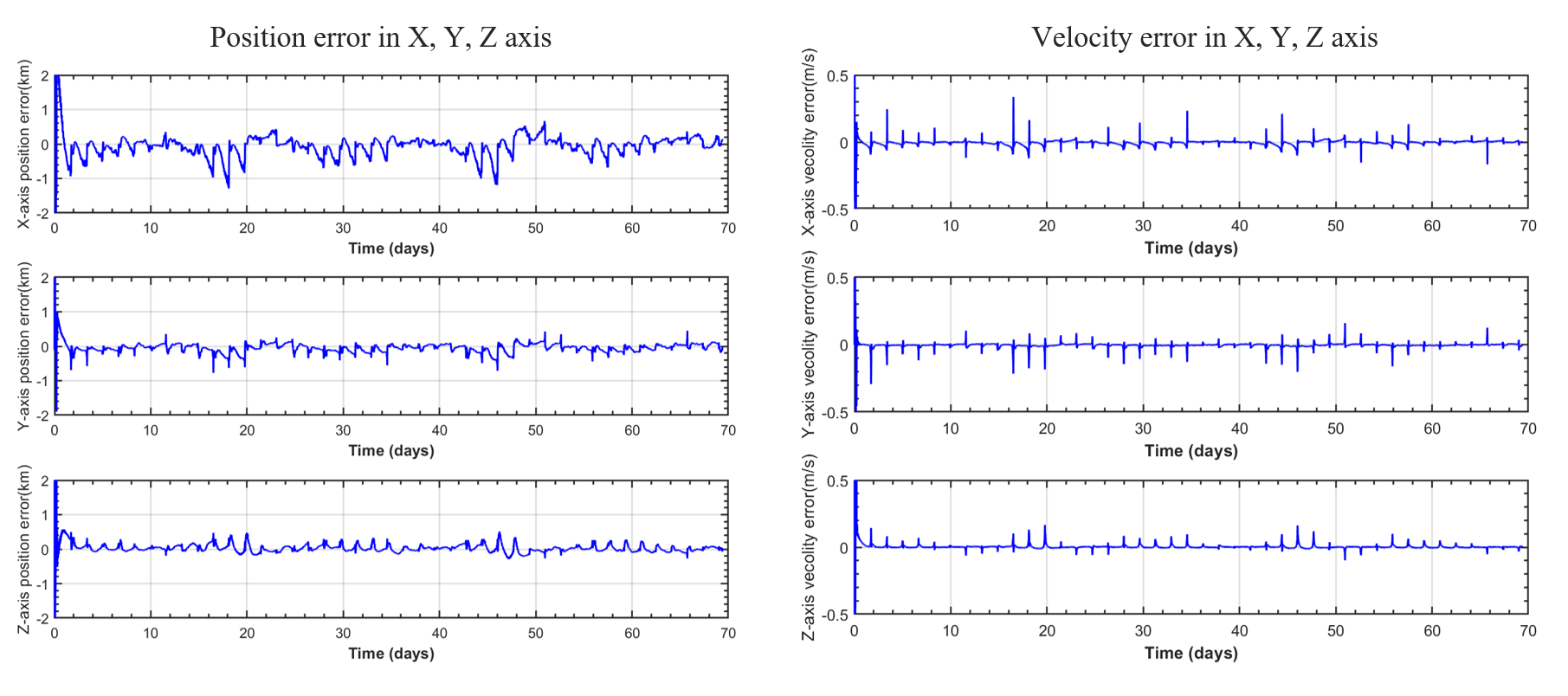}
    \caption{Position (left) and velocity (right) errors along X, Y, Z axes}
    \label{fig:pulsar4}
\end{figure*}

Figure \ref{fig:pulsar2} shows the position and velocity errors (the difference between the estimated states of the spacecraft and its real states). As shown in Figure \ref{fig:pulsar2}, the navigation performance is better than 2 km. The position error quickly converges within the first day and stabilizes below 2 km. In the same figure below, the velocity error also quickly reduces to and maintains below 1 m/s. The root mean square (RMS) of the position error and the velocity error are 0.6386 km and 0.2165 m/s, respectively. The position and velocity errors along the X, Y, Z axes in the Earth-centered Inertial Coordinate System J2000·0 are shown in Figure \ref{fig:pulsar4}. It can be seen that the position errors in each axis maintain below 1.5 km and the velocity errors in each axis maintain below 0.5 m/s after converge.

\vspace{0.2in}
$\bullet$ {\it pulsar-based time-keeping results}

In the absence of regular calibration and synchronization, onboard atomic clock timing errors exhibit progressive accumulation. The timing error $X(t)$ is modeled as:

\begin{equation}
X(t) = X_0 + V_x (t-t_0) + \frac{1}{2}G_x (t-t_0)^2 + \sigma(t)
\label{eq:timing}
\end{equation}
Where $X_0$ denotes the initial timing offset at epoch $t=t_0$, $V_x$ represents the frequency accuracy, $G_x$ indicates the frequency drift rate, and $ \sigma(t)$ characterizes red noise contributions from clock instability. For the eXTP atomic clock specifications provided, simulated timing errors grow to 0.5\, $\mu$s, 13\, $\mu$s, 90\, $\mu$s, for 1\,day, 10\,days, and 30\,days, respectively.

\begin{figure*}
    \centering
   \includegraphics[width=0.6\textwidth]{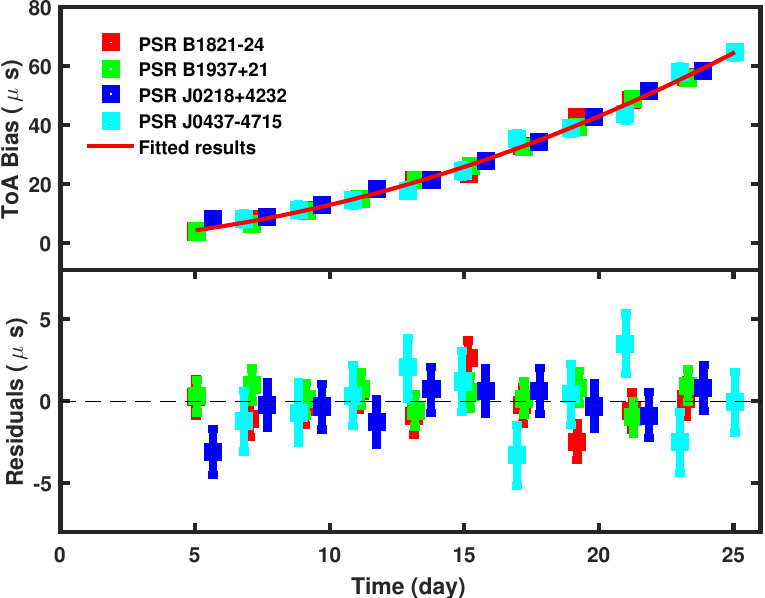}
    \caption{The ToA bias and fitting with atomic parameters. The red, green, blue and cyan squares represent PSRs B1821$-$24, B1937+21, J0218+4232 and J0437$-$4715, respectively. The red line is fitted result with atomic parameters.}
    \label{fig:clock}
\end{figure*}

We simulate the case that four MSPs (PSRs B1821$-$24, B1937+21, J0218+4232, J0437$-$4715) are observed sequentially with exposure times 20\,ks, 5\,ks, 50\,ks, and 100\,ks for each time,  achieving the TOA measurements with approximately 1\, $\mu$s precision. It is noted that these observations can be performed in an interrupted manner, provided that a sufficient amount of time is accumulated to achieve a highly accurate calculation of the ToA. 
As illustrated in Figure \ref{fig:clock}, systematic TOA biases emerge from uncorrected clock drift with $t_0=62533.0$ (UTC). By fitting the atomic clock parameters to these observations, the atomic clock bias can be corrected. The frequency accuracy is fit with  $5.2 \pm 0.5 \times 10^{-12}$ and the frequency drift rate is fit with $1.98 \pm 0.06 \times 10^{-12}~\rm day^{-1}$. These values demonstrate strong consistency with the eXTP clock's design specifications, validating the ability of pulsar-based time-keeping.

\emph{Conflict of interests.}  The authors declare that they have no conflict of interest.


\emph{Acknowledgements.} 
We thank B.\ Haskell, J.E.\ Horvath, N.\ Dutra Pire, B.B. Martins, M.G.B. de Avellar, L.M.de Sá, D.\ Vasconcelos Rodrigues for providing helpful comments. This work is supported by China's Space Origins Exploration Program. PZ acknowledges the support from the National Natural Science Foundation of China (No.\ 12273010).
S-NZ is supported by the National Natural Science Foundation of China (No. 12333007), the International Partnership Program of Chinese Academy of Sciences (No.113111KYSB20190020) and the Strategic Priority Research Program of the Chinese Academy of Sciences (No. XDA15020100).
JHF is supported by the China National Natural Science Foundation (No. 12433004). AP acknowledges support from grant PID2021-124581OB-I0, PID2024-155316NB-I00 and 2021SGR00426. YC acknowledges support from a Ramon y Cajal fellowship (RYC2021-032718-I) financed by MCIN/AEI/10.13039/501100011033 and the European Union NextGenerationEU/PRTR. BDM acknowledges support via a Ram\'on y Cajal Fellowship RYC2018-025950-I and the Spanish MINECO grant PID2022-136828NB-C44.
BDM, YC and GS acknowledge support from the Spanish MINECO (PID2023-148661NB-I00), the E.U. FEDER funds, and the AGAUR/Generalitat de Catalunya (SGR-386/2021).
LD acknowledges funding from the Deutsche Forschungsgemeinschaft (DFG, German Research Foundation) - Projektnummer 549824807. PR acknowledges support by the “Programma di Ricerca Fondamentale INAF 2023”. 
YH is supported by National Natural Science Foundation of
China (grant No. 12233002), by National SKA Program of China No. 2020SKA0120300, by the National Key R\&D Program of China (2021YFA0718500), and by the support from the Xinjiang Tianchi Program. 
C-YN is supported by a GRF grant of the Hong Kong Government under HKU 17304524.
FX is supported by National Natural Science Foundation of China (grant No. 12373041 and No. 12422306), and Bagui Scholars Program (XF).





\bibliographystyle{scpma-2}
\bibliography{ref.bib, amxps.bib}









\end{multicols}
\end{document}